\documentclass[aps,prstab,twocolumn,superscriptaddress,preprintnumbers,floatfix,nofootinbib]{revtex4-1}
\usepackage{color}
\usepackage{amssymb}
\usepackage{amsmath}

\usepackage{blindtext}
\usepackage[nodots,nocompress]{numcompress}
\usepackage{lineno}
\usepackage{upgreek}
\usepackage[colorlinks=false, pdfborder={0 0 0}]{hyperref}
\usepackage{graphicx}
\usepackage{multirow}
\usepackage{refcount}
\usepackage{upgreek}
\usepackage{dcolumn}
\usepackage{textcomp}

\usepackage{booktabs,dcolumn}

\begin{document}


\title{\hspace{2cm}Toward polarized antiprotons: Machine development for \newline spin-filtering experiments}


\author{C.~Weidemann}
\email{c.weidemann@fz-juelich.de}
\thanks{corresponding author}
\affiliation{Universit\'a di Ferrara and INFN, 44122 Ferrara, Italy}
\affiliation{Institut f\"ur Kernphysik, Forschungszentrum J\"ulich, 52425 J\"ulich, Germany}

\author{F.~Rathmann}
\author{H.J.~Stein}
\author{B.~Lorentz}
\affiliation{Institut f\"ur Kernphysik, Forschungszentrum J\"ulich, 52425 J\"ulich, Germany}
\author{Z.~Bagdasarian}
\affiliation{Institut f\"ur Kernphysik, Forschungszentrum J\"ulich, 52425 J\"ulich, Germany}
\affiliation{High Energy Physics Institute, Tbilisi State University, 0186 Tbilisi, Georgia}
\author{L.~Barion}
\affiliation{Universit\'a di Ferrara and INFN, 44122 Ferrara, Italy}
\author{S.~Barsov}
\affiliation{High Energy Physics Department, St. Petersburg Nuclear Physics Institute, 188350 Gatchina, Russia}
\author{U.~Bechstedt}
\affiliation{Institut f\"ur Kernphysik, Forschungszentrum J\"ulich, 52425 J\"ulich, Germany}
\author{S.~Bertelli}
\affiliation{Universit\'a di Ferrara and INFN, 44122 Ferrara, Italy}
\author{D.~Chiladze}
\affiliation{Institut f\"ur Kernphysik, Forschungszentrum J\"ulich, 52425 J\"ulich, Germany}
\affiliation{High Energy Physics Institute, Tbilisi State University, 0186 Tbilisi, Georgia}
\author{G.~Ciullo}
\affiliation{Universit\'a di Ferrara and INFN, 44122 Ferrara, Italy}
\author{M.~Contalbrigo}
\affiliation{Universit\'a di Ferrara and INFN, 44122 Ferrara, Italy}
\author{S.~Dymov}
\affiliation{Laboratory of Nuclear Problems, Joint Institute for Nuclear Research, 141980 Dubna, Russia}
\author{R.~Engels}
\affiliation{Institut f\"ur Kernphysik, Forschungszentrum J\"ulich, 52425 J\"ulich, Germany}
\author{M.~Gaisser}
\affiliation{Institut f\"ur Kernphysik, Forschungszentrum J\"ulich, 52425 J\"ulich, Germany}
\author{R.~Gebel}
\affiliation{Institut f\"ur Kernphysik, Forschungszentrum J\"ulich, 52425 J\"ulich, Germany}
\author{P.~Goslawski}
\affiliation{Institut f\"ur Kernphysik, Universit\"at M\"unster, 48149 M\"unster, Germany}
\author{K.~Grigoriev}
\affiliation{Institut f\"ur Kernphysik, Forschungszentrum J\"ulich, 52425 J\"ulich, Germany}
\affiliation{High Energy Physics Department, St. Petersburg Nuclear Physics Institute, 188350 Gatchina, Russia}
\author{G.~Guidoboni}
\affiliation{Universit\'a di Ferrara and INFN, 44122 Ferrara, Italy}
\author{A.~Kacharava}
\affiliation{Institut f\"ur Kernphysik, Forschungszentrum J\"ulich, 52425 J\"ulich, Germany}
\author{V.~Kamerdzhiev}
\affiliation{Institut f\"ur Kernphysik, Forschungszentrum J\"ulich, 52425 J\"ulich, Germany}
\author{A.~Khoukaz}
\affiliation{Institut f\"ur Kernphysik, Universit\"at M\"unster, 48149 M\"unster, Germany}
\author{A.~Kulikov}
\affiliation{Laboratory of Nuclear Problems, Joint Institute for Nuclear Research, 141980 Dubna, Russia}
\author{A.~Lehrach}
\affiliation{Institut f\"ur Kernphysik, Forschungszentrum J\"ulich, 52425 J\"ulich, Germany}
\affiliation{III. Physikalisches Institut B, RWTH Aachen University, 52056 Aachen, Germany}
\author{P.~Lenisa}
\affiliation{Universit\'a di Ferrara and INFN, 44122 Ferrara, Italy}
\author{N.~Lomidze}
\affiliation{High Energy Physics Institute, Tbilisi State University, 0186 Tbilisi, Georgia}
\author{G.~Macharashvili}
\affiliation{Institut f\"ur Kernphysik, Forschungszentrum J\"ulich, 52425 J\"ulich, Germany}
\affiliation{Laboratory of Nuclear Problems, Joint Institute for Nuclear Research, 141980 Dubna, Russia}
\author{R.~Maier}
\affiliation{Institut f\"ur Kernphysik, Forschungszentrum J\"ulich, 52425 J\"ulich, Germany}
\author{S.~Martin}
\affiliation{UGS Gerlinde Schulteis and Partner GbR, 08428 Langenbernsdorf, Germany}
\author{D.~Mchedlishvili}
\affiliation{High Energy Physics Institute, Tbilisi State University, 0186 Tbilisi, Georgia}
\author{H.O.~Meyer}
\affiliation{Physics Department, Indiana University, Bloomington, IN 47405, USA}
\author{S.~Merzliakov}
\affiliation{Institut f\"ur Kernphysik, Forschungszentrum J\"ulich, 52425 J\"ulich, Germany}
\affiliation{Laboratory of Nuclear Problems, Joint Institute for Nuclear Research, 141980 Dubna, Russia}
\author{M.~Mielke}
\affiliation{Institut f\"ur Kernphysik, Universit\"at M\"unster, 48149 M\"unster, Germany}
\author{M.~Mikirtychiants}
\author{S.~Mikirtychiants}
\affiliation{Institut f\"ur Kernphysik, Forschungszentrum J\"ulich, 52425 J\"ulich, Germany}
\affiliation{High Energy Physics Department, St. Petersburg Nuclear Physics Institute, 188350 Gatchina, Russia}
\author{A.~Nass}
\affiliation{Institut f\"ur Kernphysik, Forschungszentrum J\"ulich, 52425 J\"ulich, Germany}
\author{N.N.~Nikolaev}
\affiliation{Institut f\"ur Kernphysik, Forschungszentrum J\"ulich, 52425 J\"ulich, Germany}
\affiliation{L.D. Landau Institute for Theoretical Physics, 142432 Chernogolovka, Russia}
\author{D.~Oellers}
\affiliation{Universit\'a di Ferrara and INFN, 44122 Ferrara, Italy}
\affiliation{Institut f\"ur Kernphysik, Forschungszentrum J\"ulich, 52425 J\"ulich, Germany}
\author{M.~Papenbrock}
\affiliation{Institut f\"ur Kernphysik, Universit\"at M\"unster, 48149 M\"unster, Germany}
\author{A.~Pesce}
\affiliation{Universit\'a di Ferrara and INFN, 44122 Ferrara, Italy}
\author{D.~Prasuhn}
\affiliation{Institut f\"ur Kernphysik, Forschungszentrum J\"ulich, 52425 J\"ulich, Germany}
\author{M.~Retzlaff}
\author{R.~Schleichert}
\affiliation{Institut f\"ur Kernphysik, Forschungszentrum J\"ulich, 52425 J\"ulich, Germany}
\author{D.~Schr\"oer}
\affiliation{Institut f\"ur Kernphysik, Universit\"at M\"unster, 48149 M\"unster, Germany}
\author{H.~Seyfarth}
\affiliation{Institut f\"ur Kernphysik, Forschungszentrum J\"ulich, 52425 J\"ulich, Germany}
\author{H.~Soltner}
\affiliation{Zentralinstitut f\"ur Engineering und Technologie (ZEA-1), Forschungszentrum J\"ulich, 52425 J\"ulich, Germany}
\author{M.~Statera}
\affiliation{Universit\'a di Ferrara and INFN, 44122 Ferrara, Italy}
\author{E.~Steffens}
\affiliation{Physikalisches Institut II, Universit\"at Erlangen-N\"urnberg, 91058 Erlangen, Germany}
\author{H.~Stockhorst}
\affiliation{Institut f\"ur Kernphysik, Forschungszentrum J\"ulich, 52425 J\"ulich, Germany}
\author{H.~Str\"oher}
\affiliation{Institut f\"ur Kernphysik, Forschungszentrum J\"ulich, 52425 J\"ulich, Germany}
\author{M.~Tabidze}
\affiliation{High Energy Physics Institute, Tbilisi State University, 0186 Tbilisi, Georgia}
\author{G.~Tagliente}
\affiliation{INFN, Sezione di Bari, 70126 Bari, Italy}
\author{P.~Th\"orngren Engblom}
\affiliation{Universit\'a di Ferrara and INFN, 44122 Ferrara, Italy}
\affiliation{Department of Physics, Royal Institute of Technology, SE-10691, Stockholm, Sweden}
\author{S.~Trusov}
\affiliation{Institut f\"ur Kern- und Hadronenphysik, Forschungszentrum Rossendorf, 01314 Dresden, Germany}
\affiliation{Skobeltsyn Institute of Nuclear Physics, Lomonosov Moscow State University, 119991 Moscow, Russia}
\author{Yu.~Valdau}
\affiliation{High Energy Physics Department, St. Petersburg Nuclear Physics Institute, 188350 Gatchina, Russia}
\affiliation{Helmholtz-Institut f\"ur Strahlen- und Kernphysik, Universit\"at Bonn, 53115 Bonn, Germany}
\author{A.~Vasiliev}
\affiliation{High Energy Physics Department, St. Petersburg Nuclear Physics Institute, 188350 Gatchina, Russia}
\author{P.~W\"ustner}
\affiliation{Zentralinstitut f\"ur Systeme der Elektronik (ZEA-2), Forschungszentrum J\"ulich, 52425 J\"ulich, Germany}

\begin{abstract}

The paper describes the commissioning of the experimental equipment and the machine studies required for the first spin-filtering experiment with protons at a  beam kinetic energy of 49.3\,MeV in COSY. The implementation of a low-$\beta$ insertion made it possible to achieve beam lifetimes of $\tau_{\rm{b}}=8000$\,s in the presence of a dense polarized hydrogen storage-cell target of areal density $d_{\rm t}=(5.5\pm 0.2)\times 10^{13} \mathrm{atoms/cm^{2}}$. The  developed techniques can be directly applied to antiproton machines and allow the determination of the spin-dependent $\bar{p}p$ cross sections via spin filtering.
\end{abstract}


\maketitle

\section{Introduction}
\label{sec:intro}
As long ago as 1968 it was realized  that by means of a spin filter using  an internal polarized hydrogen target  polarized high-energy proton beams could be produced at the 30\,GeV ISR\footnote{{\bf I}ntersecting {\bf S}torage {\bf R}ing} at CERN~\cite{Csonka1968247}. Since more efficient methods of providing polarized beams had already been developed, the idea of using a spin filter was revisited only in 1982 to polarize antiprotons at LEAR\footnote{{\bf L}ow-{\bf E}nergy {\bf A}ntiproton-cooler {\bf R}ing} of CERN~\cite{Kilian:1982fm}. At the 1985 workshop at Bodega Bay, CA, USA, a number of different techniques were discussed for providing stored beams of antiprotons~\cite{Krisch:1986nt1}. Of these techniques, spin filtering was considered practical and promising.

Spin filtering and related mechanisms leading to a polarization build-up in a stored beam were discussed in great detail at the Daresbury workshop in 2007~\cite{Chattopadhyay:2008zz}, and in a WE-Heraeus seminar in 2008 at Bad Honnef, Germany~\cite{badhonnef}. In the framework of the FILTEX collaboration, polarization build-up in an initially unpolarized  beam was observed for the first time using 23\,MeV protons stored in the TSR\footnote{{\bf T}est {\bf S}torage {\bf R}ing} at Heidelberg, interacting with polarized hydrogen atoms in a storage-cell target~\cite{Rathmann:1993xf}. (A  detailed description of the experimental effort is given in~\cite{Rathmann:1994,Zapfe1996293,zapfe:28}, up-to-date results are summarized  in~\cite{Oellers:2009nm}.)

The renewed interest in experiments with polarized antiprotons aims to produce a polarized antiproton beam at the HESR\footnote{{\bf H}igh-{\bf E}nergy {\bf S}torage {\bf R}ing }~\cite{Lehrach:2005ji} of FAIR\footnote{{\bf F}acility for {\bf A}ntiproton and {\bf I}on {\bf R}esearch, \url{http://www.fair-center.de}} \cite{fair2006} in Darmstadt, Germany. In 2003, a Letter of Intent for a variety of spin-physics experiments with polarized antiprotons was proposed by the PAX\footnote{{\bf P}olarized {\bf A}ntiproton e{\bf X}periments, \url{http://collaborations.fz-juelich.de/ikp/pax/}} collaboration~\cite{PAX_2003}. In 2005, the PAX collaboration submitted a technical proposal to the QCD program committee of FAIR, suggesting as an upgrade for HESR a double-polarized antiproton-proton collider to study, among other subjects, the transversity distribution of the proton~\cite{Barone:2005pu,Bradamante:2005wk}.

Polarizing a stored beam by spin-flip in polarized electron-proton ($\vec{e}^-p$) or polarized positron-antiproton ($\vec{e}^+\bar{p}$) scattering~\cite{Rathmann:2004pm} presents an advantage, because in contrast to spin filtering beam particles are not lost.
Triggered by the PAX proposal, the theory of spin-flip interactions was radically revised, leading to negligibly small cross sections for proton-electron scattering~\cite{Milstein:2005bx, Nikolaev:2006gw, Nikolaev:2008zz, Milstein:2008tc}.
In a recent experiment performed at COSY\footnote{{\bf CO}oler {\bf SY}nchrotron and storage ring}\cite{Maier:1997zj}, the $e^-\vec{p}$ spin-flip cross sections were indeed shown to be too small to allow the efficient production of polarized antiprotons based on $e^+\bar{p}$ interactions~\cite{Oellers:2009nm, Oellers20146}. 

Polarizing antiprotons by spin filtering, using the spin-dependent part of the nucleon-nucleon interaction, remains as yet the only viable method experimentally confirmed for a stored beam of protons and a polarized hydrogen gas target~\cite{Rathmann:1993xf,Rathmann:1994}. Theoretical considerations for beams of antiprotons have meanwhile been extended from $\bar{p}\vec{H}$ interactions \cite{Dmitriev:2010rj,Zhou:2013ioa} to $\bar{p}\vec{D}$~\cite{PhysRevC.87.054003} and $\bar{p}^3\vec{\mathrm{He}}$~\cite{PhysRevC.84.054011}.

In order to complement the Heidelberg TSR spin-filtering experiment by a second measurement, and to commission  the experimental setup for the proposed $\bar{p}p$ experiment at the AD\footnote{{\bf A}ntiproton {\bf D}ecelerator} of CERN~\cite{AD_2009}, a spin-filtering experiment was performed in 2011 at COSY. The experiment confirmed that only $pp$ scattering contributes to the polarization build-up~\cite{Augustyniak:2012vu}. At a beam kinetic  energy of $T=49.3$\,MeV, slightly above the COSY injection energy of $T=45$\,MeV, precise $\vec{p}d$ analyzing power data for the beam polarization measurement are available~\cite{King1977151}.

The spin-filtering method exploits the spin-dependence of the total hadronic cross section~\cite{Bystricky:1976jr},
\begin{linenomath}
\begin{equation}
\sigma_{\rm tot}=\sigma_{0}\pm \sigma_{1}\cdot Q \hspace{0.1cm} ,
\label{Sigma}
\end{equation}
\end{linenomath}
where $\sigma_{0}$ is the spin-independent part, $\sigma_{1}$ the spin-dependent part, and $Q$ is the nuclear polarization of the target. The positive (negative) signs denote a parallel (antiparallel) orientation of the spins of beam and target protons.

The number of beam protons with spin orientation parallel (antiparallel) to that of the target spins is denoted by $N^{\uparrow}$ ($N^{\downarrow}$). One can safely neglect the numerically minuscule spin-flip cross section. Then the decrease of the total number of beam particles as a function of time from the initial values $N^{\uparrow}(t=0)=N^{\downarrow}(t=0)=N_{\rm tot}(t=0)/2$ is described by
\begin{linenomath}
\begin{eqnarray}
N_{\rm tot}(t) &=&N^{\uparrow}(t)+N^{\downarrow}(t)\nonumber\\
                    &=&N_{\rm tot}(0)\cdot  \exp\left(-\frac{t}{\tau_{\rm b}}\right) \cdot \cosh\left(\frac{t}{\tau_1}\right) \,,
\label{Beam}
\end{eqnarray}
\end{linenomath}
where
\begin{linenomath}
\begin{eqnarray}
 \tau_{\rm b}=(d_{\rm t}f \sigma_{\rm{b}})^{-1} \, {\rm and} \, \tau_1 = (Q  d_{\rm t} f \tilde{\sigma}_{1})^{-1}\,.
\label{eq:tau}
\end{eqnarray}
\end{linenomath}
Here $d_{\rm t}$ is the areal target gas density, $f$ is the revolution frequency determined by the beam momentum and the ring circumference, and $\tilde{\sigma}_{1}=\sigma_{1}(\Theta>\Theta_\mathrm{acc})$ represents the effective removal cross section. Furthermore, $\sigma_{\rm {b}}=\sigma_0+\sigma_\mathrm{C}$ combines $\sigma_0$ and single Coulomb scattering $\sigma_{\rm C}$ in the target, the latter for scattering angles larger than the acceptance angle $\Theta_\mathrm{acc}$  of the machine. For single Coulomb scattering and small values of $\Theta_\mathrm{acc}$, the beam lifetime $\tau_\mathrm{b} \propto \sigma_{\rm C}^{-1} \propto\Theta_\mathrm{acc}^{2} \propto \beta^{-1}$ (see section~\ref{sec:lowb}). Therefore the betatron function (or $\beta$-function) at the target should be small in order to achieve a long beam lifetime. 

The polarization build-up in the stored, circulating beam is given by
\begin{linenomath}
\begin{eqnarray}
P(t)&=&\frac{N^{\uparrow}(t)-N^{\downarrow}(t)}{N^{\uparrow}(t)+N^{\downarrow}(t)} =\tanh\left(\frac{t}{\tau_1}\right).
\label{Pol}
\end{eqnarray}
\end{linenomath}
It depends on the spin-dependent removal of particles. The effective removal cross section in (\ref{eq:tau}) depends on the machine acceptance and consequently so does the achievable beam polarization, as illustrated, \textit{e.g.}, in figure~15 of \cite{Zhou:2013ioa}.

The present paper describes the development effort, including a variety of measurements, necessary to prepare the COSY storage ring and the experimental equipment for the  spin-filtering experiments~\cite{Augustyniak:2012vu,Weidemann:2011}. The paper is organized  as follows:

\begin{itemize}
\item Section~\ref{sec:COSY} presents the essential components of the COSY ring, in particular its lattice and the electron cooler~(\ref{sec:lattice}), followed by the requirements for the low-$\beta$ insertion at the position of the polarized gas target, and its realization~(\ref{sec:lowb}).

\item Section~\ref{sec:target} describes the internal polarized hydrogen storage-cell target (\ref{sec:cell}), the coil system to produce the magnetic holding field at the storage cell (\ref{sec:holdingfield}), and the vacuum system around the polarized target (\ref{sec:Vacuum}).

\item In section~\ref{sec:tools} the equipment employed for beam diagnosis is described, comprising the beam current transformer, H$^{0}$ monitor, ionization profile monitor, beam-position monitor, movable frame system for acceptance measurements, and beam polarimeter.

\item Section~\ref{sec:tuneandorbit} describes the betatron tune mapping (section~\ref{sec:tune}) and orbit adjustment (section~\ref{sec:orbit}) to provide long beam lifetime for the spin-filtering experiments.

\item Section~\ref{sec:lowbeta} highlights the commissioning of the low-$\beta$ insertion, including   the determination of the $\beta$-function at the target.

\item Section~\ref{sec:acce} presents the measurements of the beam widths~(\ref{sec:beamwidth}) at the location of the internal target, the beam emittance~(\ref{sec:beamemittance}), and the determination of the machine acceptance and the acceptance angle at the target position~(\ref{sec:acceptancemeasurement}).

\item In section~\ref{sec:lifetime} the efforts are described to optimize the beam lifetime by means of closed orbit correction and tune adjustment. Space-charge effects~(\ref{sec:spacecharge}) and vacuum considerations are discussed as well~(\ref{sec:vacuum}).

\item In section~\ref{sec:beamprep} it is explained how the beam was set up for the experiments~(\ref{sec:setup}) and what a typical measurement cycle looked like~(\ref{sec:meascycle}). In addition, the measurement of the beam polarization lifetime~(\ref{sec:pollifetime}) and the efficiency of the RF spin flipper are described~(\ref{sec:flipefficiency}).

\item Section~\ref{sec:conclusion} summarizes the main results.
\end{itemize}

\section{COSY accelerator and storage ring}
\label{sec:COSY}
The synchrotron and storage ring COSY accelerates and stores unpolarized and polarized proton or deuteron beams in the momentum range between 0.3\,GeV/c and 3.65\,GeV/c. COSY has a racetrack design with two $180^{\circ}$ arc sections connected by 40\,m long straight sections. It is operated as a cooler storage ring with internal targets (ANKE\footnote{\textbf{A}pparatus for Studies of \textbf{N}ucleon and \textbf{K}aon \textbf{E}jectiles, \url{http://collaborations.fz-juelich.de/ikp/anke/}}, WASA\footnote{\textbf{W}ide \textbf{A}ngle \textbf{S}hower \textbf{A}pparatus, \url{http://collaborations.fz-juelich.de/ikp/wasa/}}, PAX$^6$) or with an extracted beam (see figure~\ref{fig:cosy}, bottom panel). Beam cooling, \textit{i.e.}, reducing the momentum spread of the beam and shrinking the transverse equilibrium phase space, is realized by electron cooling up to proton-beam momenta of 0.6\,GeV/c~\cite{Prasuhn:398615} and by stochastic cooling for proton momenta above 1.5\,GeV/c~\cite{Prasuhn:2000eu}.

Polarized proton and deuteron beams are routinely delivered to experiments over the whole momentum range~\cite{Eversheim:1996qu}. Polarized beams from the ion source are pre-accelerated in the cyclotron JULIC~\cite{Aldea:1981}, injected and accelerated in COSY without significant loss of polarization. Imperfection and intrinsic depolarization resonances are overcome by well-established procedures~\cite{Stockhorst:2004qj,Lorentz:2011zz,Lehrach:2003jw}.
When the polarization lifetime is by several orders of magnitude longer than the spin filtering periods required, it becomes feasible to polarize an originally unpolarized  beam by filtering, as was confirmed in a dedicated experiment~\cite{Augustyniak:2012vu}, described in section~\ref{sec:beamprep}.

\subsection{COSY lattice and electron cooler}
\label{sec:lattice}
\begin{figure}[t]
\includegraphics[width=\columnwidth]{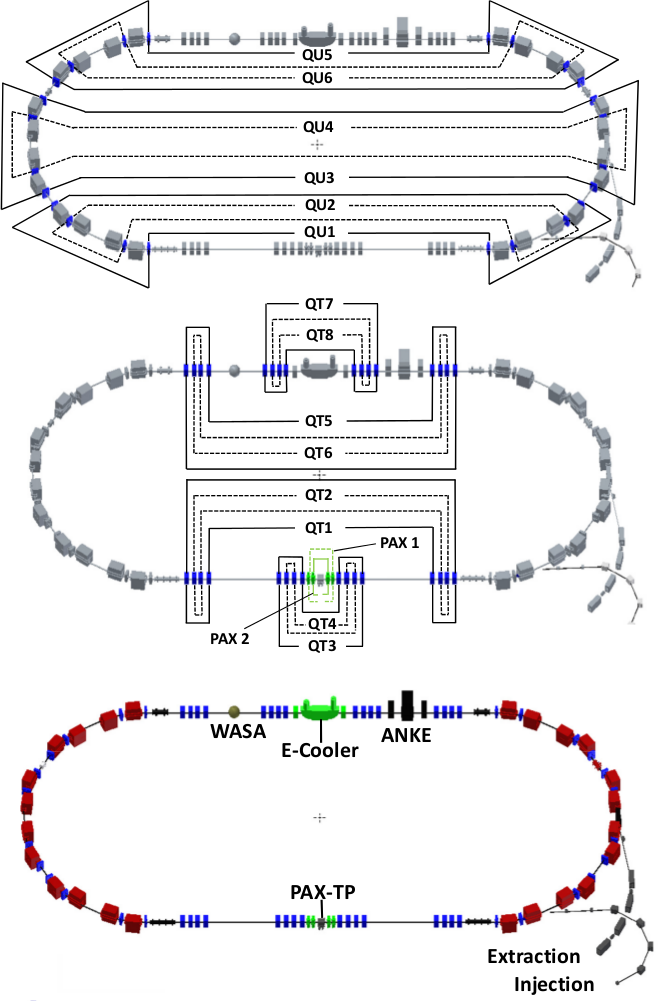}
\caption[]{Bottom panel: Floor plan of the COSY facility. The 24 dipole magnets are shown in red, the quadrupole magnets in blue except those around the PAX target point (PAX-TP) and the 100\,keV electron cooler which are shown in green. The quadrupole magnets of COSY are combined into quadrupole families, each consisting of four magnets with a common current supply. There are eight families for the telescopic straight sections (QT1 to QT8, middle panel) and six for the arcs (QU1 to QU6, top panel). The PAX quadrupoles of the low-$\beta$ insertion at the PAX-TP are combined into an outer pair (PAX1) and an inner pair (PAX2).}
\label{fig:cosy}
\end{figure}
\begin{table}[t]
\caption[Parameters of COSY.]{Main parameters of the COSY accelerator and storage ring \cite{Maier:1997zj}.} \label{tab:cosy}
\begin{ruledtabular}
\begin{tabular}{p{0.4\columnwidth} p{0.55\columnwidth}}
\multicolumn{2}{c}{\textbf{COSY}}\\
Circumference &  183.47\,m \\
Particles & (Un)polarized $p$ and $d$\\
Type of injection & H$^{-}$, D$^{-}$ stripping injection\\
Current at source exit &  Polarized:        $15$\,\textmu A\\
	  	  &  unpolarized: $100-200$\,\textmu A\\
Momentum range &  $0.3 - 3.65$ GeV/c \\
Betatron tune range &  $3.55 - 3.7$ in both planes\\
Phase-space cooling&  Electron and stochastic\\
Beam position monitors &  31 (horizontal and vertical)\\
Steerers &  23 (horizontal), 21 (vertical)\\ \hline
\multirow{4}{*}{ Straight sections} &  Length:  40\,m \\
&  $4 \times 4$ quadrupole magnets\\
&  4 sextupole magnets\\ 
&  Beam pipe diameter: 0.15\,m\\
\hline 
\multirow{5}{*}{ Arc sections} &  Length: 52\,m  \\
&  $3 \times 4$ dipole magnets \\
&  $3 \times 4$ quadrupole magnets\\
&  5 sextupole magnets\\
&  Beam pipe in dipole magnets: height: 0.06\,m, width: 0.15\,m\\
\end{tabular}
\end{ruledtabular}
\end{table}

The COSY lattice is designed to provide  flexibility with respect to ion-optical settings~\cite{Bongardt:1989hb} in order to fulfill the requirements for internal and external experiments. Each of the arcs is composed of three mirror-symmetric unit cells (U) consisting of four dipole magnets (O), two horizontally focusing  (F) and two horizontally defocusing quadrupole magnets (D). Each of the six unit cells has a DOFO-OFOD structure (see figure~\ref{fig:cosy}, top panel). The two inner (and outer) quadrupole magnets of each unit cell are connected to the inner (and outer) pair of the opposite unit cell located in the other arc, thus resulting in six quadrupole families (QU1 to QU6). A symmetric operation of all unit cells leads to a sixfold symmetry of the $\beta$-functions~\cite{Lehrach:2000vw}.

The straight sections are composed of two mirror-symmetric telescopic (T) arrangements with two quadrupole triplets, each consisting of four quadrupoles, either operated in FDDF or DFFD mode. A 2$\pi$ phase advance and 1:1 imaging over the complete straight section is thus achieved, decoupling to first order the arcs from the straight sections~\cite{Bongardt:1989hb} and providing three possible locations per straight section for internal target experiments with adjustable $\beta$-functions in the center of the triplets.
\begin{figure}[t]
 \centering
 \includegraphics[width=\columnwidth]{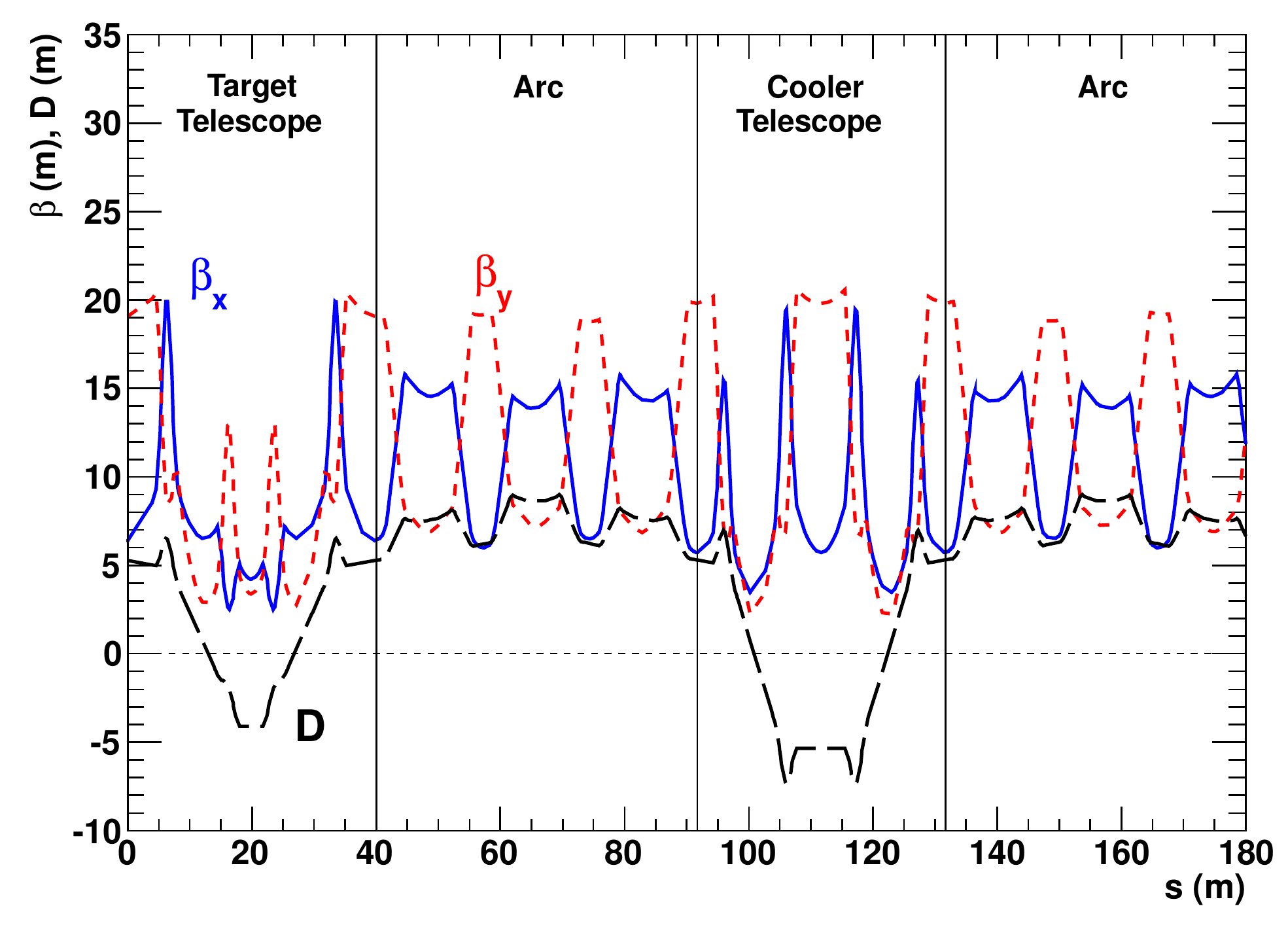}
 \includegraphics[width=\columnwidth]{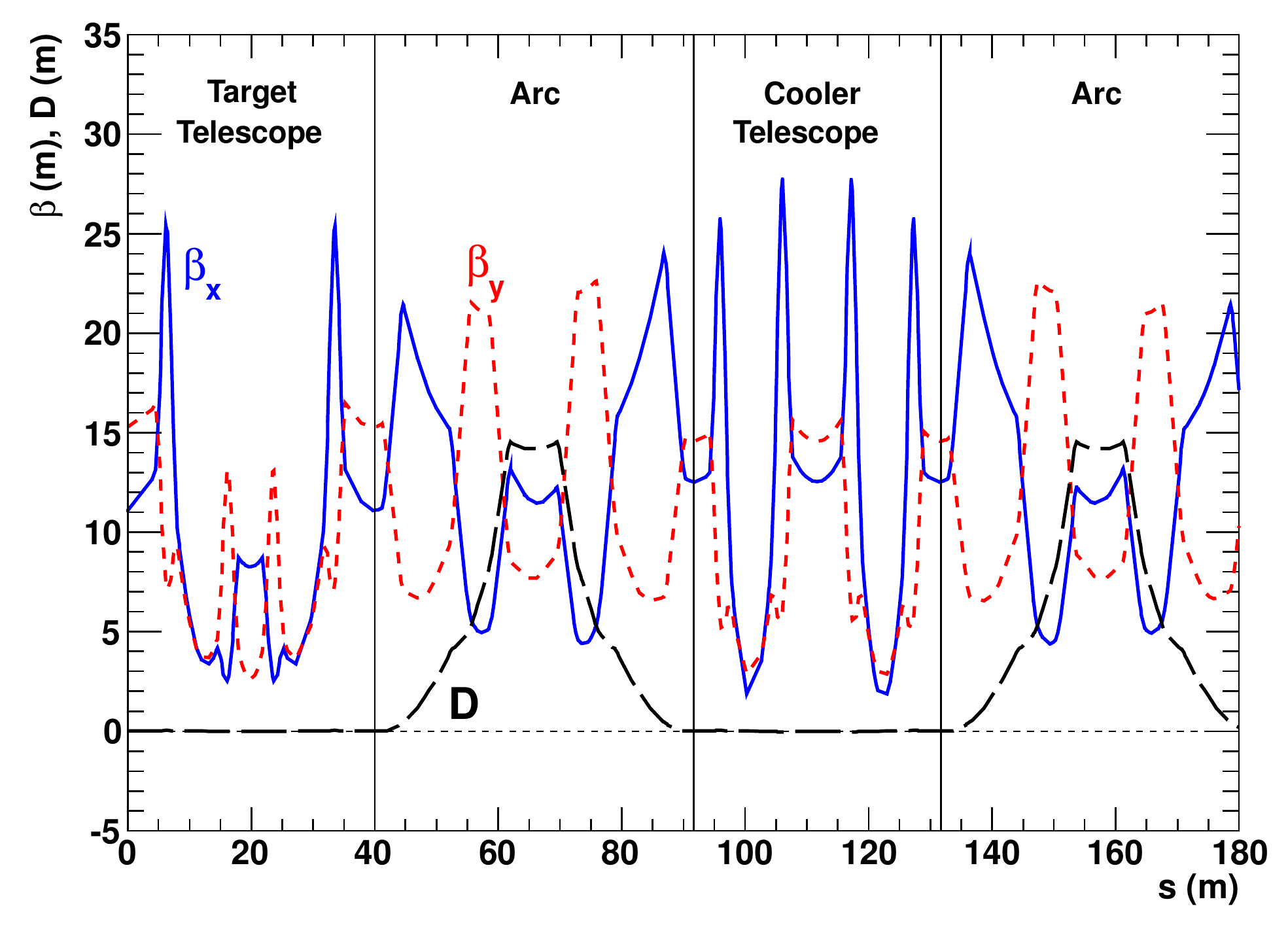}
\caption{Top panel:
Optical functions ($\beta_{x},\beta_{y}$) and dispersion $D$ along COSY for a standard setting ($D\neq 0$). In each of the arcs symmetric behavior due to the three unit cells in each section can be seen. Bottom panel: $\beta$-functions and dispersion for the $D=0$ setting through the telescopes. The PAX target is located in the center of the target telescope.}
\label{fig:betafu}
\end{figure}
 Figure~\ref{fig:betafu} (top panel) shows the horizontal ($x$) and vertical ($y$) $\beta$-functions, $\beta_{x}$ and $\beta_{y}$, and the dispersion $D$ for a typical setting of COSY used at injection. The basic parameters of COSY are listed in table~\ref{tab:cosy}.
 
The straight sections can be made free of dispersion by breaking the sixfold symmetry with a specific setting of the six arc quadrupole families (see figure~\ref{fig:betafu}, bottom panel). This dispersion-free $D=0$ setting is advantageous for the operation of the storage-cell target, and it was therefore chosen during the spin-filtering experiments. A non-zero dispersion causes a displacement of a particle with a relative momentum deviation $\frac{\Delta p}{p}$ from the reference orbit $x_{\rm{ref}}$ and the deviation from the ideal orbit is given by~\cite{Wille:2000}
\begin{linenomath} 
\begin{equation}
 x(s)_{\rm{total}}=x_{\rm{ref}}(s)+D(s)\cdot\frac{\Delta p}{p} \hspace{0.15cm} ,
 \label{eq:dispersion}
\end{equation}
\end{linenomath} 
where $s$ is the position along the reference orbit and $s=0$ is located at the beginning of the straight section where the PAX-TP is located.\\

The COSY electron cooler (see figure~\ref{fig:cosy}) is used to compensate multiple small-angle Coulomb scattering and energy loss in the target and the residual gas in the machine. It provided stable beam emittance and beam energy during the spin-filtering experiment. It was designed for electron energies up to 100\,keV, thus enabling phase-space cooling up to a proton-beam kinetic energy of 183.6\,MeV \cite{Stein:2011zc}. Its main parameters are listed in table~\ref{tab:cooler}. Two short solenoids located in the 8\,m long drift region in front of and behind the electron cooler (see figure~7 of ~\cite{6620982}) and operated with reversed polarity to that in the drift solenoid compensate phase-space coupling and avoid spin rotation in the case of polarized beams. The field strengths are adjusted such that $\int B\cdot dl$ over the cooler magnets and the compensating solenoids equals zero. The main drift solenoid was typically operated at magnetic fields of $B=50-80\,$mT.

Beams of small emittance, as produced by electron cooling, tend to develop coherent betatron oscillations which lead to  beam loss \cite{Stein:2011zc}. The transverse feedback system of COSY \cite{Kamerdzhiev:2003, Kamerdzhiev:2004ha} was used to avoid these instabilities. 
\begin{table}
\begin{ruledtabular}
\caption[Design parameters of the electron cooler. ]{Parameters of the  electron cooler at COSY~\cite{Stein:2011zc}.} \label{tab:cooler}
\begin{tabular}{p{0.62\columnwidth} p{0.38\columnwidth}}
\multicolumn{2}{c}{\textbf{Electron Cooler}}\\
\hline
Electron energy 	& $20-100$\,keV\\
Typical electron beam current	& 0.25\,A\\
Magnetic field strength	& $50 - 150$\,mT\\
Length of drift solenoid  	& 2.00\,m\\
Bending radius in the toroids & 0.60\,m\\
Effective length of cooling	& 1.50\,m\\
Effective length of solenoidal field &  3.20\,m\\
Effective length of compensation&  0.5\,m\\
Solenoids &\\
Diameter of COSY beam tube	& 0.15\,m\\
Diameter of electron beam	& 0.025\,m\\
Typical $\beta$-functions at the e-cooler& $\beta_{x}=6\,$m, $\beta_{y}=20\,$m\\
Diagnosis			& H$^{0}$ profile monitor \\ & and count rate\\
\end{tabular}
\end{ruledtabular}
\end{table}
\subsection{Low-$\beta$ insertion}
\label{sec:lowb}
In a storage ring, the geometrical machine acceptance\footnote{Throughout this paper \textmu m is used as a unit of machine acceptance and beam emittance, equivalent to mm\,mrad.}~\cite{Wille:2000} is defined by \begin{linenomath}
\begin{equation}
A_{x,y}=\left (\frac{a_{x,y}^2}{\beta_{x,y}}\right )_{\rm {min}} ,
\label{eq:acceptance}
\end{equation}
\end{linenomath}
and the acceptance angle $\Theta_{\mathrm{acc}}$ \cite{Madsen:2000ep}, by
\begin{linenomath}
\begin{equation}
\frac{1}{\Theta_{ \mathrm{acc}}^{2}}=\frac{1}{2\Theta_{x}^{2}}+\frac{1}{2\Theta_{y}^{2}} \hspace{0.5cm} {\rm with} \hspace{0.5cm} \frac{1}{\Theta_{x,y}^{2}}=\frac{\beta_{x,y}}{A_{x,y}} \hspace{0.15cm} ,
\label{eq:acc}
\end{equation}
\end{linenomath}
where $x$ and $y$ indicate the horizontal and vertical direction, respectively, and $a$ is the free aperture along the ring.
At the kinetic energy of $T_{p}=49.3\,$MeV of the spin-filtering experiment, the beam lifetime (see (\ref{eq:tau})) is dominated by the Coulomb scattering loss on the target gas and the residual gas in the ring; the hadronic losses amount to about 10\% of the total loss cross section $\sigma_{\rm b}$ (see section~\ref{sec:vacuum}).  
The Coulomb-loss cross section can be derived by integration of the differential Rutherford cross section for scattering angles larger than $\Theta_{\mathrm{acc}}$ \cite{Poth:1990pg},
\begin{linenomath}
\begin{equation}
 \sigma_{\mathrm C}=\int\limits_{\Theta_{\mathrm{acc}}}^{\Theta_{\mathrm{max}}}\int\limits_{0}^{2\uppi}\frac{d \sigma}{d \Omega}d \phi\sin\Theta d \Theta=4\uppi\frac{Z_{\mathrm{gas}}^{2}Z_{\rm i}^{2}r_{\rm i}^{2}}{\beta_{\rm L}^{4}\gamma_{\rm L}^{2}} \cdot \frac{1}{\Theta_{\mathrm{acc}}^{2}} \hspace{0.15cm}  .
\label{eq:sc}
\end{equation}
\end{linenomath}
$Z_{\mathrm{gas}}$ and $Z_{\rm i}$ are the atomic numbers of the target (or residual) gas and the ion beam, respectively, $\beta_{\rm L}$ and  $\gamma_{\rm L}$ are the relativistic Lorentz factors, and $r_{\rm i}=r_{\rm e}m_{\rm e}/m_{\rm i}$ is the classical ion radius.
The beam lifetime due to single Coulomb scattering
\begin{linenomath}
\begin{equation}
  \tau_{\rm b} \approx \tau_{\rm C}=\frac{1}{\sigma_{\mathrm C} d_{\mathrm t}f}
=\frac{\beta_{\rm L}^{4}\gamma_{\rm L}^{2}}{4\pi Z_{\mathrm
{gas}}^{2}Z_{\rm i}^{2}r_{\rm i}^{2}} \cdot{\frac{\Theta_\mathrm{acc}^{2}}{d_{\mathrm t}\cdot f}}
\propto \frac{1}{d_{\mathrm t}\cdot\beta} \hspace{0.15cm} ,
\label{eq:appB}
\end{equation}
\end{linenomath}
is inversely proportional to the $\beta$-function and the gas density. Therefore, especially the $\beta$-functions at the PAX-TP should be small, because of the high densities.
\begin{figure}[t]
\centering
\includegraphics[width=\columnwidth]{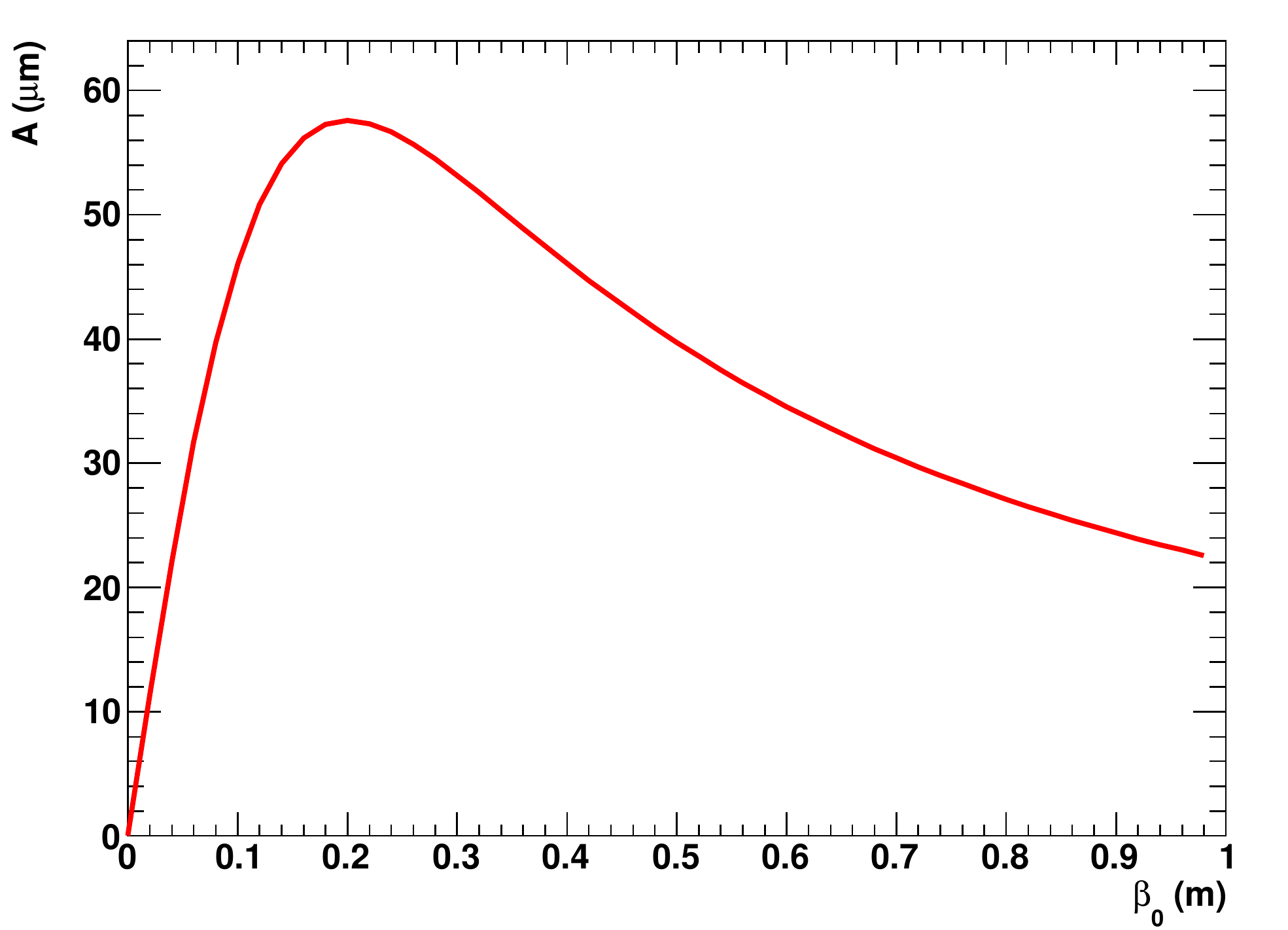}
\caption[]{Machine acceptance $A(\beta_0)$ using (\ref{eq:Aofbeta_0}) for a storage cell with diameter $d=9.6$\,mm and length $l=400$\,mm as a function of $\beta_0$ at the target center. $A$ reaches a maximum for $\beta_{0}=l/2=0.2\,$m.}
\label{fig:optbeta}
\end{figure}

It turns out that for a given target-gas cell there is an optimal value for the $\beta$-function at the cell center. The $\beta$-function in a symmetric drift space is described by
\begin{linenomath}
\begin{equation}
 \beta(s')=\beta_{0}+\frac{s'^2}{\beta_{0}}\hspace{0.15cm},
\end{equation}
\end{linenomath}
where $s'=s-s_{0}$ is the distance from the cell center $s_{0}$ and $\beta_0$ is the $\beta$-function at the center. The machine acceptance for a storage cell of diameter $d$ and length $l$ as a function of $\beta_0$  is therefore given by
\begin{linenomath}
\begin{equation}
A(\beta_{0})=\frac{\left(d/2\right)^2}{\beta_{0}+\frac{\left(\displaystyle l/2\right)^2}{\displaystyle \beta_{0}}}\hspace{0.15cm},
\label{eq:Aofbeta_0}
\end{equation}
\end{linenomath}
 and $A(\beta_0)$ reaches a maximum for $\beta_{0}=l/2$.
 A storage cell of $d=9.6$\,mm and $l=400$\,mm is used to maximize the target areal density in the experiment (see section~\ref{sec:cell}). 
 For the specified cell the maximum acceptance is \mbox{$A(\beta_{0}=0.2\,\rm{m})\approx58\,$\textmu m} (see figure~\ref{fig:optbeta}). 
 The standard COSY lattice ($D\neq0$) provides geometrical acceptances of about $A_x\approx(75\, \rm{mm})^2/25\,\rm{m}=225\,$\textmu m and $A_y\approx(30\, \rm{mm})^2/20\,\rm{m}=45\,$\textmu m (see figure~\ref{fig:betafu} and table~\ref{tab:cosy}), thus with the smallest $\beta$-functions of about 3\,m, the given storage cell would restrict the machine acceptance to $A(\beta_{0}=3\,\rm{m})\approx8$\,\textmu m.
 
\begin{figure*}[t]
\centering
\includegraphics[width=0.8\textwidth]{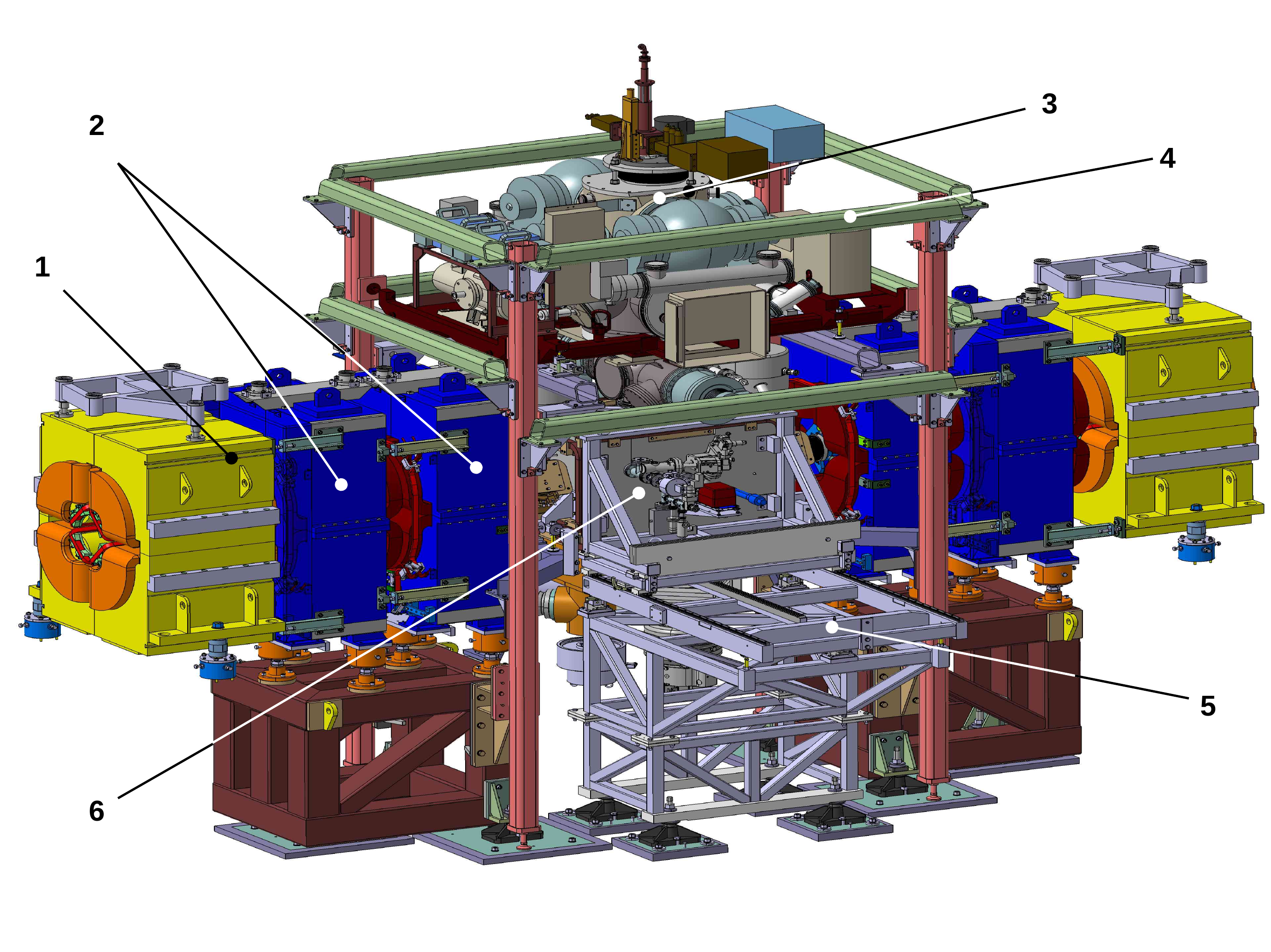}
 \caption[PAX low-$\beta$ section.]{View of the PAX installation at COSY from the interior of the ring (the beam comes from the left). 1: COSY quadrupole magnet, 2: two of the four PAX quadrupoles (formerly used at CELSIUS \cite{Ekstrom:1988if}) forming the low-$\beta$ insertion by doublet focusing (DF-FD), 3: atomic beam source (ABS), 4: support system, 5: flange supporting rail system, 6: PAX target chamber housing the storage cell. The Breit-Rabi polarimeter (BRP) and the target-gas analyzer (TGA) are mounted towards the outside of the ring. The horizontal distance between the inner faces of the two COSY quadrupole magnets is 3.75\,m, the height of the beam-tube center from the ground is 1.80\,m.}
\label{fig:beta}
\end{figure*}
To obtain the required small $\beta$-functions, a low-$\beta$ insertion consisting of four additional quadrupole magnets (blue in figure~\ref{fig:beta}), formerly used at CELSIUS \cite{Ekstrom:1988if}, was installed in the drift space in front of and behind the target. The quadrupole magnets are arranged in a doublet structure (DF-FD), where the D and F magnets are powered by separate power supplies. When  the doublets are operated, the four regular COSY quadrupole families in this straight section are reduced in strength to maintain its telescopic nature. Thus the other magnets in the machine do not require any readjustment.

Precise positioning of the beam inside the storage cell was provided by horizontal and vertical steerer coils, which because of space restrictions were mounted on the yokes of the adjacent quadrupole magnets upstream and downstream of the low-$\beta$ insertion.
\begin{figure}[t]
\centering
\includegraphics[width=\columnwidth]{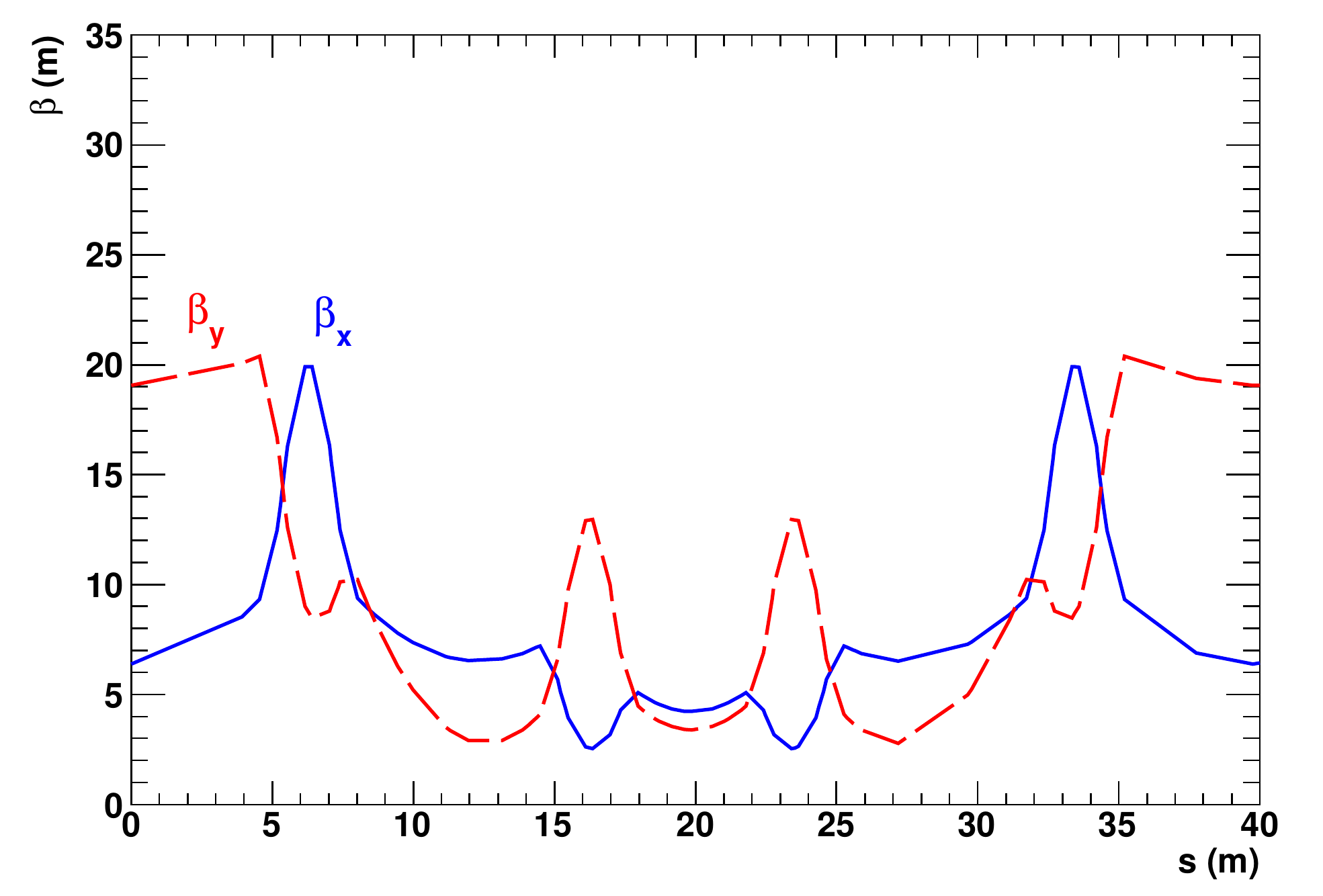}
\includegraphics[width=\columnwidth]{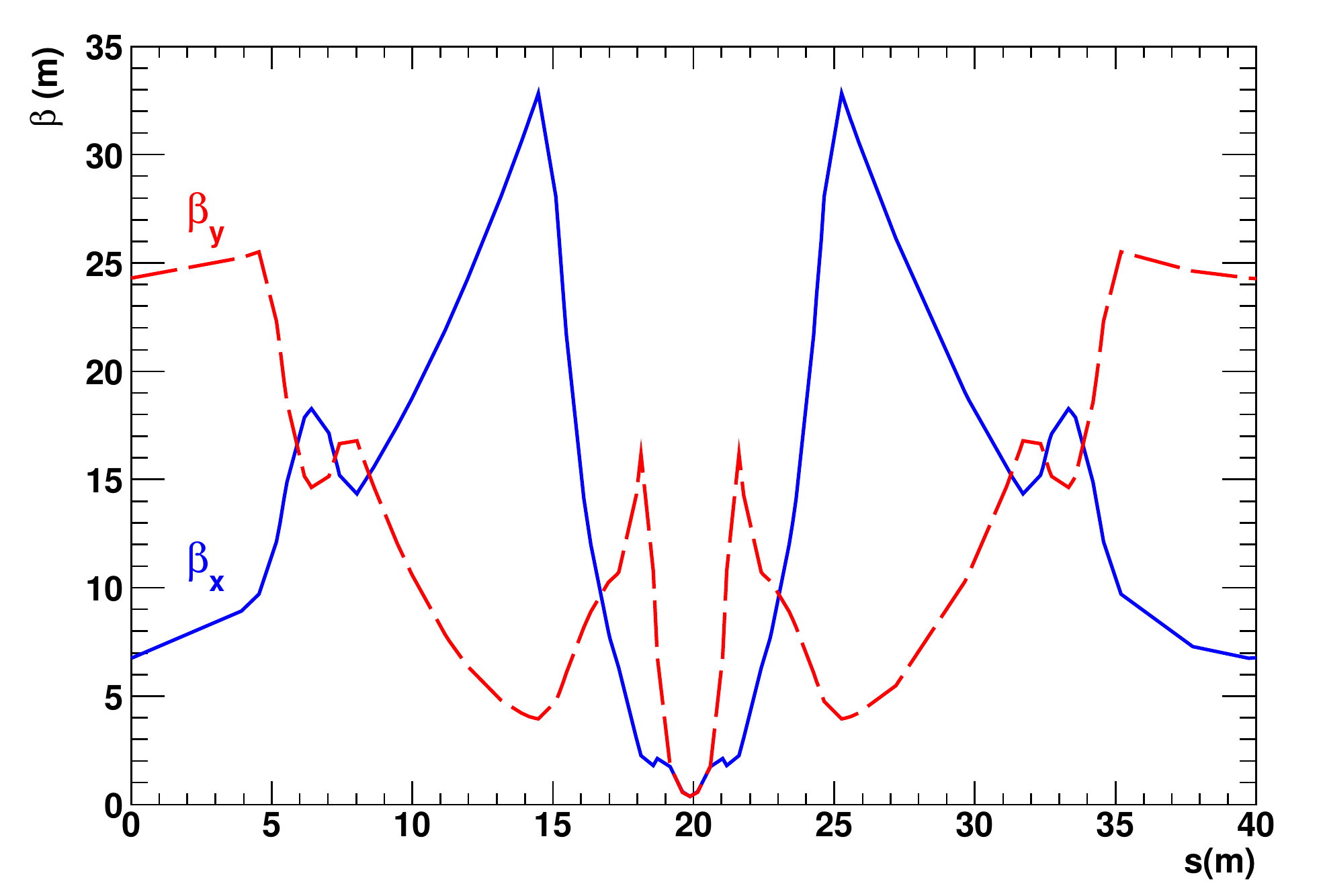}
\caption[]{Model calculation of the $\beta$-functions for the standard COSY setting ($D\neq0$) with PAX magnets switched OFF (top panel) and ON (bottom panel), indicating that minimal values of $\beta_{x,y} \approx 0.3$\,m can be reached at the target point at $s=19.87$\,m. s = 0 is located at the beginning of the target straight section }
\label{fig:betafunc}
\end{figure}

Based on the COSY lattice using the standard magnet settings, the optical functions were calculated with the MAD\footnote{\textbf{M}ethodical \textbf{A}ccelerator \textbf{D}esign} program, version 8~\cite{Grote:1991zp}.
The results obtained with the PAX magnets switched ON and OFF are shown in figure~\ref{fig:betafunc}, indicating that $\beta_{x}$ and $\beta_{y}$ at the target point can be reduced by more than one order of magnitude, with minimal values of $\beta_{x,y}\approx 0.3$~m. The commissioning of the low-$\beta$ section, including the measurement of $\beta_{x}$ and $\beta_{y}$, is described in section~\ref{sec:lowbeta}.

Reduced $\beta$-functions at the target, however, are accompanied by increased  $\beta$-functions up- and downstream, reaching values of about  $33$\,m (see figure~\ref{fig:betafunc}, bottom panel). Therefore, excellent vacuum conditions also have to be maintained in these regions to avoid adversely affecting the beam lifetime.

\section{Polarized target}
\label{sec:target}
\subsection{Polarized atomic beam source and storage cell}
\label{sec:cell}
The polarized internal target (PIT) consists of the atomic beam source (ABS), which was developed for the TSR spin-filtering experiment~\cite{Stock:1994vv,zapfe:28}, later used in the HERMES experiment at DESY~\cite{Nass2003633,Airapetian:2004yf} and now modified for spin-filtering at COSY, a storage cell~\cite{Ciullo:2011zz}, a Breit-Rabi polarimeter (BRP)~\cite{Baumgarten:2001ym}, and a target gas analyzer (TGA)~\cite{Baumgarten:2003rp}. H$^{0}$ atoms in a single hyperfine state are prepared in the ABS and injected into a thin-walled storage cell.
A fraction of the gas diffuses from the cell through a side tube into the diagnostic system, where the BRP determines the atomic polarization and the TGA  the relative fraction of atoms and molecules. A magnetic guide-field system defines the quantization axis for the target polarization, which can be oriented along the $x$ (outward), $y$ (up), or $s$ (along beam) direction, or any superposition thereof (see section~\ref{sec:holdingfield}).

The gas load into the target chamber and the neighboring sections causes beam losses due to the interaction of beam particles with the residual gas. A dedicated pumping system, described in section~\ref {sec:Vacuum}, was  developed to minimize these losses.

\begin{figure}[ht]
\includegraphics[width=\columnwidth]{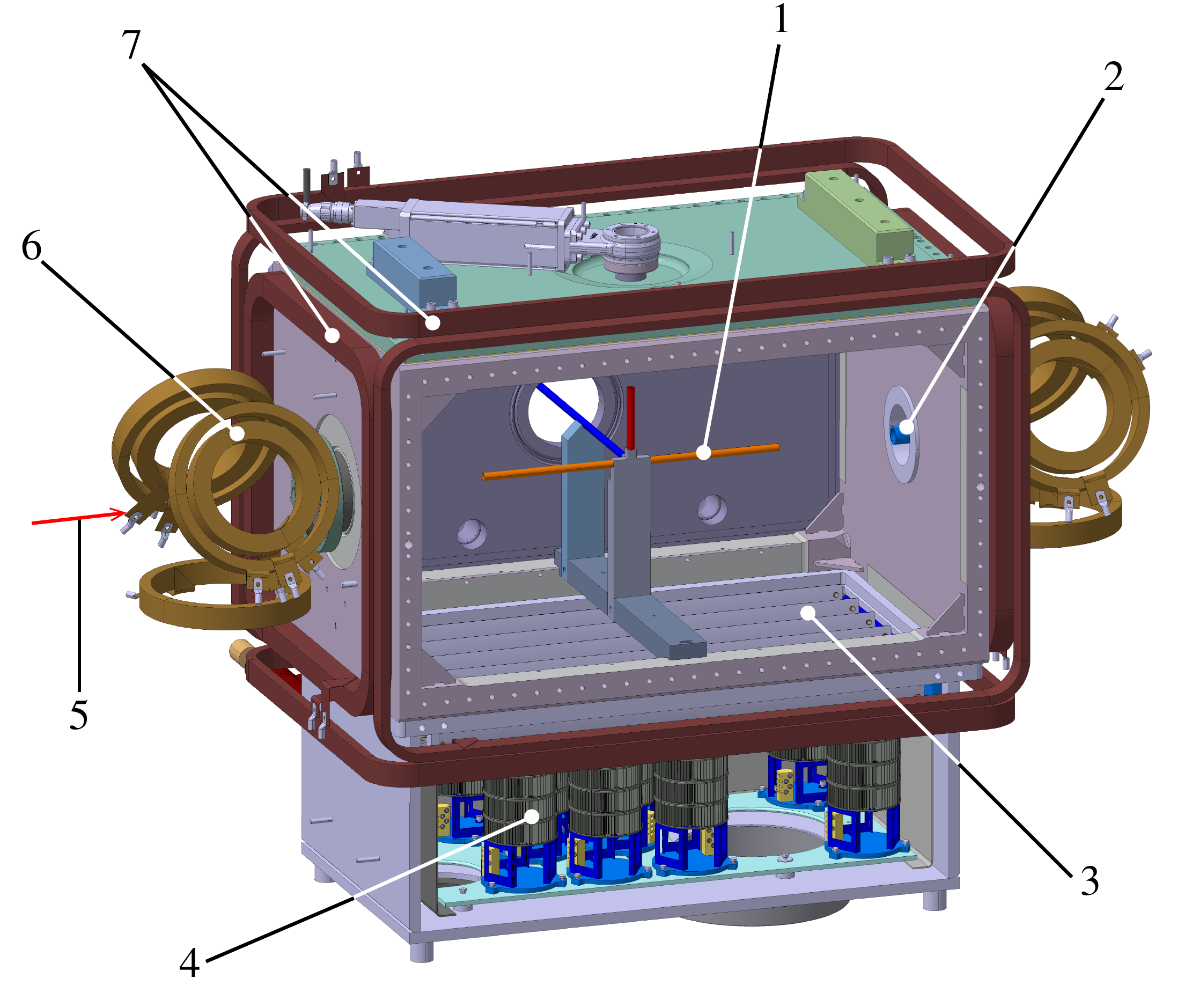}
\caption[PAX target chamber]{Section view of the PAX target chamber. The labels denote the storage cell (1) with feeding tube to the ABS (vertical), and extraction tube to the BRP (to the backside), flow limiters (2) of 19\,mm diameter and 80\,mm length, jalousie (3) to protect the cell from heat radiation during activation of the NEG pumps (4), COSY beam (5), guide field compensation coils (6), and magnetic guide field coils (7).}
 \label{fig:coils}
\end{figure}
The storage cell (see figure~\ref{fig:coils}, label 1) increases the dwell time of the polarized atomic gas in the interaction region with the beam and enhances the areal target density compared to a free atomic jet by about two orders in magnitude. The cell was made from aluminum and coated with Teflon\footnote{Teflon (polytetrafluoroethylene) coating was done by Rhenotherm, Kunst\-stoffbeschichtungs GmbH, Kempen, Germany, \url{http://www.rhenotherm.de/}} to reduce depolarization and recombination \cite{Price1994321}.
Assuming linear decrease of the gas density from the center to the open ends, the areal target-gas density is given by
\begin{linenomath}
\begin{equation}
 d_{\rm t}=\frac{1}{2}\cdot\frac{l\cdot I}{C_{\rm tot}}\hspace{0.15cm},
\end{equation}
\end{linenomath}
where $I$ [s$^{-1}$] is the intensity of the injected beam from the ABS, $l$ [cm] the total length of the storage tube, and $C_{\rm tot}$ the total conductance of the storage cell. The conductance [$\ell /s$] of a circular tube of diameter $d_i$ [cm] and length $l_i$ [cm] can be written as~\cite{Roth:1990}
\begin{linenomath}
\begin{equation}
 C_i=3.81\sqrt{\frac{T}{M}}\cdot\frac{d_i^{3}}{l_i+1.33\cdot d_i}\hspace{0.15cm},
\label{eq:conductance}
\end{equation}
\end{linenomath}
where $T$ [K] is the temperature and $M$ [u] the molar mass.

The total conductance $C_{\rm tot}$ of the storage cell is given by the sum of all conductances with respect to the cell center. For a storage-cell tube ($l=400\,\rm{mm}$, $d=9.6\,$mm), a feeding tube from the ABS (\mbox{$l=100$\,mm}, \mbox{$d=9.6$\,mm}), and the extraction tube to the target polarimeter (\mbox{$l=380$\,mm}, $d=9.6$\,mm), the conductance of the storage cell  yields $C_{\rm{tot}}=2\cdot C_{\rm{\frac{1}{2}cell}}+C_{\rm{feed}}+C_{\rm{extract}}=12.15\,\ell$/s. With an intensity from the ABS injected into the feeding tube of $I=3.3\cdot10^{16}\,\rm{s}^{-1}$~\cite{Nass2003633}, an areal density of $d_{\rm t}=5.45\cdot 10^{13}\,\rm{cm}^{-2}$ is expected. During the spin-filtering experiment, in good agreement with the estimate given above, a target density   of ~\cite{Augustyniak:2012vu}
\begin{linenomath}
\begin{equation}
 d_{\rm t}=(5.5\pm 0.2)\cdot10^{13}\,\rm{cm}^{-2},
\label{eq:targetdensity}
 \end{equation}
\end{linenomath}
was deduced from the shift of the orbit frequency of the coasting beam caused by the energy loss in the target gas (see section~\ref{sec:BPMs}) \cite{Zapfe1996293,Stein:2008ga}. 

\subsection{Holding field coil system}
\label{sec:holdingfield}
The operation of the polarized target requires a coil system providing guide fields of about $1\,\rm mT$~\cite{0034-4885-66-11-R02} in order to define the orientation of the target polarization and allowing it to be reversed in short sequence. The polarization of the gas atoms is known to be fully reversed within about $10\,\rm ms$  after switching the polarity of the magnetic field (see figure~11 of~\cite{Rathmann:1998zz}). A system of coils, providing fields in transverse ($x$, $y$) and longitudinal ($s$) directions, was installed on the target chamber (see figure~\ref{fig:coils}). 

\begin{figure}[t]
\includegraphics[width=\columnwidth]{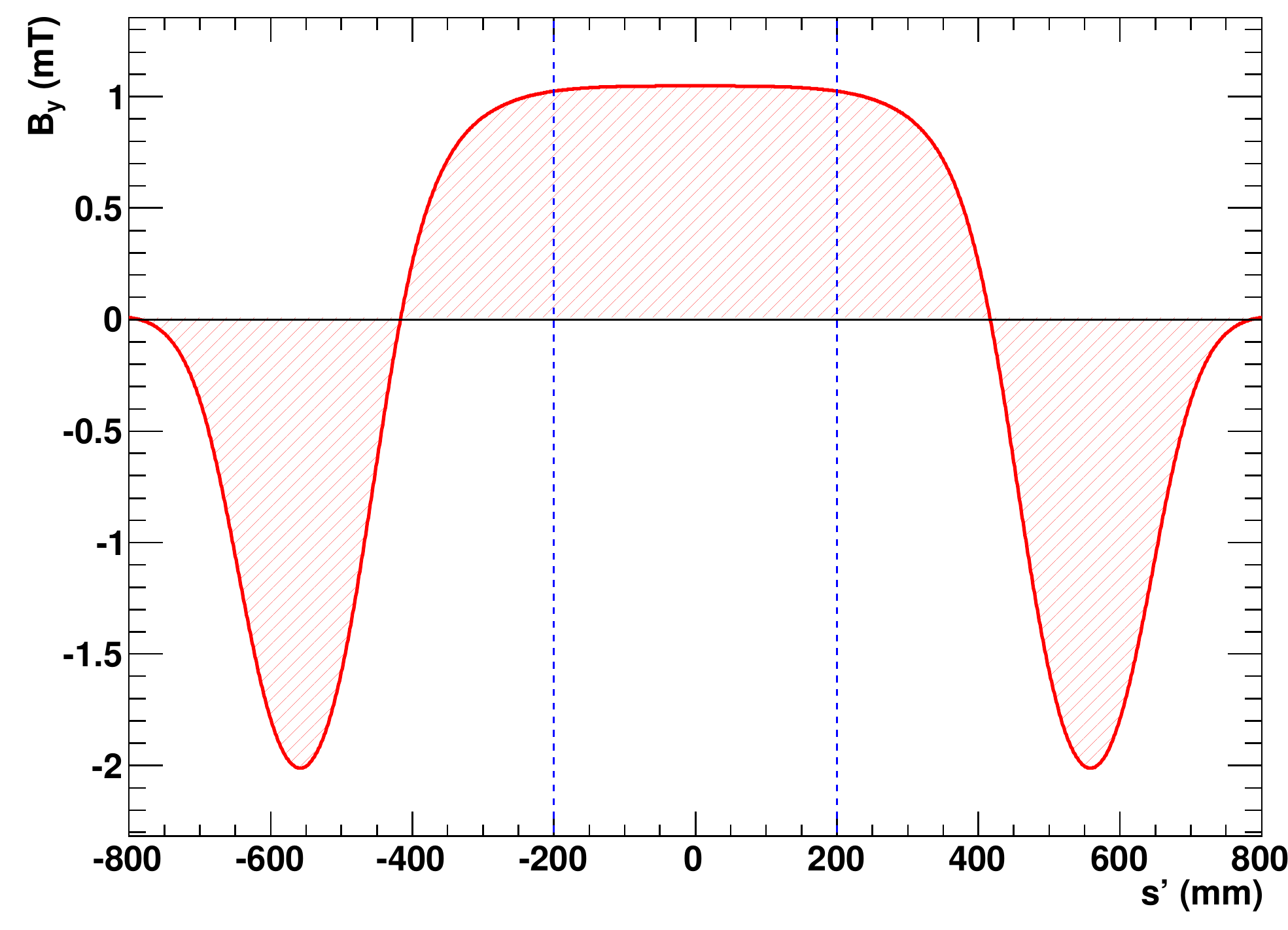}
\caption[]{Calculated vertical magnetic flux density $B_{y}$ along the $s'$-direction for the coil configuration of figure~\ref{fig:coils}. In the target cell region ($s'=-200\,$mm to $+200\,$mm), indicated by the vertical dashed bars, the magnetic field is about $1\,$mT.}
 \label{fig:holdingfield}
\end{figure}

Additional coils installed on the up- and downstream ends of the target chamber (see figure~\ref{fig:coils}) made sure that the horizontal and sideways field integrals $\int B_{x,y} ds$ vanish (see figure~\ref{fig:holdingfield}), thereby preventing the beam positions in the rest of the machine from being affected. Holding field and compensation coils require only a single power supply.

A measurement of the magnetic field $B_{y}$ in the center of the target chamber using a Hall probe yielded \mbox{$ B_{y\downarrow}=-1.08\pm0.03\,\rm mT$} and $B_{y\uparrow}=1.10\pm0.03\,\rm mT$,  pointing downward and upward, respectively. This result is in good agreement with the calculated magnetic field of $1.0$ to $1.1\,\rm mT$ inside the storage cell based on the coil geometry shown in figure~\ref{fig:coils}, using the Amperes\footnote{Integrated Engineering Software (IES), Winnipeg, Manitoba, Canada, \url{http://www.integratedsoft.com}} program.

The vertical magnetic guide field causes a deflection of the proton beam in the horizontal direction. 
According to $\vec{F}_{x}=q(\vec{v}_{s}\times \vec{B}_{y})$, for a beam at experiment energy the expected change of the beam position at the target center between the two polarities ($B_{y}=\pm 1$mT) is $\Delta x\approx 0.28\,$mm. A measurement of the beam displacement using the movable frame system (see section~\ref{sec:movableframe}) resulted in $\Delta x=x_{B_{y\uparrow}}-x_{B_{y\downarrow}}=(0.33\pm0.04)\,$mm, confirming independently the magnetic holding field strength of $|B_{y\uparrow,\downarrow}| \approx 1\,$mT.

\begin{figure}[b]
\includegraphics[width=\columnwidth]{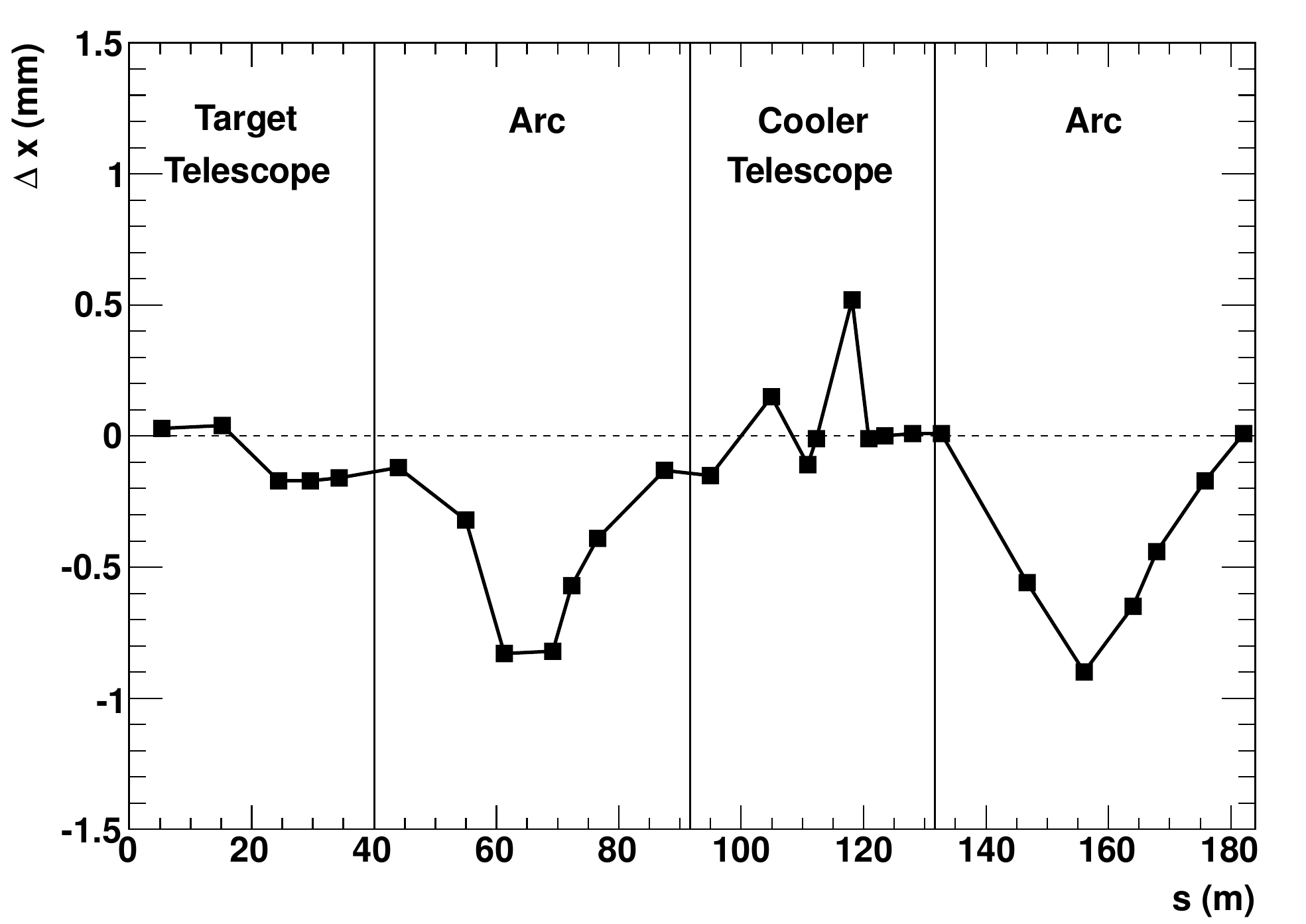}
\caption[Change of the horizontal orbit position for $\pm B_{y}$]{Difference of the horizontal orbit position $\Delta x=x_{B_{y}\uparrow}-x_{B_{y}\downarrow}$ along COSY for reversed vertical holding fields.}
\label{fig:orbitdev}
\end{figure}

The quality of the magnetic compensation scheme was determined using the dispersion-free setting ($D=0$) of the telescopes by measuring the horizontal orbit difference $\Delta x=x_{B_{y\uparrow}}-x_{B_{y\downarrow}}$ for reversed vertical magnetic holding fields ($B_{y\uparrow}$ and $B_{y\downarrow}$) using the beam position monitors (see section~\ref{sec:tools}).  Small orbit differences in the arcs of $\Delta x \leq 0.9$\,mm and in the straight sections of $\Delta x \leq 0.2$\,mm were observed (see figure~\ref{fig:orbitdev}), yielding satisfactory stability of the beam position in the machine.

The largest orbit displacements occur in the arcs, where the dispersion reaches values of $D\approx 15$\,m (see figure~\ref{fig:betafu}, bottom panel). It is interesting to note that according to (\ref{eq:dispersion}) the observed orbit difference in the arcs apparently corresponds to a relative momentum change of $|\Delta p/p|\approx 10^{-5}$, which is probably due to a change of the proton-beam position inside the electron cooler beam when the magnetic holding field changes from $B_{y\uparrow}$ to $B_{y\downarrow}$. 

\subsection{Vacuum system around the target}
\label{sec:Vacuum}
The atomic beam source injected about $3.3\cdot10^{16}$ $\rm H^{0}$/s (one hyperfine state) into the target chamber, thus generating a significant gas load in the region around the PAX target. In the up- and downstream areas where the betatron functions are large (see section~\ref{sec:lowb}), and therefore the acceptance angles are small, single scattering on the residual gas causes beam losses that limit the beam lifetime. In order to minimize these losses, a complex vacuum system was installed. It consists of

\begin{enumerate}
\item ten NEG cartridges\footnote{SAES getter pump GP 500 MK5, a type of vacuum pump manufactured by SAES GETTERS (Deutschland) GmbH, Cologne, Germany, [\url{http://www.saesgetters.com}], sorbs active gases with a nonevaporable getter (NEG) material (Zr-V-Fe alloy).} installed below the target chamber, providing a nominal pumping speed of $10 \times 1900\,\rm \ell/s$ for H$_{2}$ (see figure~\ref{fig:coils}),
\item NEG coating of the beam pipes up- and downstream of the target region with a nominal pumping speed of $2 \times 5000\,\rm \ell/s$~\cite{Chiggiato2006382},
\item flow limiters with an inner diameter of $19\,\rm mm$ and a length of $80\,\rm mm$ (see figure~\ref{fig:coils}) installed at the entrance and exit of the target chamber in order to minimize the gas flow from the target into the adjacent sections without restricting the machine acceptance, and
\item one turbo pump\footnote{HiPace 1800, Pfeiffer Vacuum GmbH, Asslar, Germany, \url{http://www.pfeiffer-vacuum.de/} } with a nominal pumping speed of $1200\,\rm \ell/s$ for H$_{2}$ installed below the target chamber, primarily used during the activation of the NEG pumps.
\end{enumerate}

The NEG coating and the NEG cartridges were activated by heating to $230\,^{\circ}\rm C$ and $450\,^{\circ}\rm C$, respectively, exploiting the fact that the entire low-$\beta$ section is bakeable. Assuming a gas flow of about $3.3\cdot10^{16}\,\rm H^{0}/s$ during operation of the target, approximately one activation per week is required. A jalousie with mirror plates is mounted above the NEG cartridges in order to minimize the heat radiation into the target chamber during activation. The jalousie is closed during heating and opened for pumping. In addition, fast closing valves \footnote{VAT fast closing valve, series 750: DN-100-CF, VAT Deutschland GmbH, Grasbrunn, Germany, \url{http://www.vatvalve.com/} } were installed at the up- and downstream ends of the target chamber, which are capable of sealing the section off from the rest of the ring during bake-out, or in case of a sudden vacuum break.

The vacuum system enabled a base pressure of $2\cdot 10^{-10}$\,mbar in the target chamber and less than $10^{-11}$\,mbar in the adjacent sections when the polarized target is switched off. During operation of the polarized target the pressure never exceeded about $10^{-7}$\,mbar in the target chamber and $10^{-9}\,$mbar in the adjacent NEG-coated vacuum tubes.

\section{Beam diagnostic tools}
\label{sec:tools}
Various beam diagnostics systems available at COSY were used to perform the studies described in this paper.

\subsection{Beam current transformer}
\label{sec:BCT}
A beam current transformer (BCT) measures the current of the circulating ion beam. The BCT electronics are based on the DCCT principle (DC current transformer) \cite{Unser:1981fh} and can be set to deliver $1\,$V or alternatively $0.1\,\rm V$ output signal for $1\,\rm mA$ of beam current. The BCT signal forms the basis for the measurement of the beam lifetime, which was determined from a continuous record of the beam current as function of time, fitted by an exponential.

\subsection{Beam position monitors}
\label{sec:BPMs}
The beam position monitors (BPM) at COSY are of the electrostatic type.    
Each BPM consists of two pairs of electrodes, providing sensitivity along the $x$ and $y$ direction. The electrodes, diagonally cut from a cylindrical or rectangular stainless steel tube, are matched to the size of the beam tubes in the straight and arc sections (see table~\ref{tab:cosy})~\cite{Maier:1990cr}.  

A bunch of charged particles passing through the device induces a voltage change that depends on the distance of the beam to the electrodes. The voltage difference at the two electrodes  $\Delta=U_{1}-U_{2}$ divided by the voltage sum $\sum=U_{1}+U_{2}$ determines the beam position. 
A Fourier analysis of $\Delta$ as a function of time allows the transverse Fourier components of the beam spectrum to be extracted, which are used to determine the betatron tunes $Q_{x}$ and $Q_{y}$ (described in more detail in section~\ref{sec:tune}).

The sum signal $\sum$ recorded with an unbunched beam was used to determine the longitudinal Fourier components of the beam spectrum, from which the revolution frequency $f$ and the momentum spread $\Delta p$ were obtained. 

The beam-energy loss, caused by the interaction of the beam with the residual gas in the machine and the target gas, leads to a change of the revolution frequency per unit of time and is used to determine the target density (see (7) of \cite{Stein:2008ga}). 

\subsection{Stripline unit}
\label{sec:stripline}
The stripline unit of COSY uses four electrodes mounted azimuthally at 45\,$^{\circ}$ with respect to the $x$ and $y$ direction to excite coherent betatron oscillations \cite{Dietrich:1998qz}. 
The unit is powered with a frequency-swept sine wave voltage. 
The coherent betatron oscillations of the beam as a function of the exciting frequency are recorded with a BPM and Fourier-analyzed to yield the fractional betatron tune, as described in section~\ref{sec:tune}.

\subsection{Movable frame system}
\label{sec:movableframe}
\begin{figure}[t]
 \centering
 \includegraphics[width=0.8\columnwidth]{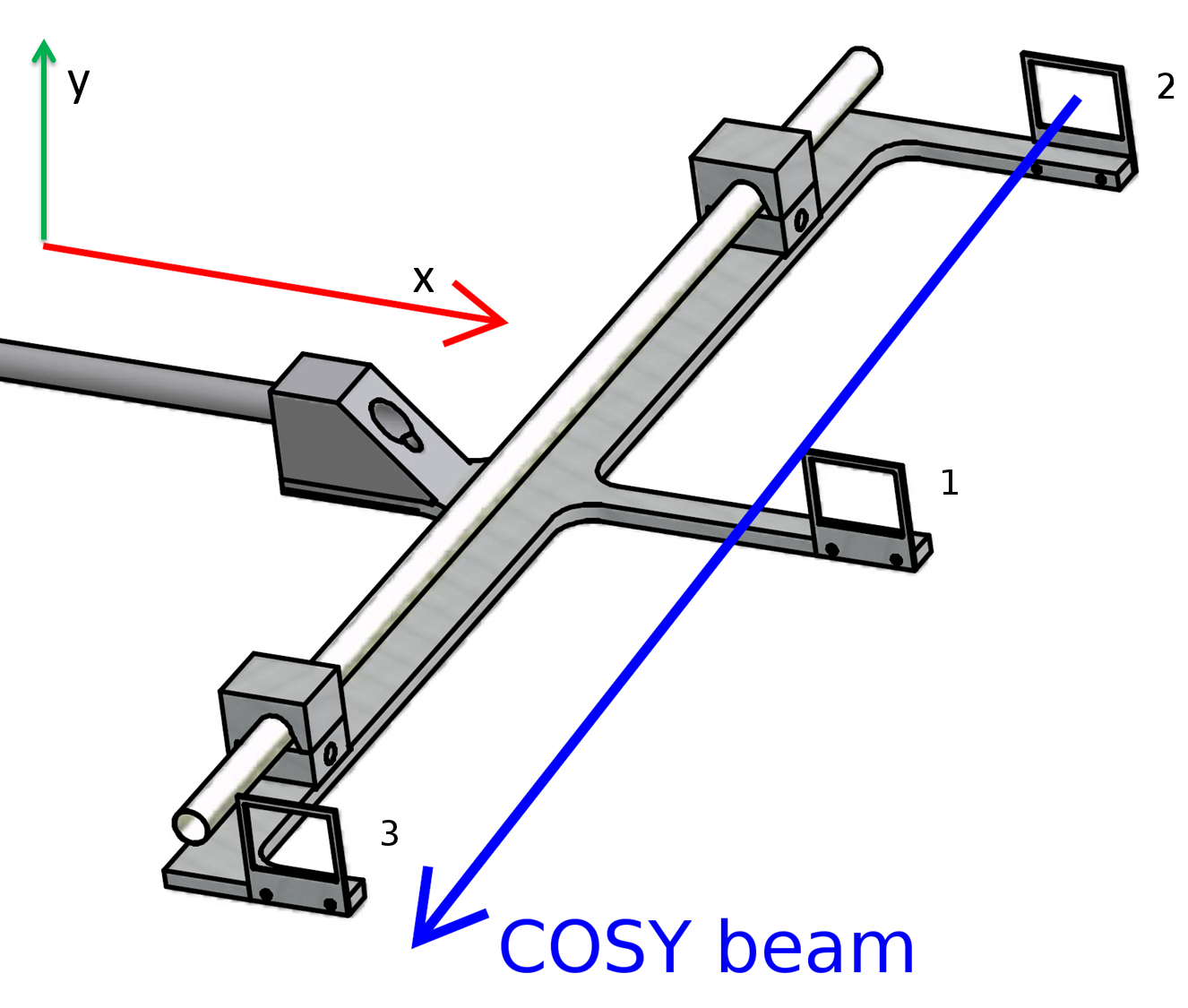}
\caption[Movable frame system]{Movable system with three frames of orifice cross section $w_{x}\times w_{y}\approx 25$\,mm$\,\times\,20\,\rm mm$, at the upstream (2), center (1), and downstream position (3) of the storage cell, and one tube of $9.6\,\rm mm$ inner diameter and $400$\,mm length. The system is movable in the horizontal ($x$) and vertical ($y$) direction perpendicular to the beam while the beam passes through one of the apertures.}
\label{fig:frame}
\end{figure}

A frame system was installed at the PAX target position consisting of three frames and a tube (see figure~\ref{fig:frame})~\cite{Grigoryev:2009zza}. The widths of each frame were determined with a precision of 1\,\textmu m by a coordinate measuring machine. The machine acceptance angles $\Theta_{x}$ and $\Theta_{y}$ at the upstream end, the center, and the downstream end of the storage cell were determined with the beam passing through one of the orifices by moving the system along the $x$ or the $y$ direction and by simultaneously measuring the beam lifetime (see section~\ref{sec:acceptancemeasurement}). The tube was utilized to precisely align the proton beam at the target prior to the installation of the storage cell.    

\subsection{Ionization profile monitor}
\label{sec:IPM}
\begin{figure}[t]
 \centering
 \includegraphics[width=\columnwidth]{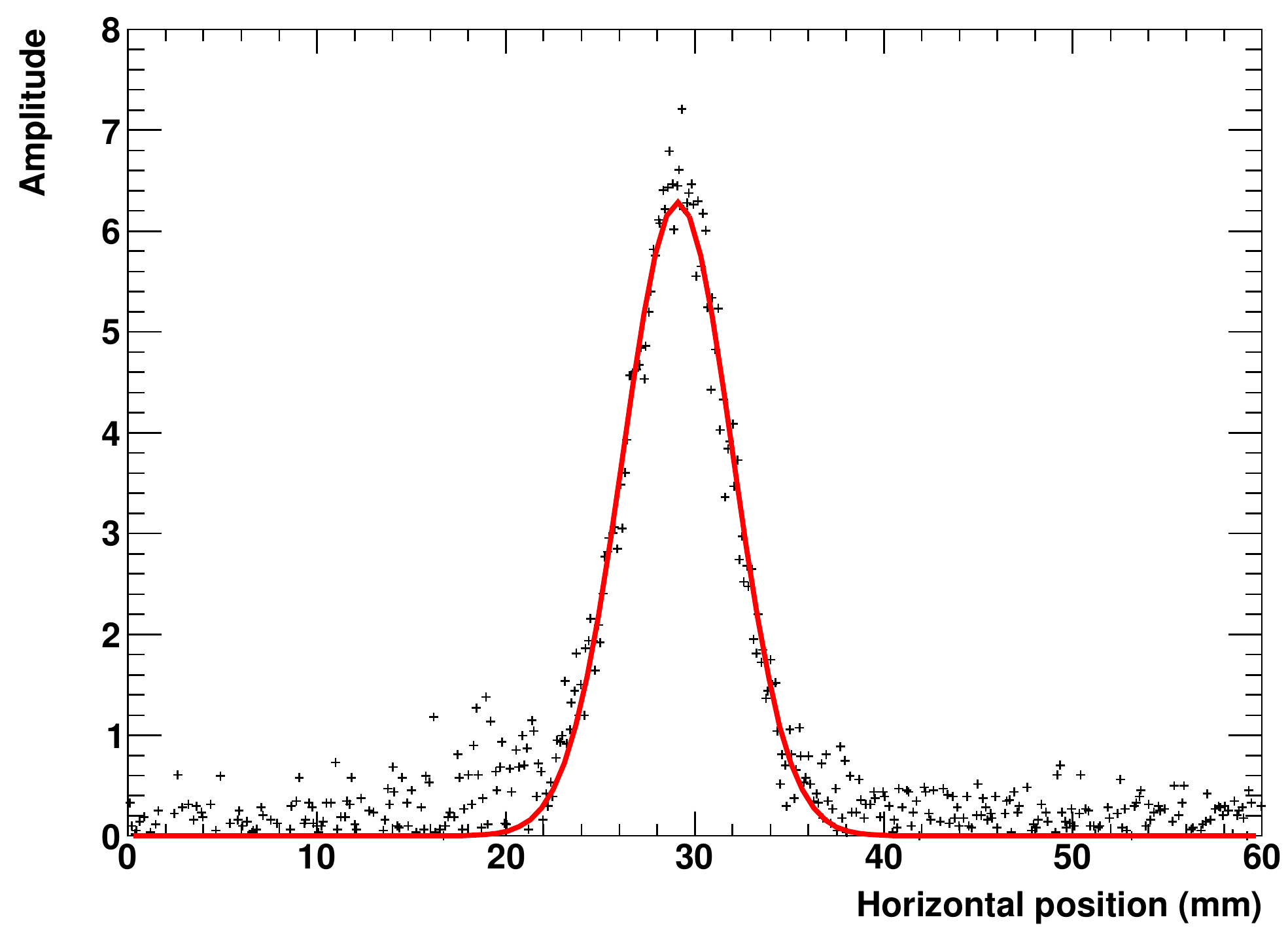}
\caption[IPM profile]{Beam profile measurement in the horizontal plane using the ionization profile monitor. Fitting the particle  distribution by a Gaussian provides the beam width.}
 \label{fig:beamprofile}
\end{figure}

An ionization profile monitor (IPM), developed in cooperation with GSI\footnote{Gesellschaft f{\"u}r Schwerionenforschung mbH, Darmstadt, Germany, \url{https://www.gsi.de/}},
provides a fast and reliable non-destructive beam profile and position measurement~\cite{Boehme:2009}. The interaction of the stored beam with the residual gas produces  ions which are guided to a position-sensitive detector by transverse electric fields. The ion detection is based on an arrangement consisting of microchannel plates (MCP), where secondary electrons are produced, a phosphor screen to produce light, and a CCD camera to detect the light. The system enables a continuous recording of the beam width during the cycle with a resolution of $0.1\,\rm mm$~\cite{Forck:2010zz}. The measured distribution of ions is fitted by a Gaussian (see figure~\ref{fig:beamprofile}). The resulting beam widths $2\sigma_{x,y}$ are used to calculate the $2\sigma$ beam emittances,
  \begin{linenomath}
  \begin{equation}
  \epsilon_{x,y}=\frac{(2\sigma_{x,y})^{2}}{\beta_{x,y}}\hspace{0.15cm},
  \label{eq:emit}
  \end{equation}
  \end{linenomath}
  where $\beta_{x,y}$ represent the $\beta$-functions in the horizontal ($x$) and vertical ($y$) plane at the location of the IPM.
\subsection{H$^{0}$ monitor}
A small fraction of protons and electrons recombines in the electron cooler to neutral H$^{0}$ atoms, which are not deflected in the magnetic elements.
The H$^{0}$ monitor~\cite{Stein:2004}, located at the end of the cooler straight section, records the H$^{0}$ beam profile using a multiwire proportional chamber, while scintillators are used to determine the intensity of the H$^{0}$ beam.
In particular, the H$^{0}$ beam intensity provides an indispensable tool to properly set up the electron cooler and to monitor its performance.

\subsection{Beam polarimeter (ANKE)}
\label{sec:beampolarimeter}
The beam polarization after spin filtering was measured using $\vec{p}d$ elastic scattering, described  in detail  in~\cite{Oellers:2009nm, Augustyniak:2012vu}. The ANKE deuterium cluster-jet target~\cite{Khoukaz:1999} provides target densities of about $1.5\cdot10^{14}$\, deuterons per cm$^{2}$. Elastically scattered particles were detected in the silicon tracking telescopes (STTs)~\cite{Schleichert:2003ec} located left and right of the cluster target at the ANKE interaction point (see figure~\ref{fig:cosy}), allowing the determination of the beam polarization from the measured left-right asymmetry and the analyzing power of $\vec{p}d$ elastic scattering~\cite{King1977151}.

\section{Betatron tune and orbit adjustment}
\label{sec:tuneandorbit}
Before actual commissioning of the low-$\beta$ section, suitable betatron tune settings and corrections to the machine orbit had to be carried out in order to provide good starting conditions for further optimization of the machine with respect to the beam lifetime (section~\ref{sec:tune}).

In the following section, in particular the mapping of the betatron tunes under different conditions and the coupling of the horizontal and vertical phase space are discussed. The implemented closed orbit correction procedures aimed at a reduction of the local acceptance limitations in the machine in order to optimize the beam lifetime (section~\ref{sec:orbit}).

\subsection{Betatron tune mapping}
\label{sec:tune}
The particles circulating in COSY with frequency $f$ perform betatron oscillations in the horizontal ($x$) and vertical ($y$) plane which are induced by the focusing strength of the quadrupole magnets in the ring. To first order the betatron motion constitutes a sinusoidal wave with frequency $f_{\beta_{x,y}}=f\cdot Q_{x,y}$, where $Q_{x,y}$ denotes the betatron tunes (or working point). The number of betatron oscillations per turn is given by
\begin{linenomath}
\begin{equation}
 Q_{x,y}=\frac{\Delta \psi_{x,y}}{2\uppi}=\frac{1}{2\uppi}\oint\frac{ds}{\beta_{x,y}(s)}\hspace{0.15cm}.
\label{eq:tune}
\end{equation}
\end{linenomath}
Here $\Delta\psi_{x,y}=\psi_{x,y}(s+C)-\psi_{x,y}(s)$ is the phase change per revolution, and $C$ the ring circumference. 

At COSY, in order to analyze the betatron tune of the machine, a network analyzer is used to induce coherent transverse betatron oscillations of the beam by powering the stripline unit (see section~\ref{sec:stripline}) with a frequency-swept sine wave voltage, covering the frequency range of a sideband. These oscillations are detected by a position-sensitive pickup and the output signals are analyzed with a spectrum analyzer. The resulting spectrum consists of a series of lower ($-$) and upper ($+$) betatron sidebands at each revolution harmonic $n$ with center frequencies of
\begin{linenomath}
\begin{equation}
 f_{-}=(n-q_{x,y})f \hspace{0.5cm} \mathrm{and} \hspace{0.5cm} f_{+}=(n+q_{x,y})f \, ,
\label{eq:betafrequency}
\end{equation}
\end{linenomath}
where $f$ denotes the average revolution frequency. Since the betatron motion is sampled by the pickup once per turn, the measured spectrum provides only information about the fractional tune $q_{x,y}=\mathrm{frac}(Q_{x,y})$, where $Q_{x,y} = \mathrm{int}(Q_{x,y}) + q_{x,y}$. The fractional tune is deduced from the peak value of both sideband frequencies, and the revolution frequency is found by adding them up. Inserting the resulting value for $f$ into (\ref{eq:betafrequency}) yields $q_x$ and $q_y$.

Because of the symmetry in a synchrotron such as COSY, the magnetic structure after each full turn merges into itself. Consequently, the forces on the beam recur periodically, and, therefore, the betatron tunes should be irrational numbers in order to avoid  betatron resonances that can lead to an expansion of the beam or even to beam loss. The resonance condition is given by
\begin{linenomath}
\begin{equation}
 mQ_{x}\pm nQ_{y}=l \hspace{1.3cm} m,n,l \in \mathbb N
\label{eq:resonance}.
\end{equation}
\end{linenomath}

In order to increase the beam lifetime, a search for the optimal betatron tunes was performed for several machine settings, and, to this end, different tune combinations $(Q_x,Q_y)$ were investigated. In this procedure, called tune-mapping, the currents in the quadrupole magnet families QU1-3-5 and QU2-4-6 were varied in the range of $\pm 3\%$, while the beam lifetime was determined from an exponential fit to the beam current using the BCT signal (see section~\ref{sec:BCT}). 

\begin{figure}[t]
 \includegraphics[width=\columnwidth]{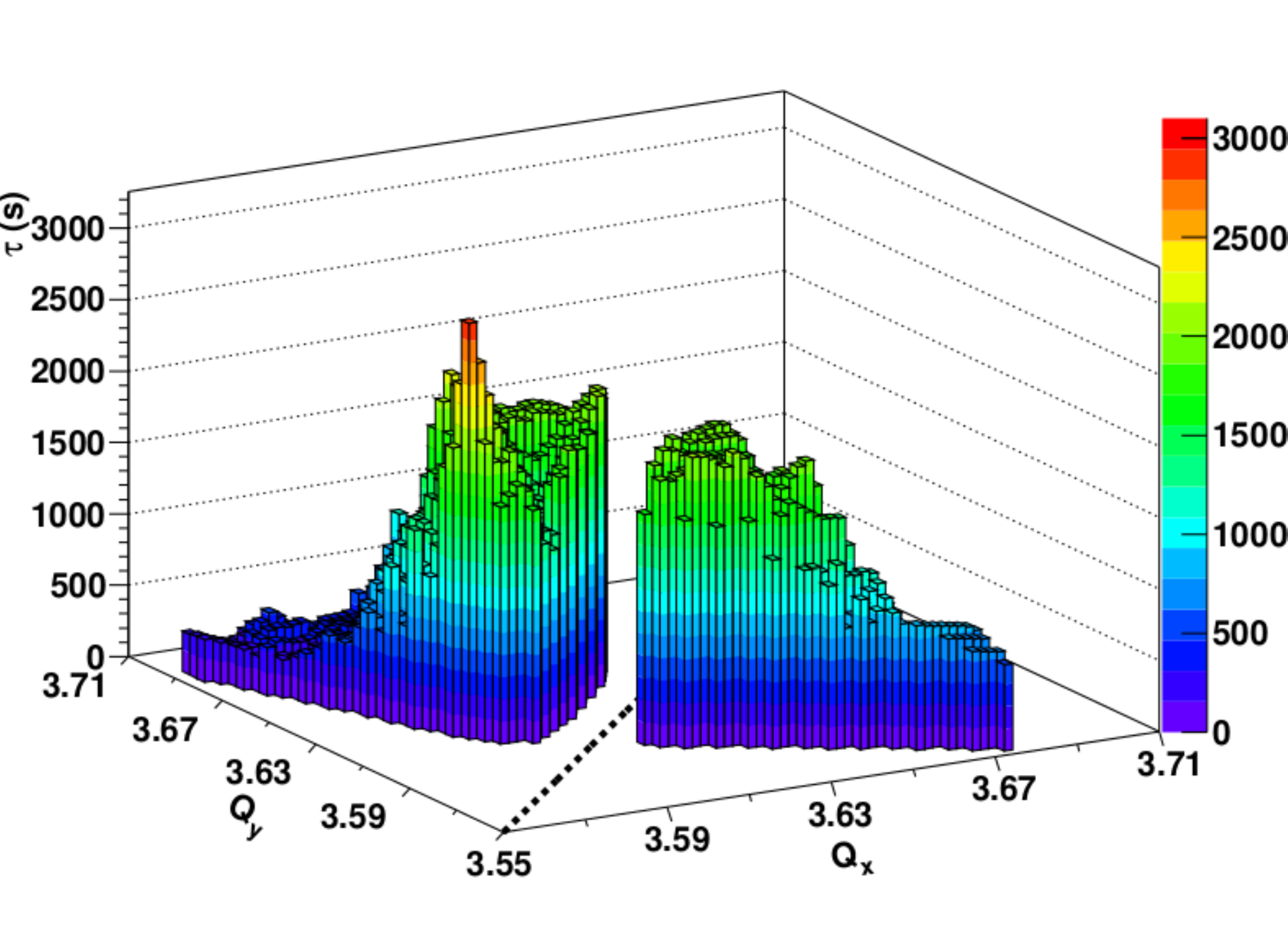}
 \caption[Beam lifetime as a function of the working point.]{Beam lifetime as a function of the working point (Q$_{x}$, Q$_{y}$). While the beam lifetime increases with decreasing distance to the difference tune $Q_{x}=Q_{y}$ (dashed line), $\Delta Q^{\rm{split}}=0$, cannot, however, be achieved because of coupling.}
 \label{fig:eqtunes}
\end{figure}

\begin{figure}[t]
 \includegraphics[width=\columnwidth]{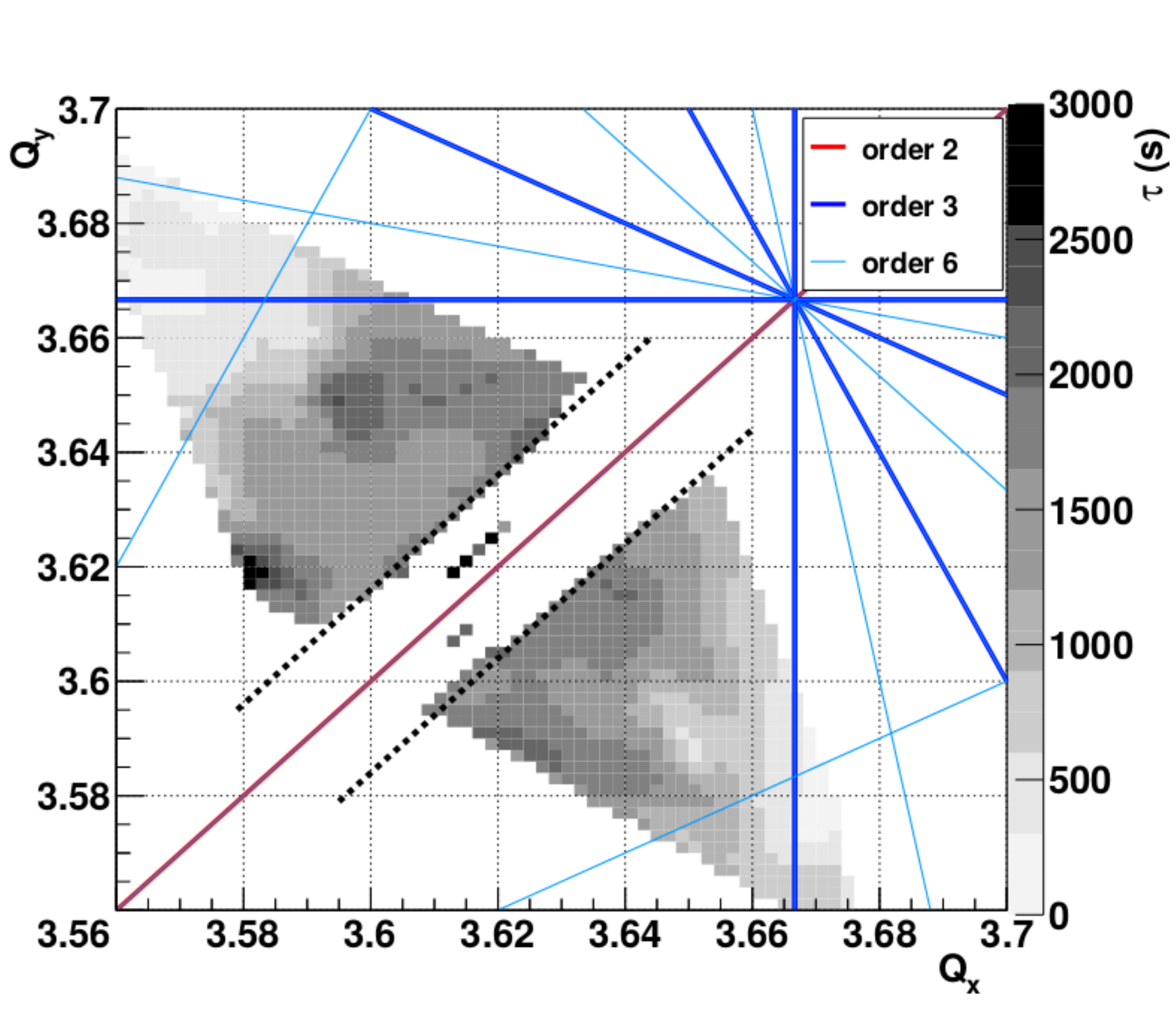}
   \caption{Beam lifetime as a function of the betatron tunes $Q_x$ and  $Q_y$. The data from figure~\ref{fig:eqtunes} are shown here again together with additional data points in the central region close to $Q_x = Q_y$ (low coupling), which were obtained by an adjustment of sextupole magnets.  The dashed lines correspond to $|\Delta Q^{\rm{split}}|=0.014$. Sum and difference resonances of the second, third, and sixth order are shown, representing the strongest multipole components of dipole and quadrupole magnets  in the machine.}
   \label{fig:eqtunes2}
\end{figure}

The betatron tune scans, carried out with $D \neq 0$ setting of COSY, showed a large variation in the beam lifetime by a factor of six in a rather small region of betatron tunes (see figure~\ref{fig:eqtunes}). Maximum beam lifetimes were observed close to the standard COSY working point of $Q_x=3.58$ and $Q_y=3.62$. This is in good agreement with tracking calculations carried out for COSY using MAD-X \cite{Grote:1991zp}.
The impact of the third and sixth order machine resonances on the beam lifetime is clearly visible, as shown in figure~\ref{fig:eqtunes2}.

An early investigation of the COSY beam lifetime as a function of the betatron tunes $(Q_x,Q_y)$ confirmed that the beam lifetime increased with decreasing tune split $\Delta Q^{\rm{split}}= Q_{x}-Q_{y}$ (see figure~\ref{fig:eqtunes}), as mentioned in~\cite{Cappi:2001aq}. Coupling between the horizontal and vertical betatron oscillations leads to a rotation of the eigenvectors of the transverse oscillations, thus the difference resonance \mbox{$\Delta Q^{\rm{split}}=0$} cannot be reached.
Betatron motions can be coupled through solenoidal and skew-quadrupole fields. The latter arise, for instance, from quadrupole rolls and feed-downs from higher-order multipoles caused by an off-axis beam orbit \cite{Lee:2004}. 
The observed tune split $\Delta Q^{\rm{split}}=0.014$ (shown in figure~\ref{fig:eqtunes2}) cannot be attributed to phase-space coupling induced by the main solenoid and the two compensation solenoids of the electron cooler, because they were operated in compensation mode (see section~\ref{sec:lattice}).

Applying additional corrections by using the COSY sextupole magnets of proper polarity led to a reduced coupling and yielded $\Delta Q^{\rm{split}}\approx 0.006$. The 7 additional data points originating from this correction are shown in the central corridor in figure~\ref{fig:eqtunes2}. This indicated that the coupling might originate from sextupole components in the fields of the dipole magnets which affect the beam in an off-axis position. This conclusion was confirmed in later measurements, performed to commission the low-$\beta$ insertion (see figure~\ref{fig:betaf}), which showed that a comparably small $\Delta Q^{\rm{split}}$ could be reached without sextupole corrections by applying instead a closed orbit correction. An independent measurement at COSY with a $232.8$\,MeV deuteron beam~\cite{jedi:2012} arrived at the same conclusion. Starting with a distorted orbit at the acceptance limit yielded $\Delta Q^{\rm{split}}= 0.011$, and by applying a careful closed orbit correction, the coupling was decreased by about a factor of four to $\Delta Q^{\rm{split}} =0.003$.

The achieved tune splits correspond to a small linear coupling in the machine, which is neglected in later considerations.

\subsection{Closed orbit correction}
\label{sec:orbit}
Due to misalignment or field errors of magnets, the real orbit in a machine deviates from the ideal one. In regions where the $\beta$-functions are large, these deviations lead to local restrictions of the machine aperture and thus reduce the lifetime of the beam. A closed orbit correction scheme, based on the orbit response matrix (ORM), was implemented to increase the machine acceptance and to improve the beam lifetime~\cite{Talman1988, Welsch:2009}. In addition, the orbit correction allows one to specify boundary conditions such as the beam position at the target or the electron cooler.

The entries ${\rm R}_{s,i}^{u}$ of the ORM reflect changes of the orbit deviation $u(s)$ ($u=x$ or $y$) measured with a BPM at a position $s$ in the ring, which is caused by a change in the deflection strength $\Theta_{u}(i)$ of a correction-dipole magnet at a position $i$ affecting the beam in the horizontal ($u=x$) or vertical ($u=y$) direction. For $x$ or $y$ these quantities are connected by the relation
\begin{equation}
u(s)={\rm R}_{s,i}^{u}\cdot \Theta_{u}(i),
\label{relation1}
\end{equation}
where 
\begin{equation}
{\rm R}_{s,i}^{u}=\sqrt{\beta_{u,i}\beta_{u,s}}\cdot\frac{{\rm cos}(\pi Q_{u}-\psi_{u,s \rightarrow i})}{2{\rm sin}(\pi Q_{u})}
\end{equation}
depends on the transverse tune $Q_{u}$, on the $\beta$-function at the beam position monitors and correction-dipole magnets, and on the phase advance between the positions $s$ and $i$, denoted by $\psi_{u,s \rightarrow i}$. The ORM can either be calculated for the beam optics of the ring or measured. Here, the latter method was applied. When M horizontal ($x$) and vertical ($y$) BPMs and N$_{x}$ and N$_{y}$ correcting elements are installed, then (\ref{relation1}) is replaced by
\begin{equation}
\vec{u}={\rm} R^{u}\cdot \vec{\Theta}_{u},
\label{relation2}
\end{equation}
where $\vec{\Theta}_{u}$ is a vector of N$_{x}$ or N$_{y}$ components, $\vec{u}$ is a vector of M components, and R$^{u}$ is a M$\times$N$_{x}$ or M$\times$N$_{y}$ matrix with the calculated elements R$_{s,i}^{u}$.

For M $\ge$ N$_{ x}$, N$_{y}$, which was fulfilled in the present studies, the horizontal and vertical closed orbit corrections were derived by varying the $\Theta_{u}(i)$ kick angles to find the minimum quadratic residual $\mid{\rm R}^{u}\cdot \vec{\Theta}_{u}-\vec{u}\mid ^{2}$~\cite{Talman1988, Welsch:2009,Chao:1999}. This method was also used in the present studies. Another possibility uses the inversion of the ORM, where the appropriate settings are calculated from $\vec{\Theta}_{u}=R^{-1}\vec{u}$. This method is usually faster, though it should be noted that an inversion of the matrix $R$ is not always possible.

The closed orbit correction procedure for COSY was tested for the first time in January 2009 within the framework of a PAX beam time and has been further optimized since then with the aim of achieving longer beam lifetimes at injection energy. The measurement of the ORM made use of up to N$_{y}=17$ vertical orbit correction dipole magnets for the measurement of the vertical ORM. 20 horizontal orbit correction dipole magnets, two horizontal back-leg windings at the ANKE dipole magnets, and both compensation dipole magnets next to the electron cooler toroid magnets were used for the determination of the horizontal ORM, i.e., N$_{x}=24$. Depending on their availability, up to $\rm{M}=31$ beam position monitors were employed. The M $\ge$ N$_{ x}$, N$_{y}$ required above was always fulfilled. Phase-space coupling was neglected in these measurements. The beam was deflected in both transverse planes by changing the current of a particular correction dipole magnet by about $5\%$. The orbit changes at the BPMs, 
normalized to 
the variation of the current, correspond to the entries of the ORM. In spite of the longer 
computation time, a $\chi^{2}$ minimization was used to determine the correction angle kicks $\vec{\Theta}_{u}(i)$.
\begin{figure}[t]
\centering
 \includegraphics[width=\columnwidth]{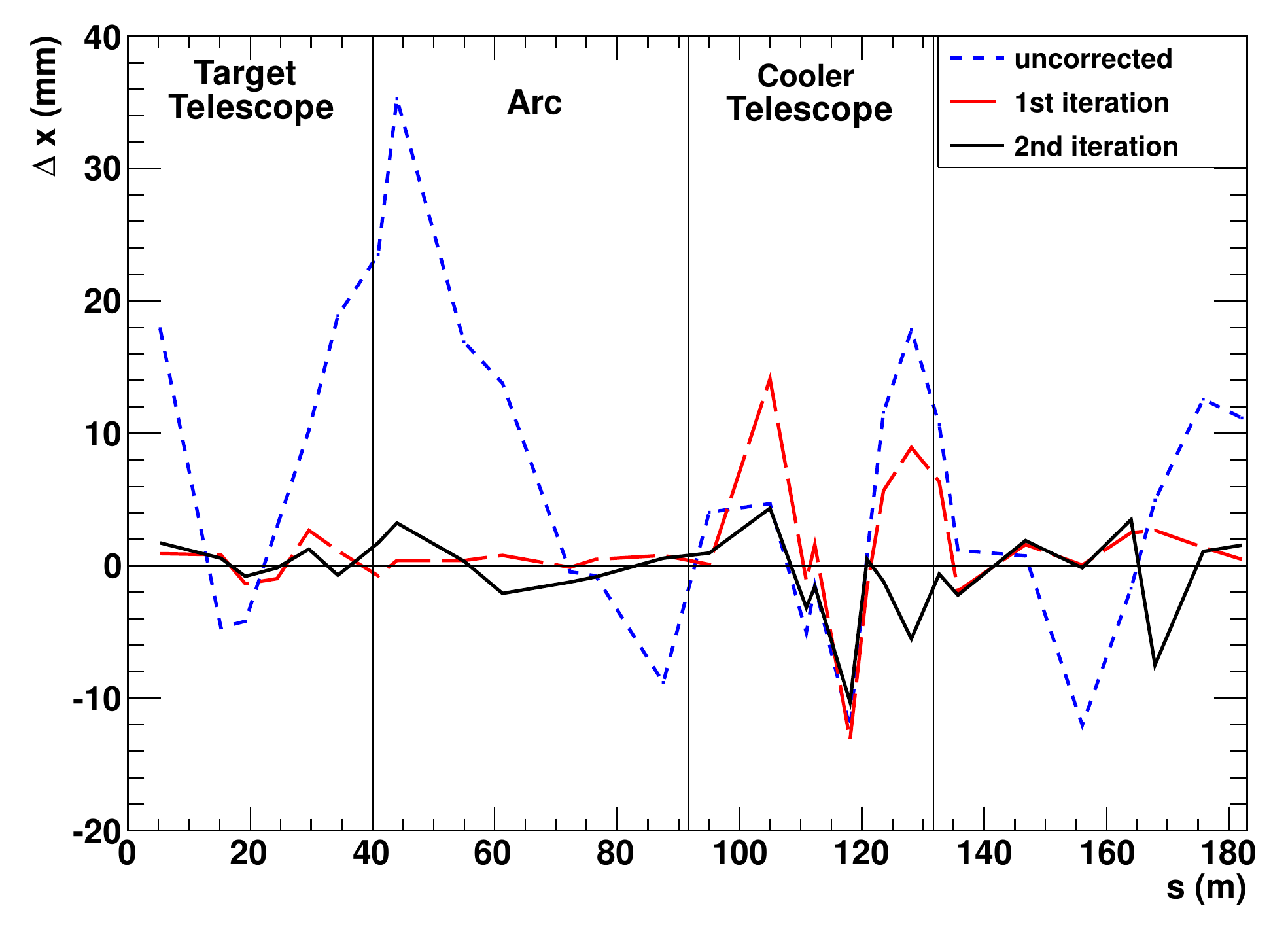}
 \includegraphics[width=\columnwidth]{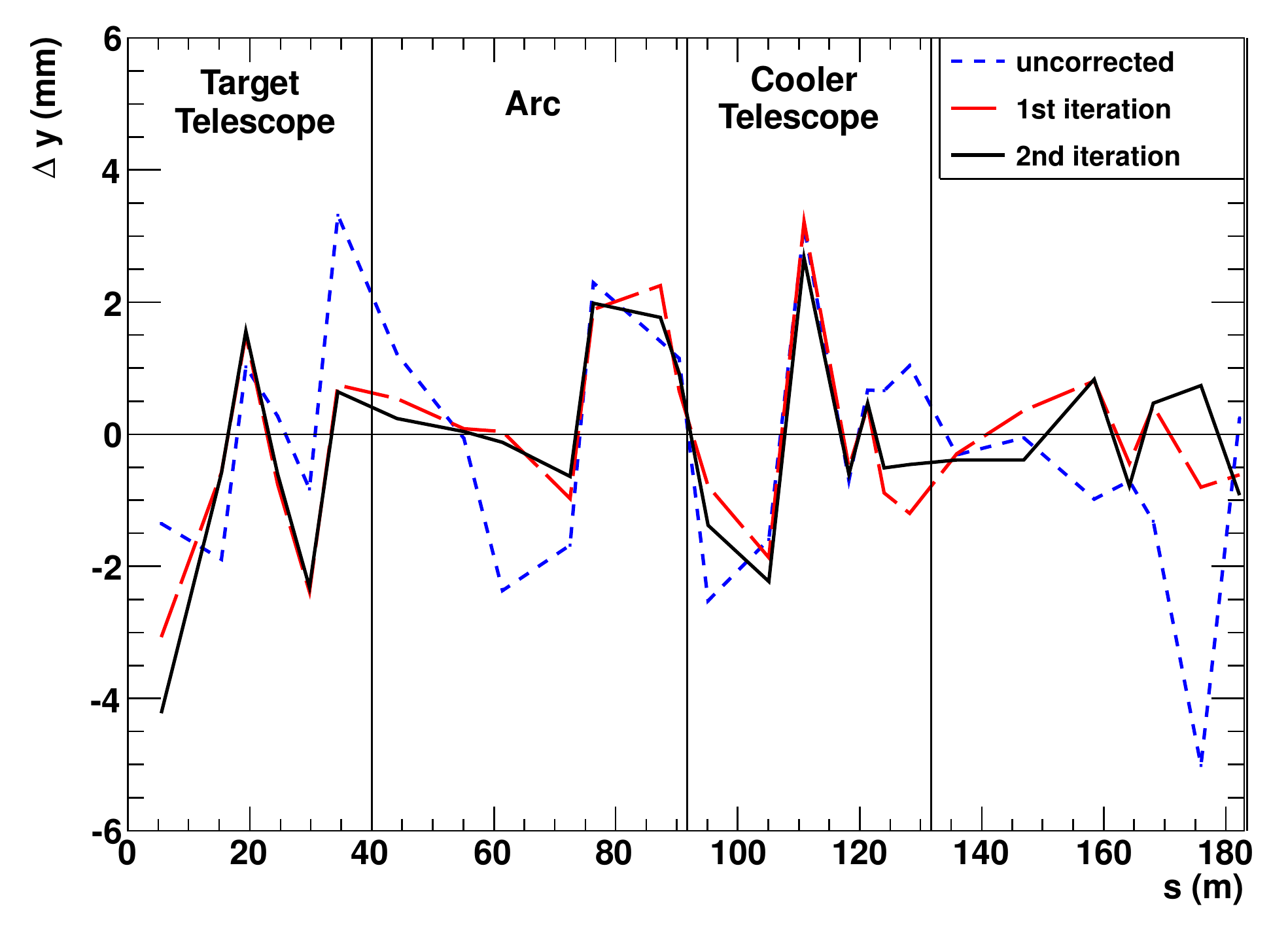}
\caption[Closed orbit correction]{Effect of horizontal (top) and vertical (bottom) closed orbit correction for COSY at injection energy (\mbox{$T_{p}=45\,\rm{MeV}$}).  The initial vertical orbit deviations are in general smaller than the horizontal ones.}
 \label{fig:orbit}
\end{figure}

A typical example of a closed orbit correction with two iterations is shown in figure~\ref{fig:orbit}. The vertical COSY orbit displays smaller deviations than the horizontal one. For the horizontal orbit correction, the initial deviations of up to $35\,\rm mm$ were decreased to less than $10\,\rm mm$. Recent studies, carried out in 2011, show that with the presently available instrumentation at COSY the limit of the orbit correction procedure is  
$\Delta x= \Delta y\approx 3\,\rm mm$. 

\section{Commissioning of low-$\beta$ insertion}
\label{sec:lowbeta}
Prior to the polarization build-up measurements, the low-$\beta$ insertion (see section~\ref{sec:lowb}) was commissioned in a dedicated beam time. The aim was to achieve betatron amplitudes at the target center of about $\beta_{x,y}\approx 0.3$\,m without significant reduction of the beam lifetime. MAD calculations~\cite{Grote:1991zp} verified that the PAX low-$\beta$ quadrupoles have to provide 10 to 40 times larger focusing strengths than the regular COSY quadrupole magnets in order to achieve the required small $\beta$-functions at the target. Horizontal or vertical displacements of the beam in the strong low-$\beta$ magnets would cause large orbit excursions along the ring. Therefore, a careful closed orbit correction (see section~\ref{sec:orbit}) and selection of a reasonable working point (see section~\ref{sec:tune}) were carried out prior to the commissioning to avoid beam losses when the low-$\beta$ quadrupole magnets are operated.

\begin{figure}[t]
 \centering
 \includegraphics[width=\columnwidth]{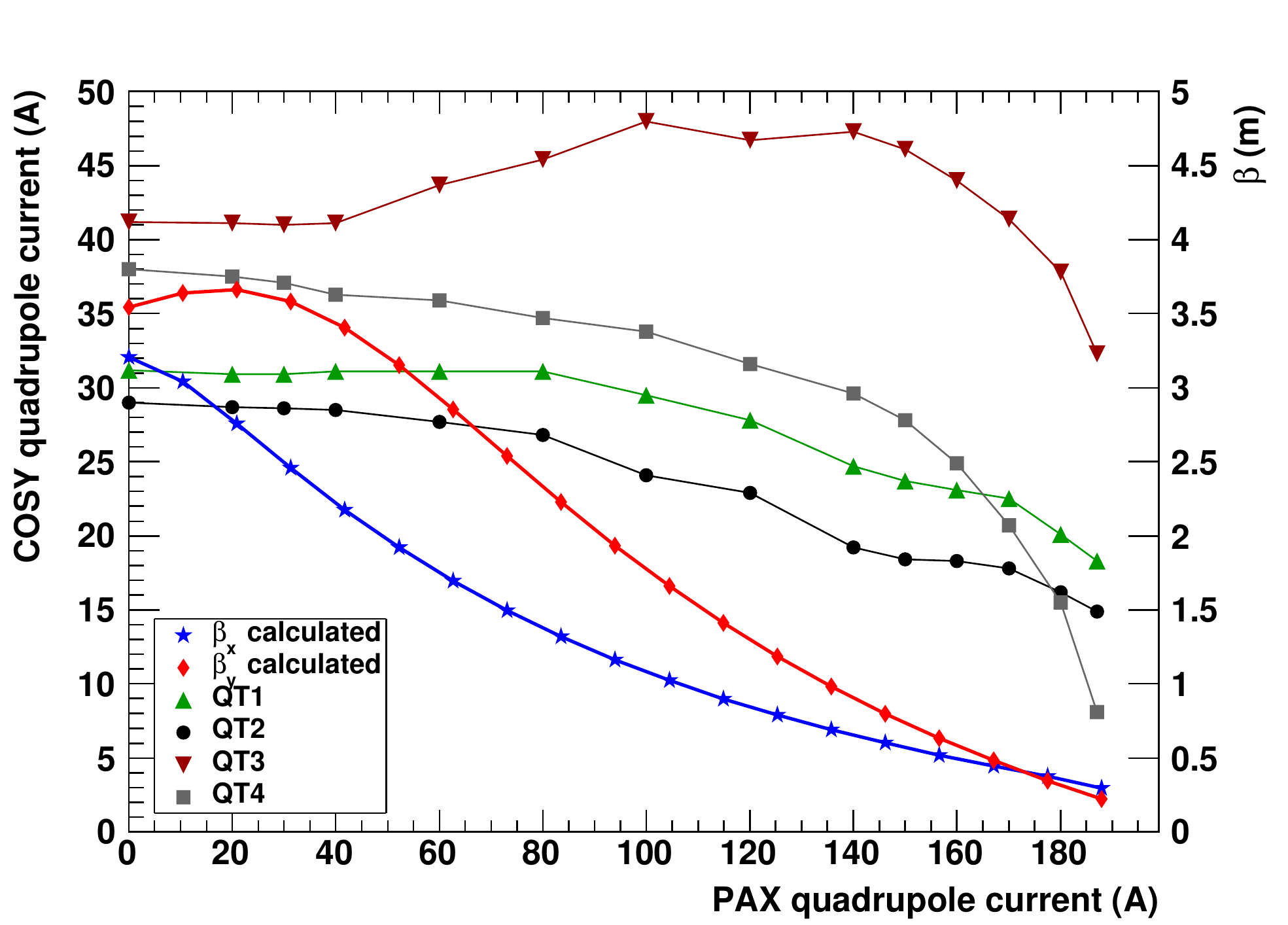}
\caption[]{Currents in the COSY quadrupoles and the betatron amplitudes vs current of the PAX low-$\beta$ quadrupoles. With increasing strength of the low-$\beta$ magnets, the betatron amplitudes $\beta_x$ (in blue) and $\beta_{y}$ (in red)(right scale) decrease. The currents in the COSY quadrupoles (QT1-QT4) were reduced to keep the tune constant, using a $D \neq 0$ setting.}
\label{fig:quadstrength}
\end{figure}

The goal of operating the low-$\beta$ insertion  while maintaining the telescopic features of the straight section was accomplished using as a starting point a regular COSY optics setting at $T_{p}=45$\,MeV, with dispersion $D\neq 0$ and low-$\beta$ section switched off. Subsequently, the fields of the low-$\beta$ quadrupole magnets were increased in strength stepwise, while those of the COSY quadrupoles in the same straight section were reduced in strength such that the betatron tunes remained constant. Figure~\ref{fig:quadstrength} shows the current in the COSY quadrupole families QT1-QT4 vs the current in the PAX low-$\beta$ magnets found in this process. 
The MAD model was used to calculate the $\beta$-function  at the center point of the insertion (see figure~\ref{fig:quadstrength}, right scale).  The strengths of the low-$\beta$ PAX quadrupole magnets were reduced in the calculation by an empirical value of 4\% to achieve stable solutions in the lattice calculations.

In order to verify the validity of the lattice model, the $\beta$-functions at the PAX quadrupoles were experimentally determined by changing the quadrupole strength and measuring the tune change of the machine.
\begin{figure}[t]
\centering
\includegraphics[width=\columnwidth]{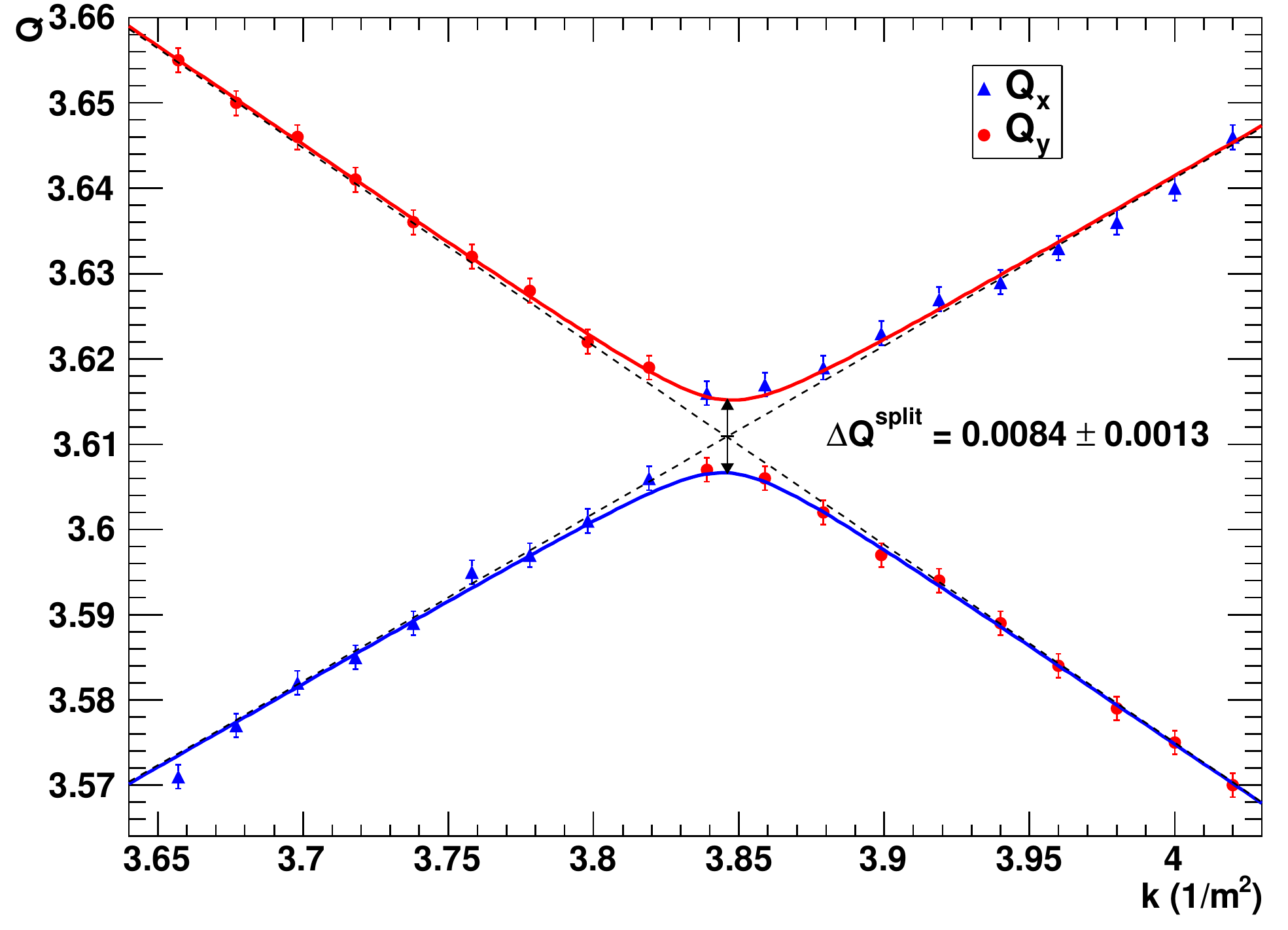}
\includegraphics[width=\columnwidth]{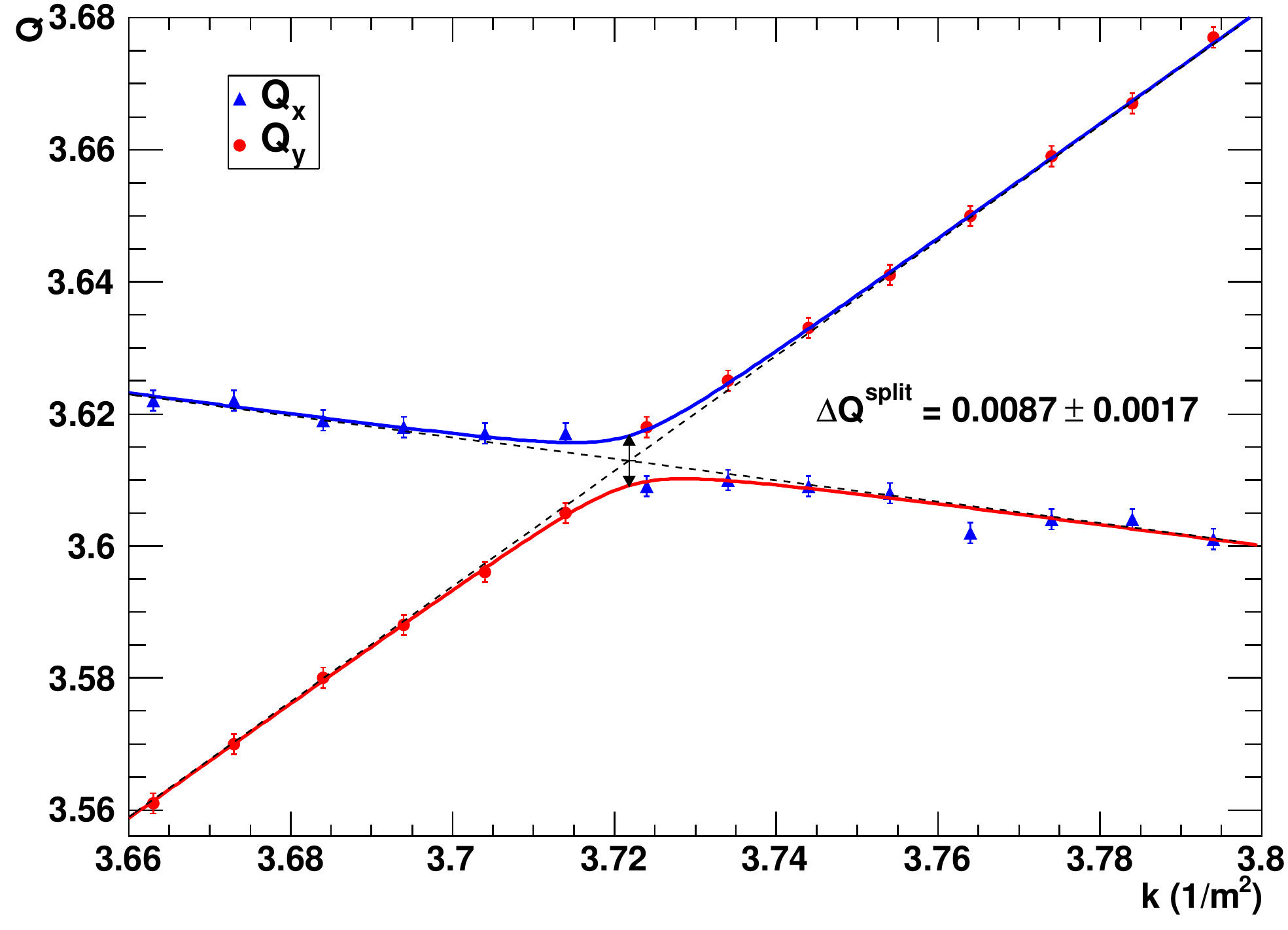}
\caption[Betatron tune change vs focusing strength $k_{x,y}$]{Betatron tunes $Q_{x}$ (horizontal) and $Q_{y}$  (vertical) as a function of the focusing strength $k_{x,y}$ for the inner (top panel) and outer (bottom panel) pair of the PAX low-$\beta$ quadrupoles. The data were fitted with a hyperbola, and the slopes of the asymptotes $\left|\Delta Q_{x,y}/\Delta k\right|$ were used to determine the $\beta$-functions.
 $\Delta Q^{\rm{split}}$ is a measure of coupling in the machine. }
\label{fig:betaf}
\end{figure}
The quadrupole focusing strength $k=\frac{1}{B\rho}\frac{\partial B_y}{\partial x}=\frac{1}{B\rho}\frac{\partial B_x}{\partial y}$ is given by the magnetic rigidity $B\rho=0.977$\,Tm for the chosen kinetic energy of $T_{p}=45$\,MeV and the magnetic field gradient. The latter is  expressed by $\frac{\partial B_y}{\partial x} = \frac{\partial B_x}{\partial y} =g \cdot I$, where $g=0.0197$\,Tm$^{-1}$A$^{-1}$ denotes the current-specific gradient and $I$ is the operating current. The four PAX quadrupole magnets are powered pairwise. Therefore, the tunes are measured either as a function of the focusing strength, i.e., the operating current of the inner pair (PAX2, figure~\ref{fig:betaf}) or of the outer pair (PAX1). The current of the inner pair was modified in steps of $1\,\rm A$ from $181.4\,\rm A$ to $199.4\,\rm A$, corresponding to the range $k=3.658\,\rm m^{-2}$ to $k=4.021\,\rm m^{-2}$. The values for the outer pair are steps of 
$0.5\,\rm A$ from $181.7\,\rm A$ to $188.2\,\rm A$, corresponding to the range $k=3.664\,\rm m^{-2}$ to $k=3.795\,\rm m^{-2}$.

In figure~\ref{fig:betaf} the measured tunes $Q_{x}$ and $Q_{y}$ are displayed as a function of the quadrupole strength of the outer pair (PAX1, bottom panel) and the inner pair (PAX2, top panel). According to \cite{Bryant:1992kc}, the functional form of $Q_{x,y}(k)$ is described by a hyperbola. The hyperbolic fits also yield the tune split of $\Delta Q^{\rm{split}}= 0.0085\pm0.0010$, obtained from a weighted average using the outer and the inner quadrupole pair. This constitutes independent evidence for the presence of slight coupling in the machine, as already discussed in section~\ref{sec:tune}.
The crossing points of the asymptotes at $Q=3.611$ for the inner pair and $Q=3.613$ for the outer pair agree within the error of $Q^{\rm split}$ as expected.

\begin{figure}[t]
\centering
\includegraphics[width=\columnwidth]{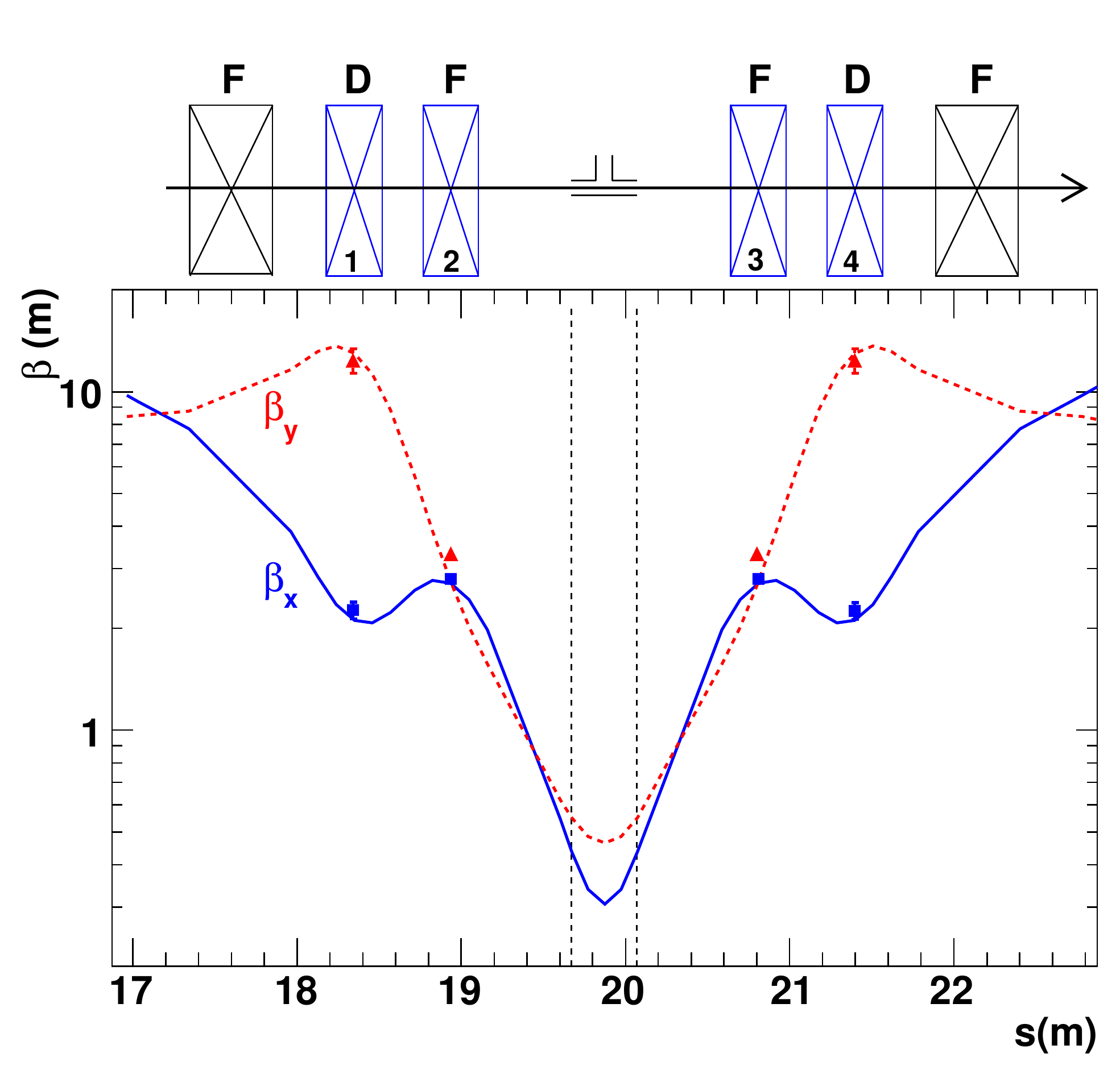}
\caption[Measurement of the $\beta$-function]{Model calculation of the $\beta$-functions at the PAX-TP and measured values of $\overline{\beta}_{x}$ and $\overline{\beta}_{y}$ at the magnet positions. The four new PAX quadrupole magnets are shown in blue. Magnets 1 and 4 form the defocusing (D) pair (PAX1) and magnets 2 and 3  the focusing (F) pair (PAX2), where each pair is operated with a single power supply. In addition, the storage cell and the beam direction are shown.}
\label{fig:betafun}
\end{figure}

The ion-optics matrix formalism for a change of the quadrupole focusing strength $\Delta k$ yields a tune shift \cite{Wille:2000,Courant:1997rq}
\begin{linenomath}
\begin{equation}
\Delta Q_{x,y}=\frac{1}{4\pi}\int\limits_{s_0}^{s_{0}+l}\Delta k \beta_{x,y}(s) ds\hspace{0.1cm},
\end{equation}
\end{linenomath}
where $\beta_{x,y}(s)$ is the position-dependent $\beta$-function and $l$ is the effective length of the field of the quadrupole magnet. For small $\Delta k$, $\beta_{x,y}(s)$ can be replaced by $\overline{\beta}_{x,y}$, which yields  
\begin{linenomath}
\begin{equation}
\overline{\beta}_{x,y}=\frac{4\pi}{l}\left|\frac{\Delta Q_{x,y}}{\Delta k}\right|.
\label{eq:betaf}
\end{equation}
\end{linenomath}
The absolute value takes into account that the $\beta$-function has to be positive, remembering that a quadrupole focuses in one plane ($\Delta Q_{x,y} > 0$ for $\Delta k >0$) and defocuses in the other plane ($\Delta Q_{y,x} < 0$ for $\Delta k >0$). 
To determine the average values $\overline{\beta}_{x}$ and $\overline{\beta}_{y}$ in the magnets of the inner and outer pair using \ref{eq:betaf}, the values of $\mid \Delta Q/\Delta k \mid$ are the absolute values of the four slopes of the asymptotes of the hyperbolas of figure~\ref{fig:betaf}. The effective length of a single PAX quadrupole magnet, measured as 0.442\,m, for each of the pairs, yields $l=0.884$\,m. The resulting $\overline{\beta}_{x}$ and $\overline{\beta}_{y}$ are shown in figure~\ref{fig:betafun} together with the result of the model calculation, which yields reasonable agreement (see table~\ref{tab:beta}) with the measured data and $\beta_{x}=0.31$\,m and $\beta_y=0.46$\,m at the center of the target. 
From a comparison of measured and calculated betatron functions, an uncertainty of about 10\% is estimated for the $\beta$-functions obtained from the MAD model.
 
\section{Beam size, beam emittance, machine acceptance, and target acceptance angle}
\label{sec:acce}
The polarization build-up cross section $\tilde{\sigma}_{1}$ depends on the acceptance angle $\Theta_{\rm{acc}}$ at the target location, as explained in section~\ref{sec:intro}. Therefore, in order to determine $\tilde{\sigma}_{1}$, it is necessary to measure $\Theta_{\rm{acc}}$.
The measurement made use of the fact that when an object is placed at a distance smaller than the maximum allowed extension of the local phase-space ellipse, the  machine acceptance is reduced, and therefore the beam lifetime as well~\cite{Ross1993424,Grigoryev:2009zza}. 
\begin{table}[t]
\centering
\begin{ruledtabular}
\caption[Results of the $\beta-$function measurement.]{Measured and calculated betatron functions $\beta_{x}$ and $\beta_{y}$ from the MAD model at the position of the PAX quadrupole magnets (outer pair: PAX1, inner pair: PAX2). The calculated $\beta$-functions at the target center are given in column six.}
\begin{tabular}{cccccc}
  & \multicolumn{2}{c}{Measurement} & \multicolumn{3}{c}{Model calculation} \\
  & PAX1 & PAX2 & PAX1 & PAX2 & center \\
  \hline
 $\beta_{x}$ (m) & \,\,\,2.31 $\pm$ 0.13 & 2.80 $\pm$ 0.04 & \,\,\,2.11 & 2.71 & 0.31\\
 $\beta_{y}$ (m) & 12.41 $\pm$ 1.01 & 3.31 $\pm$ 0.05 & 12.99 & 2.74 & 0.46\\
 \end{tabular}
  \label{tab:beta}
\end{ruledtabular}
\end{table}

In the subsequent section, we first describe the determination of the beam width at the target, since it may have some bearing on the machine acceptance extracted from a measurement with the scraper system, described in section~\ref{sec:movableframe}. The actual acceptance measurements, including the determination of $\Theta_{\rm{acc}}$ and a discussion of possible systematic errors, are discussed in section~\ref{sec:acceptancemeasurement}.

\subsection{Measurement of the beam widths at the target}
\label{sec:beamwidth}
\begin{figure}[t]
 \centering
 \includegraphics[width=\columnwidth]{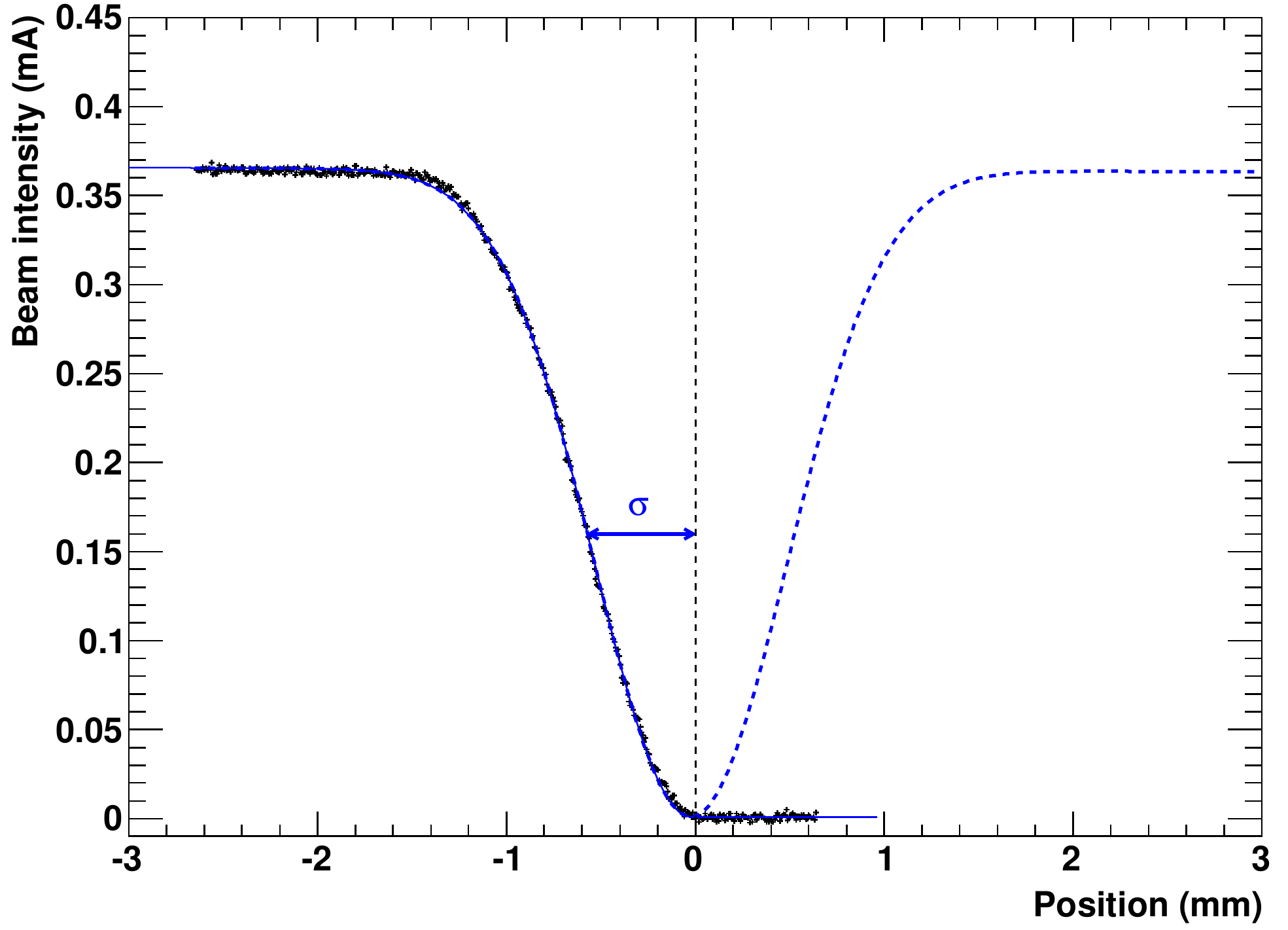}
 \caption[Determination of the beam width.]{Measured beam intensity as a function of frame position obtained by moving the frame through the beam. The resulting beam profile (black points) constitutes half of an inverted Gaussian from which $\sigma$ as a measure of the beam width is obtained by fitting using (\ref{eq:Iofx}) (dashed blue line). The beam intensity with the frame in the nominal position of $I_{0}=0.35$\,mA corresponds to about $4.67\cdot10^{9}$ protons at injection energy. }
 \label{fig:beamsize}
\end{figure}

The beam widths along the PAX target were determined by moving each of the three rectangular frames (shown in figure~\ref{fig:frame}) at constant speed through the proton beam. The decrease of the beam current was recorded with the BCT (see section~\ref{sec:BCT}). A typical result of such a frame scan is shown in figure~\ref{fig:beamsize}. The remaining beam intensity as a function of the frame position is obtained by converting the measured time into the distance from the start position, using the constant velocity of the frame movement of
\begin{linenomath}
\begin{equation}
 v_{x}=v_{y}=(1.65\pm 0.02)\,\rm mm/s.
\label{eq:speed}
 \end{equation}
\end{linenomath}

The measured beam profile constitutes half of an inverted Gaussian when the beam itself has a Gaussian profile \cite{Potter:1984}. Assuming no coupling in the machine (see section~\ref{sec:tune}), a scraper moving along the $x$ (or $y$) direction removes only those particles from the $(x,x')$ (or $(y,y')$) phase space for which the betatron amplitudes are  larger than the distance from the beam center to the edge of the scraper (see figure~1 of \cite{Grigoryev:2009zza}).

A cooled and stored beam exhibits a two-dimensional Gaussian distribution in transverse phase space where the density distribution of the betatron amplitude $\rho_{\beta}$ in  \textit{e.g.}, the $(x,x')$ plane \cite{Potter:1984, Hinterberger:1989bf} is given by
\begin{linenomath}
\begin{equation}
 \rho_{\beta}(x)=\frac{I_{0}}{\sigma_{x}^{2}}\cdot x\cdot \exp\left (
-\frac{x^{2}}{2\sigma_{x}^{2}}\right )\,.
\end{equation}
\end{linenomath}
The measured beam intensity as a function of frame position can be written as \cite{Grigoryev:2009zza},
\begin{linenomath}
\begin{eqnarray}
 I_{\rm{frame}}(x)&=&\int\limits_{0}^{x-\mu_{x}}\rho_{\beta}(x)\cdot dx\nonumber\\
     &=& I_{0}\left [1-\exp\left (
-\frac{(x-\mu_{x})^{2}}{2\sigma_{x}^{2}}\right ) \right ].
\end{eqnarray}
\end{linenomath}
Here $I_{0}$ is the beam intensity with the frame in the nominal position, $\mu_{x}$ is the beam center, and $\sigma_{x}$ describes the beam width in the $x$-direction. 
Because the beam intensity decreases exponentially before intersecting the frame, the following function
\begin{linenomath}
\begin{eqnarray} 
 I(x)&=I_{\rm{frame}}(x) \cdot \exp{\left(-\frac{\displaystyle x}{\displaystyle \tau_{\rm b}\cdot v_{x}}\right)}
\label{eq:Iofx}\end{eqnarray}
\end{linenomath}
was fitted to the measured beam intensity dependence, shown in figure~\ref{fig:beamsize}, in order to determine $\sigma_{x}$ and $\sigma_{y}$ by the same procedure. Although with coupling or dispersion at the frame position, the functional form is more complicated~\cite{Jansson:2003}, good agreement with the data was achieved using (\ref{eq:Iofx}).

The beam widths were determined for all three frames of the scraper system with the $D=0$ setting (see section~\ref{sec:lattice}) at $T_{p}=45\,\rm{MeV}$.
Horizontally, the frames could be moved in the positive and negative direction, while vertical measurements were only feasible by moving the frames upward, because in the case of the downward movement the beam could not be completely removed due to space limitation. 

The beam widths 2$\sigma_{x}$ and 2$\sigma_{y}$ for each frame were determined by averaging the results of two independent measurements. In the case of the horizontal measurement, 2$\sigma_{x}$ additionally includes averaging the results from both $x$-direction measurements. The results are listed in table~\ref{tab:beamsiz}. Unfortunately, the vertical measurement at the target center ($ s'=0\,$mm) showed distortions that made the result inconsistent.
The measurements confirm that the beam width 2$\sigma_{x}$ is smallest at the cell center and, as expected knowing the $\beta$-functions, increases symmetrically toward the up- and downstream ends of the storage cell. The appropriate $\beta$-functions at the location of each frame were obtained from the validated MAD model (see section~\ref{sec:lowbeta}) and are given in table~\ref{tab:beamsiz}.

The averaged horizontal and vertical beam widths are 2$\sigma_x=1.03 \pm 0.01$\,mm and 2$\sigma_y=0.67 \pm 0.02$\,mm. In terms of these beam widths, the walls of the storage cell ($r_\mathrm{cell}=4.8$\,mm) are at least ten standard deviations away from the center of the beam. 

\begin{table}[t]
\centering
\begin{ruledtabular}
\caption[]{Beam widths (in mm) determined at three positions, center ($s'=0$\,mm) and upstream ($s'=-200$\,mm) and downstream ($s'=+200$\,mm) ends of the PAX storage cell (see figure~\ref{fig:frame}).}
  \begin{tabular}{c|ccc}
   Frame  & 2 & 1 & 3\\
\hline
 Position ($s'$) & $-200\,$mm & $ 0\,$mm & $+200\,$mm\\
$ 2\sigma_{x}$ & $1.04 \pm 0.02$\,\,\, & $0.91 \pm 0.04$\,\,\,& $1.04 \pm 0.01$ \\
$ 2\sigma_{y}$ & $0.66 \pm 0.02$\,\,\,  & $-$ & $0.67 \pm 0.02$\\
$\beta_{x} \, (\mathrm{m})$ & $0.62 \pm 0.06$ & $0.55 \pm 0.06$ & $0.62 \pm 0.06$\\
$\beta_{y} \, (\mathrm{m})$ & $0.48 \pm 0.05$ & $0.38 \pm 0.04$ & $0.48 \pm 0.05$\\
 \end{tabular}
 \label{tab:beamsiz}
\end{ruledtabular}
\end{table}

\subsection{Determination of the beam emittance}
\label{sec:beamemittance}
 The values of the $\beta$-functions allow one to determine the $2\sigma$ beam emittance for each measurement from (\ref{eq:emit}). Weighted averaging of the resulting three horizontal emittances yields
\begin{linenomath}
\begin{equation}
\epsilon_{x}=\frac{(2\sigma_{x})^{2}}{\beta_{x}}=(1.71 \pm 0.17)\,\text{\textmu m}\,,
  \end{equation}
\end{linenomath}
and of the two vertical emittances yields
\begin{linenomath}
\begin{equation}
\epsilon_{y}=\frac{(2\sigma_{y})^{2}}{\beta_{y}}=(0.92 \pm 0.15)\,\text{\textmu m}\,.
\end{equation}
\end{linenomath}
The given uncertainties arise from the uncertainty of the frame velocity, the statistical errors of the fit, and the estimated uncertainty of 10\% on the $\beta$-functions, given in table~\ref{tab:beamsiz}.

\subsection{Determination of $A_x$, $A_y$, and $\Theta_\mathrm{acc}$ at the target}
\label{sec:acceptancemeasurement}
\begin{figure}[t]
 \centering
 \includegraphics[width=\columnwidth]{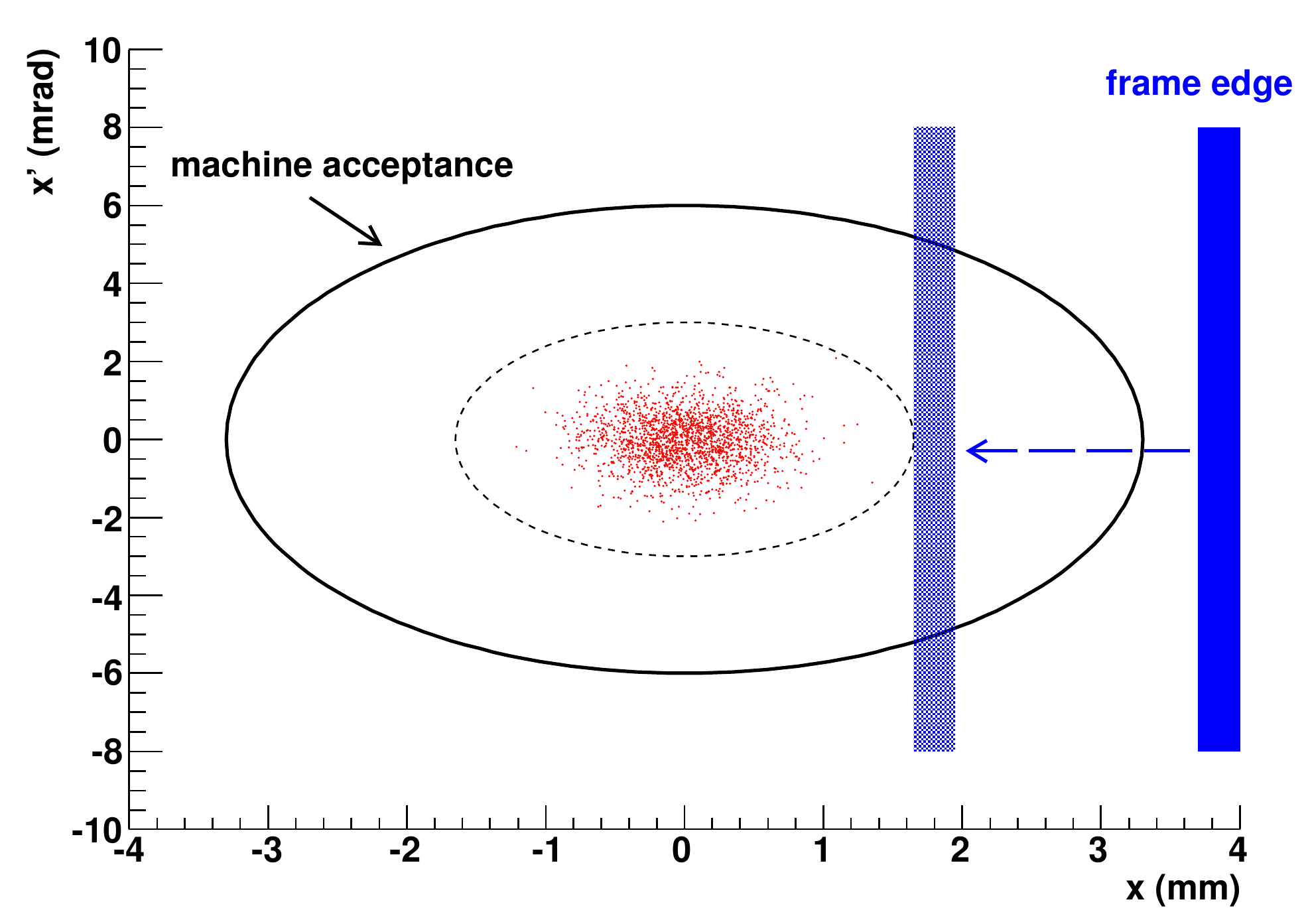}
 \caption[]{Horizontal phase-space distribution at the PAX target position from a Monte Carlo simulation. A typical machine acceptance at COSY ($A_x=20\,$\textmu m, $A_y=15\,$\textmu m) is indicated by the large ellipse. Moving a frame into the machine acceptance decreases both $x$ and $x'$, thus reducing the beam lifetime, and allowing $A_{x}$ and $\Theta_{x}$ to be determined.}
 \label{fig:acce}
\end{figure}

The acceptance of a storage ring is defined in (\ref{eq:acceptance}). At every point in the ring, the acceptance $A_{x,y}$ corresponds to a (horizontal and vertical)  phase-space ellipse~\cite{Lee:2004}. When at some point along the orbit, a restriction (frame) is moved into the machine acceptance, \textit{e.g.}, in the horizontal ($x$) direction, the maximum $(x,x')$ phase-space ellipse, representing the machine acceptance at that location, is intersected, and accordingly the beam lifetime is reduced (see figure~\ref{fig:acce}). Every particle orbits on an individual phase-space ellipse in $(x,x')$ and $(y,y')$, and all ellipses at a specific location in the ring have the \textit{same shape}~\cite{Chao:1999}. While the insertion of the frame initially only presents a limitation of the $x$ coordinate, because of the betatron motion, the $x'$ coordinate is also affected. Therefore, the beam lifetime as a function of the frame position was measured in order to determine the machine acceptance and the acceptance 
angle at the target.

The total beam lifetime due to single Coulomb scattering is found to be (see (\ref{eq:appB}) and (\ref{eq:acc}))~\cite{Madsen:2000ep},
\begin{linenomath}
\begin{eqnarray}
\tau'_{\rm b}(A_{x},A_{y})=c\cdot\Theta_{\rm{acc}}^2&=&2c\cdot\left(\frac{1}{\Theta_{x}^2}+\frac{1}{\Theta_{y}^2}\right)^{-1}\nonumber\\
&=&2c\cdot\left(\frac{\langle\beta_x\rangle}{A_x}+\frac{\langle\beta_y\rangle}{A_y}\right)^{-1},
\label{eq:lifetime}
\end{eqnarray}
\end{linenomath}
where $c$ is a constant during the measurement, $\langle\beta_x\rangle$ and $\langle\beta_y\rangle$ are, respectively, the average horizontal and vertical $\beta$-functions along the ring, and the $x$- and $y$-acceptance is either given by the ring acceptance $A^{\mathrm{ring}}_{x,y}$ or the acceptance is defined by the frame position $A^{\mathrm{frame}}_{x,y}=a_{x,y}^2/\beta_{x,y}$ (see (\ref{eq:acceptance})), whichever is smaller.
Here $a_{x,y}$ are the distances of the restriction to the beam center and $\beta_{x,y}$ are the $\beta$-functions at the location of the frame.

\begin{figure}[t]
 \centering
\includegraphics[width=\columnwidth]{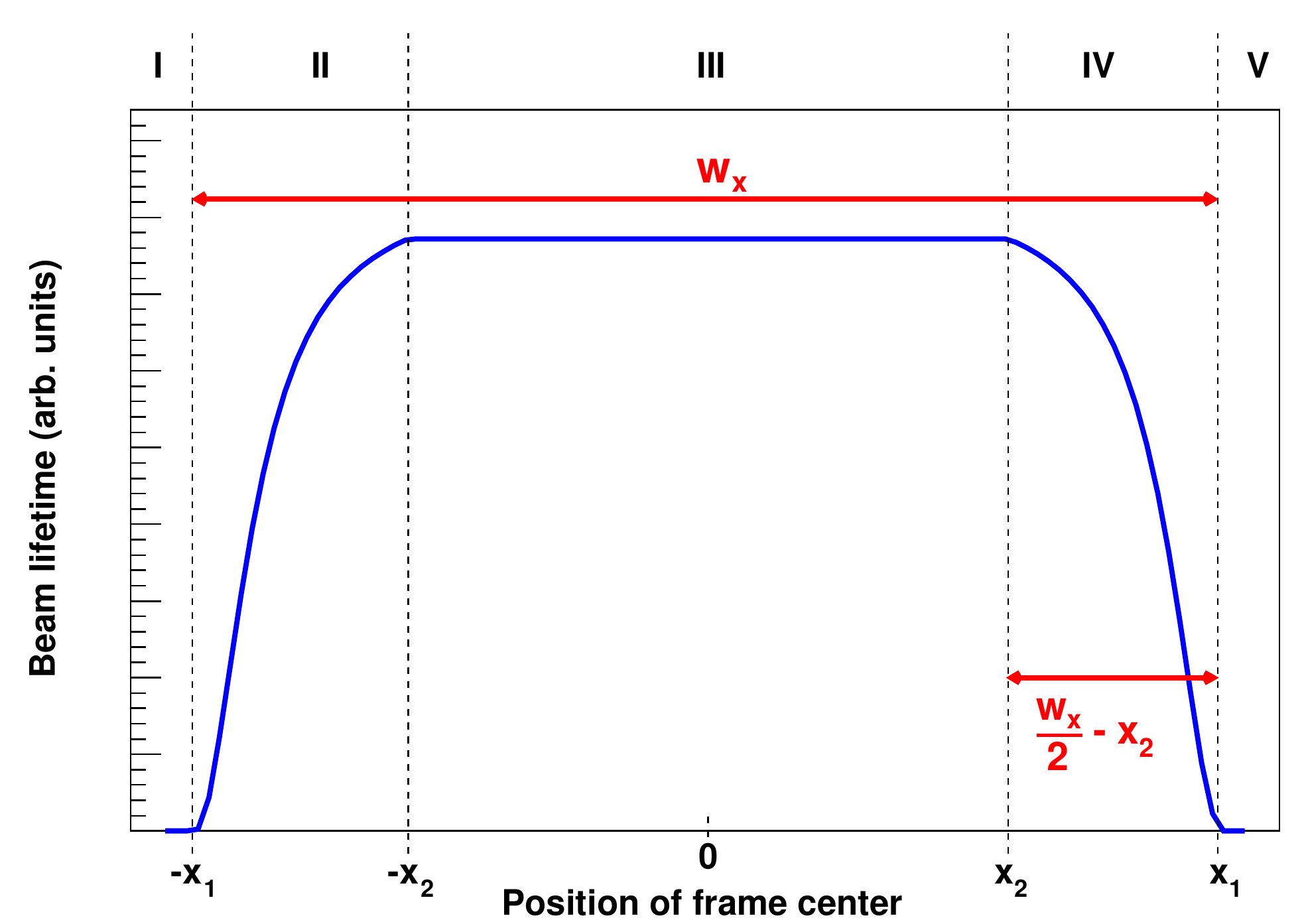}
\caption[]{Schematic of the beam lifetime as a function of the position of the frame during an acceptance measurement. As long as the frame is outside the machine acceptance (part III), the beam lifetime is unchanged. When the frame limits the acceptance the beam lifetime drops (parts II and IV) according to (\ref{eq:piecewise}), and when it intersects the beam itself, the beam lifetime vanishes (parts I and V). The acceptance is then defined by the frame width $w_{x}$ and the position where the frame enters the machine acceptance $x_{2}$.
}
 \label{fig:intersection1}
\end{figure}

In the following, the acceptance measurement in $x$-direction is exemplified (see figure~\ref{fig:intersection1}). The measurement begins with the frame horizontally and vertically centered on the beam ($x=0$). During the horizontal movement of the frame $A_{y}$ is constant. As long as the frame does not limit the machine acceptance ($|x|\leq |x_{2}|$), the beam lifetime is not affected (part III in figure~\ref{fig:intersection1}). 
When the frame moves into the machine acceptance ($|x_{2}|\leq|x|\leq |x_{1}|$), $A_{x}$ and therefore the beam lifetime become smaller (parts II and IV).
When it reaches a position of $|x|\geq|x_{1}|$ the measured beam lifetime vanishes (parts I and V). Theoretically, the beam lifetime should vanish to zero when the frame reaches the center of the beam, corresponding to a position of $|x|=w_{x}/2$, where $w_{x}$ is the measured frame width (see section~\ref{sec:movableframe}). 

Based on these considerations, the following fit function is formulated, using (\ref{eq:lifetime}),  

\begin{linenomath}
\begin{equation}
\tau_{\rm{b}}(x) = \begin{cases}
\begin{array}{lc}
0 & \quad\\
\tau'_{\rm{b}}(A_{x}^{\mathrm{frame}},A_{y}^{\mathrm{ring}})&\quad\\
\tau'_{\rm{b}}(A_{x}^{\mathrm{ring }},A_{y}^{\mathrm{ring}})&\quad\\
\tau'_{\rm{b}}(A_{x}^{\mathrm{frame}},A_{y}^{\mathrm{ring}})&\quad\\
0 & \quad \\
\end{array}
\hspace{-0.2cm}
\begin{array}{rrclc}
 \text{if} & x&\leq&-x_{1} & \textbf{I} \\
 \text{if} & -x_{1}&\leq&x\leq -x_{2}&\textbf{II}\\
 \text{if} & -x_{2}&\leq&x \leq x_{2}&\textbf{III}\\
 \text{if} & x_{2}&\leq&x \leq x_{1}&\textbf{IV}\\
 \text{if} & x&\geq&x_{1}&\textbf{V}
\end{array}
\end{cases} 
\label{eq:piecewise}
\end{equation}
\end{linenomath}

where $A_{y}^{\mathrm{ring}}$, $x_{1}$, and $x_{2}$ are fit parameters. 
The machine acceptance is determined from the distance between $x_{2}$ and the beam center by
\begin{linenomath}
\begin{equation}
A_{x}=\frac{(w_x/2-x_{2})^2}{\beta_{x}}\,.
\end{equation}
\end{linenomath}
The offset of the beam with respect to the center of the frame can be determined with a typical uncertainty of 0.1\,mm. For clarity, the offset parameter has been omitted in (\ref{eq:piecewise}), but is taken into account in the actual fitting function. 

Monte Carlo simulations of an acceptance measurement using realistic phase-space distributions at the PAX target position showed good agreement between simulated data and the fit function (see (\ref{eq:piecewise})) for typical beam sizes at the target (see section~\ref{sec:beamwidth}). 

\begin{figure*}[t]
 \centering
 \includegraphics[width=0.49\textwidth]{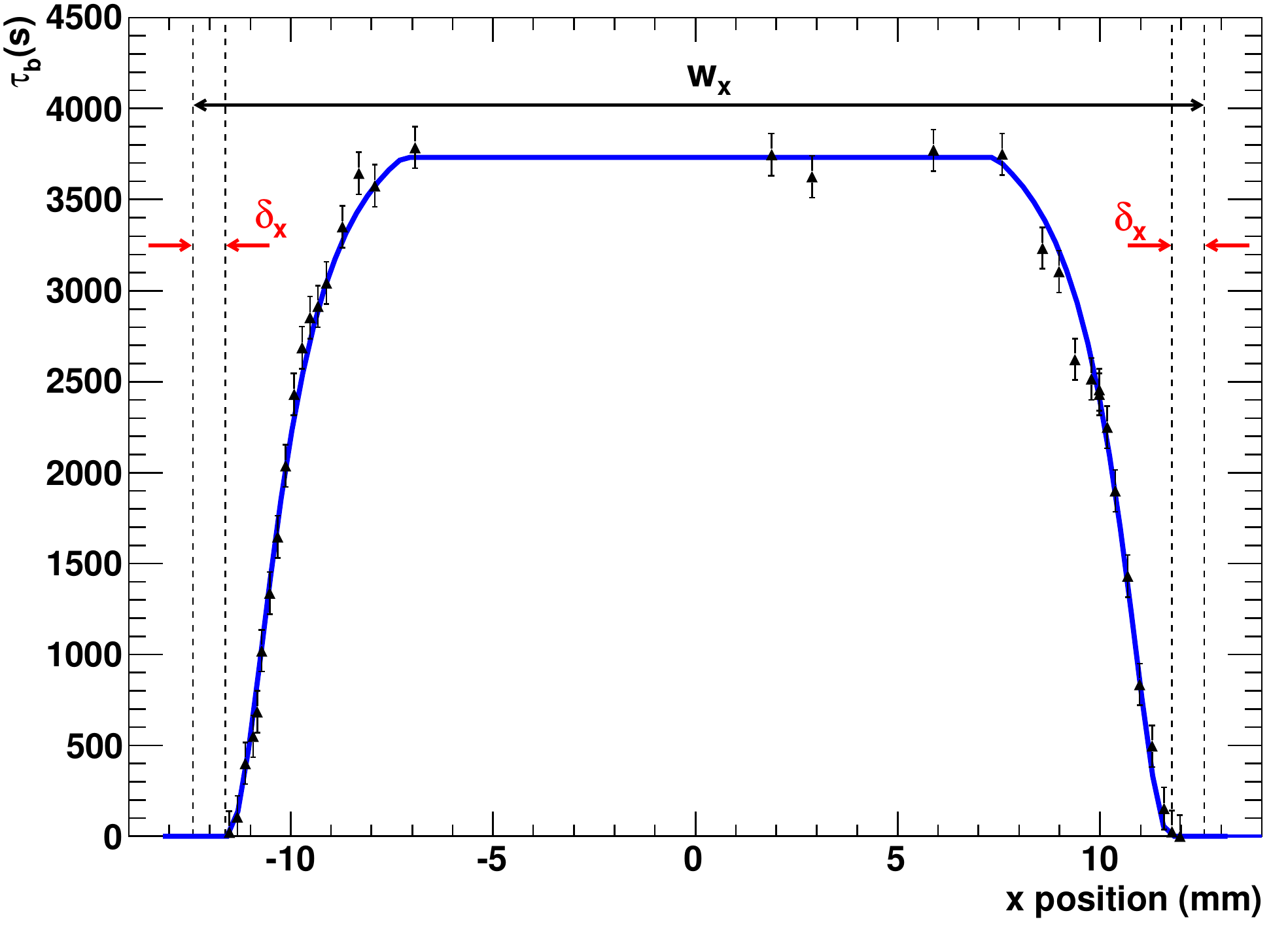}
 \includegraphics[width=0.49\textwidth]{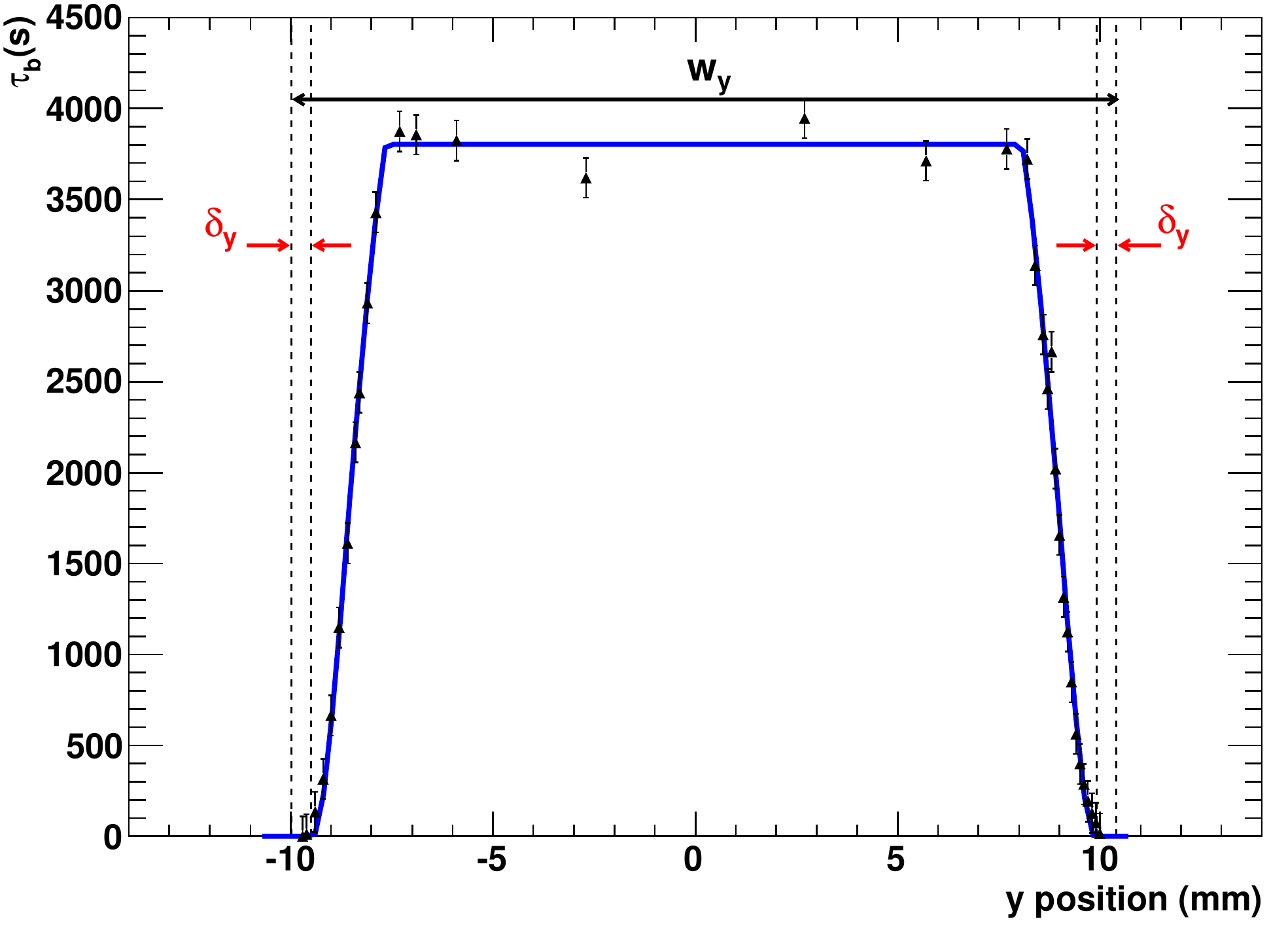}
\caption[]{Recorded beam lifetime as a function of the horizontal ($x$, left panel) and vertical ($y$, right) position of frame 1 (PAX target center), and fit of $\tau_{\rm{b}}(x)$ using (\ref{eq:piecewise}). The fits indicate that the observed widths at the base ($\tau_{b}=0$) were smaller than the corresponding frame widths ($w_x$, $w_y$), where the discrepancies $\delta_{x}=1.0\pm0.1$\,mm and $\delta_{y}=0.5\pm0.1$\,mm.}
 \label{fig:accmeas}
\end{figure*}

\begin{table*}[t]
\centering\small
\begin{ruledtabular} 
\caption[]{Acceptance measurements with the movable frame system, listing the acceptances $A_x$ and  $A_y$, the acceptance angles $\Theta_{x}$ and $\Theta_{y}$, and  $\Theta_\mathrm{acc}$, using (\ref{eq:acceptance}), (\ref{eq:acc}), and the $\beta$-functions given in table~\ref{tab:beamsiz}. One measurement with frame 1 was carried out with the cluster target switched on.
The weighted averages are given in the bottom row. Results were rounded to one decimal place, while three decimal places were used for calculation and averaging.}
\begin{tabular}{ccccccc}
Pos (m) & Frame  &  $A_{x}(\text{\textmu m})$ & $A_{y} (\text{\textmu m})$ & $\Theta_{x}
(\rm mrad)$ & $\Theta_{y} (\rm mrad)$ & $\Theta_\mathrm{acc}(\rm mrad)$\\
\hline
$\,-0.2$ & 2 & $27.1 \pm 4.0$ & $16.8 \pm 3.2$ & $6.6 \pm 0.6$ & $5.9 \pm 0.6$ &
$6.2 \pm 0.4$\\
$\,\,\,\,\,0.0$  & 1  &  $49.7 \pm 10.3$ & $14.1 \pm 6.0$ & $9.5 \pm 1.1$ & $6.1
\pm 1.3$ & $7.3 \pm 1.2$ \\
$\,\,\,\,\,0.2$ & 3  &  $31.5 \pm 4.7$ & $19.6 \pm 3.7$ & $7.1 \pm 0.6$ & $6.4 \pm
0.7$ & $6.7 \pm 0.5$ \\
 $\,\,\,\,\,0.0$ &1 (target on) &  $33.0 \pm 5.1$ & $12.4 \pm 3.1$ & $7.8 \pm
0.7$ & $5.7 \pm 0.8$ & $6.5 \pm 0.6$\\
 Average & & $31.2 \pm 2.5$ & $15.7 \pm 1.8$ & $7.3 \pm 0.3$ &
$6.0 \pm 0.4$ & 6.45 $\pm$ 0.27\\
 \end{tabular}
 \label{tab:results}
\end{ruledtabular}
\end{table*}
The acceptance measurements with the movable frame system (see section~\ref{sec:movableframe}, figure~\ref{fig:frame}) were carried out for all four edges of each of the three rectangular frames. Moving each frame individually into the machine acceptance, while recording the beam lifetime, allowed one to determine the machine acceptance angles at the entrance of the storage cell ($s'=-200\,\rm mm$), at the center (\mbox{$s'=0\,\rm mm$}), and at the exit ($s'=+200\,\rm mm$).
A measurement carried out in the presence of the ANKE cluster target (see section~\ref{sec:beampolarimeter}) showed good agreement of the resulting acceptances.

The acquired dataset enabled a precise determination of the machine acceptance, the acceptance angle in the horizontal and vertical direction, and of the total acceptance angle $\Theta_\mathrm{acc}$ (see \ref{eq:acc}) at the target. During the measurements, the beam intensity was in the range of $(7.5-10)\cdot 10^{9}$ circulating unpolarized cooled protons at the injection energy of $45$\,MeV, with the PAX low-$\beta$ section switched on and an initial beam lifetime of about $3700\,\rm s$.

During injection, the frame was horizontally and vertically centered on the beam. After injection and cooling, the frame was moved in the horizontal (vertical) direction and the resulting beam lifetime was recorded.
An example of a measurement with frame 1, located at the target center, is shown in figure~\ref{fig:accmeas}. 
The uncertainties of the beam lifetimes $\tau_{\rm b}$ are of the order of 100\,s, chosen to yield reduced $\chi^{2}$ of approximately unity for the fits. 

All fits indicate that the beam lifetime actually vanishes before the frame edge intersects the beam center. This is equivalent to stating that the observed width at the base ($\tau_{b}=0$) is smaller than the frame width (see section~\ref{sec:movableframe}), thus $|x_{1}|+\delta_{x} = w_{x}/2$ and $|y_{1}|+\delta_{y} =  w_{y}/2$ (see figure~\ref{fig:accmeas}), where the discrepancy $\delta_{x}$ ($\delta_{y}$) is of the order of $1.0\pm0.1$\,mm ($0.5\pm0.1$\,mm).
Possibly, small beam oscillations of unknown origin are responsible for this observation. It should be noted that the approach of measuring the machine acceptance with a rectangular frame is sensitive to such effects, while this is not the case for a single-sided scraper measurement. Therefore, in the latter case, the machine acceptance might be underestimated. 

The results for $A_{x}$, $A_{y}$, $\Theta_{x}$, $\Theta_{y}$, and $\Theta_\mathrm{acc}$ using (\ref{eq:acceptance}) and (\ref{eq:acc}) are listed in table~\ref{tab:results}. The total acceptance angle at the target position amounts to 
\begin{linenomath}
\begin{equation}
 \Theta_\mathrm{acc}=(6.45 \pm 0.27)\,\rm mrad\,.
 \label{eq:thetaa}
\end{equation}
\end{linenomath}
The given uncertainty includes the error of the fit as well as an estimated 10\% uncertainty of the $\beta$-functions.  

The determined horizontal and vertical machine acceptances of $A_{x}=31.2\pm2.5\,$\textmu m and $A_{y}=15.7\pm1.8\,$\textmu m (see table~\ref{tab:results}) are significantly smaller than the simple geometrical acceptances estimated from the standard COSY lattice and the dimensions of the beam pipe (see section~\ref{sec:lowb}). This is the case, because the beam lifetime is likewise impaired by dynamic effects through processes that act on long time scales, caused by nonlinear external fields~\cite{Guignard:185921}.
Therefore, the method discussed here determines the relevant machine acceptance for spin-filtering experiments. 
   
\section{Beam lifetime optimization}
\label{sec:lifetime}
This section describes further machine investigations carried out at COSY aiming at an enlargement of the beam lifetime toward $\tau_\mathrm{b}\approx 10000$\,s, which is necessary to determine the spin-dependent cross section $\tilde{\sigma}_{1}$ of the polarization build-up during a few weeks of beam time. The starting point of the optimization is marked by a beam lifetime of $\tau_{\rm b}=800\,s$, reached in 2007 for an electron-cooled proton beam at injection energy without a target~\cite{Stein:2004}.

Different processes contribute to the beam lifetime, such as betatron resonances, the Coulomb interaction with the residual gas and the target, intrabeam scattering, and hadronic interactions. 
Particle loss due to betatron resonances can be minimized by the choice of a suitable working point, also required for the commissioning of the low-$\beta$ insertion (see section~\ref{sec:tune}).  Coulomb interactions on the target and the residual gas comprise
\begin{itemize}
 \item energy loss, causing particle losses at the longitudinal acceptance,
 \item emittance growth due to multiple small-angle scattering, causing losses at the transverse acceptance, and
 \item immediate loss of ions in a single collision where the scattering angle is larger than the transverse acceptance angle of the machine.
\end{itemize}
Energy loss and emittance growth can to a large extent be compensated by electron cooling (see section~\ref{sec:lattice}).
The beam lifetime due to single Coulomb losses was improved by the closed orbit correction procedure (see section~\ref{sec:orbit}). Investigations of beam lifetime restrictions caused by space-charge effects are discussed here in section~\ref{sec:spacecharge}, while the contributions to beam lifetime from the residual gas in the machine and from the  target  are elucidated in section~\ref{sec:vacuum}.

\subsection{Space-charge effects}
\label{sec:spacecharge}
When a machine is optimized for maximum beam lifetimes, space-charge effects as fundamental collective processes in beams of high intensity usually have to be considered. In the presence of electron cooling, where small emittances are achieved, space-charge effects are, however, already visible at low beam intensities.

Studying space-charge effects and their impact on particle losses implies studying the effect of the beam emittance on the beam lifetime. For a constant beam intensity, the space charge decreases with increasing beam emittance. The beam emittance was manipulated by decreasing the cooling performance of the electron cooler. Both the horizontal and vertical electron beam steerers at the drift solenoids of the cooler were used to tilt the electron beam relative to the proton beam, whereby the cooling force was reduced. 

\begin{figure}[t]
 \centering
 \includegraphics[width=\columnwidth]{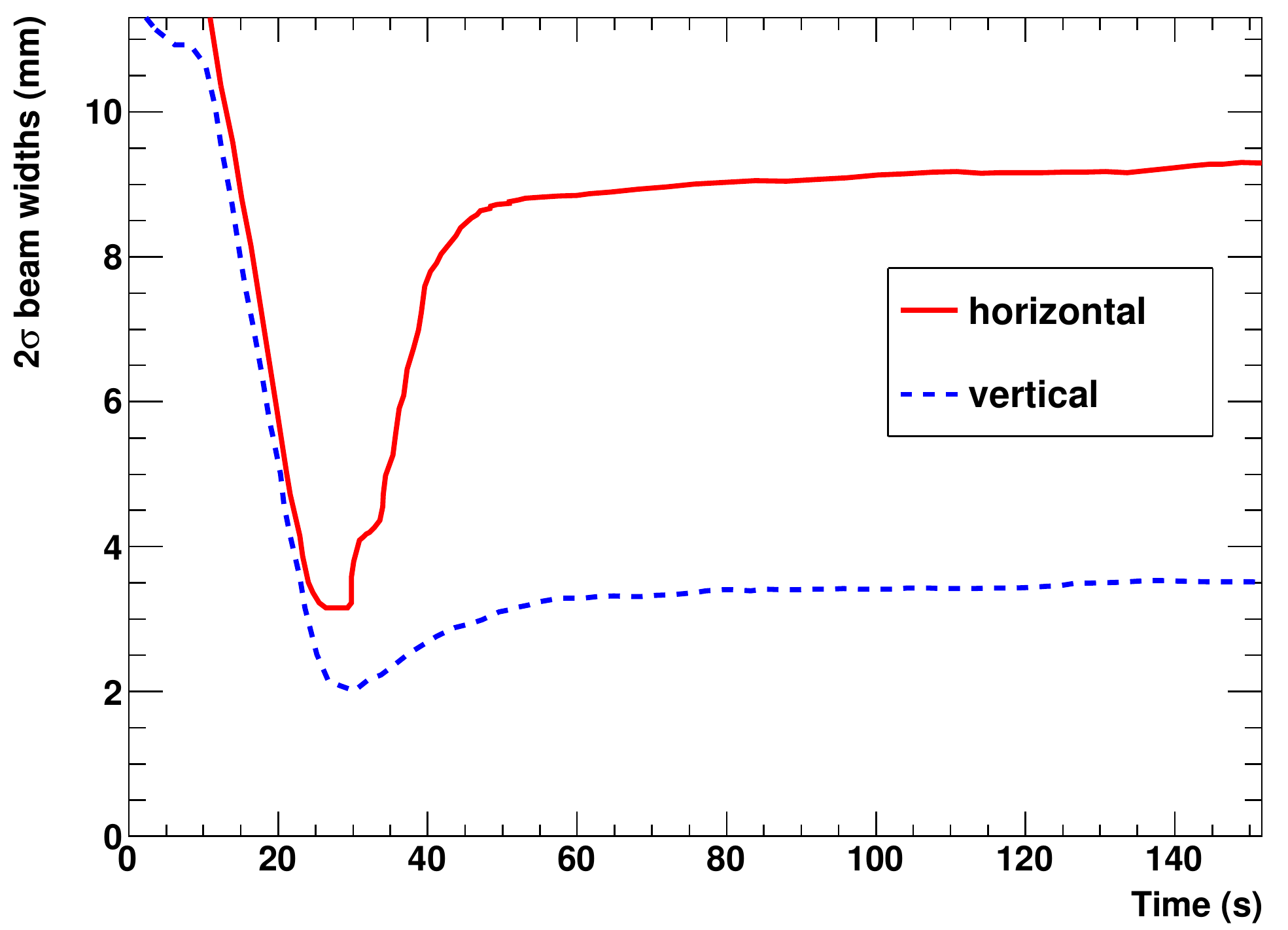}
\caption[]{Widths 2$\sigma$ from fits to beam profiles measured with the ionization profile monitor (IPM) vs the time after injection to COSY ($<25$\,s: electron cooling, $25-50$\,s tilting the electron beam with respect to the proton beam, $>50$\,s increased equilibrium beam size). The corresponding horizontal beam profile is shown in figure~\ref{fig:beamprofile}.}
\label{fig:timedev}
\end{figure}

The beam emittance was determined using the ionization profile monitor (IPM) (section~\ref{sec:IPM}), located in one of the COSY arcs. The detected beam profiles, shown in figure~\ref{fig:beamprofile}, were fitted by a Gaussian, providing the beam widths.
In figure~\ref{fig:timedev} the expansion of the beam size is illustrated. The beam was  completely cooled to widths of about $2\sigma_{x}=3.2\,\rm mm$ (continuous red line) and $2\sigma_{y}=2.0\,\rm mm$ (dashed blue line) and then expanded to a larger equilibrium beam size by tilting the electron beam. 

\begin{figure}[t]
 \centering
 \includegraphics[width=1.05\columnwidth]{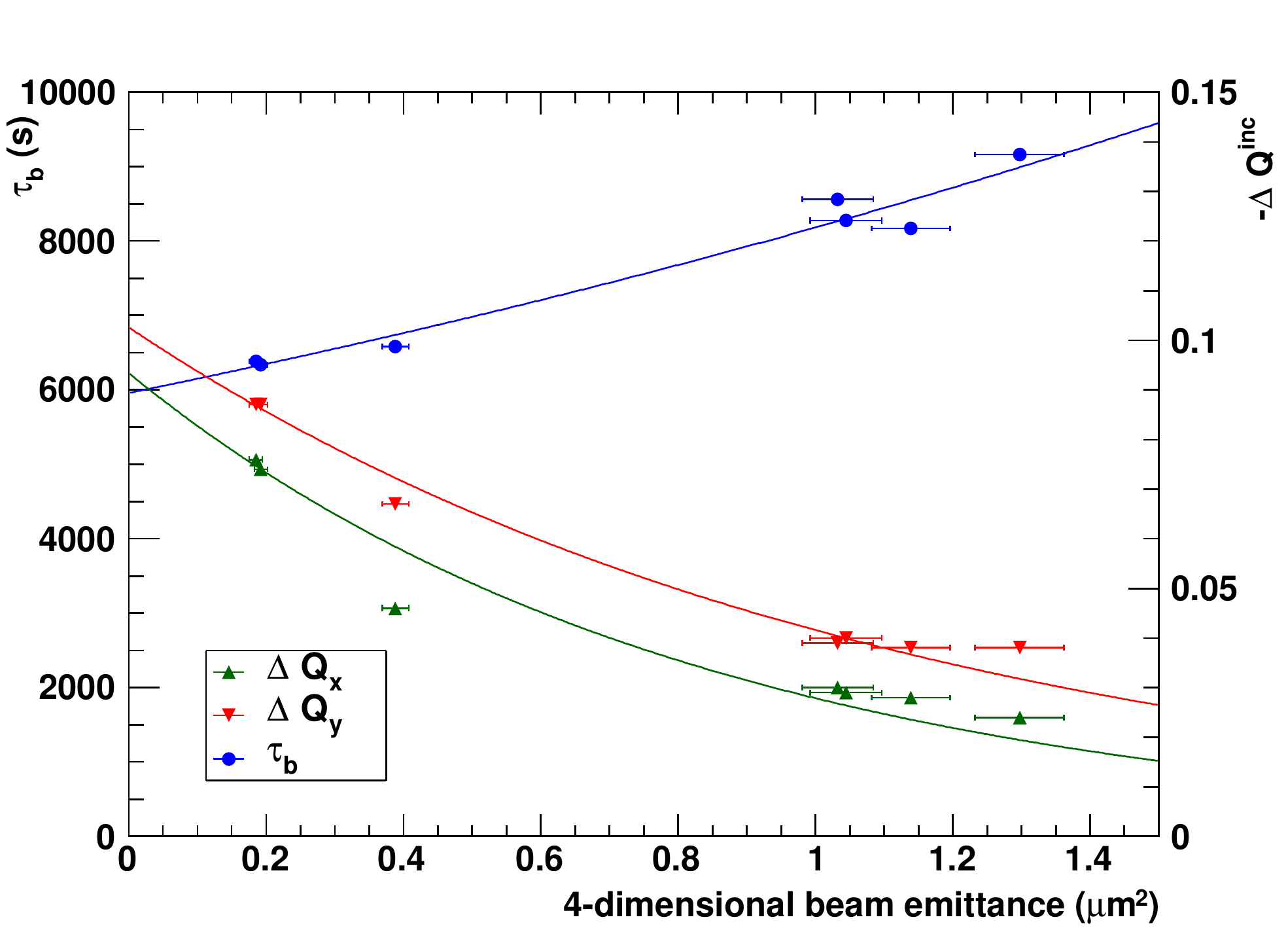}
 \caption[]{Measured beam lifetime $\tau_{\rm b}$ and calculated maximal incoherent tune shift $-\Delta Q^{\rm{inc}}_{x,y}$ from (\ref{eq:tunespread}) as a function of the four-dimensional beam emittance (see (\ref{eq:emittance})). The beam lifetime (blue dots) increases from 6300\,s to 9200\,s with increasing beam emittance. The lines are shown to guide the eye.}
 \label{fig:tauvsemit}
\end{figure}

Using the appropriate $\beta$-functions from the MAD model at the location of the IPM ($\beta_{x}= 12.6\,\rm m$ and $\beta_{y}= 9.6\,\rm m$), allows one to determine the 2$\sigma$ beam emittance $\epsilon_{x}$ and $\epsilon_{y}$ using (\ref{eq:emit}).
The obtained beam lifetimes are plotted in figure~\ref{fig:tauvsemit} (blue symbols) vs the four-dimensional beam emittance~\cite{Buon:1993dr} (see footnote \footnotemark[11]) 
\begin{linenomath}
\begin{equation}
\epsilon=\epsilon_{x}\cdot\epsilon_{y}\, ,
\label{eq:emittance}
\end{equation}
\end{linenomath}
where it should be noted that the actual definition of the combined beam emittance is of minor importance. 

The beam lifetime increased with increasing beam emittance and an improvement from $\tau_\mathrm{b}=6300$\,s to 9200\,s was achieved. 
For emittances $\epsilon > 3\,\text{\textmu m$^{2}$}$, corresponding to electron beam tilt angles of $\geq0.3$\,mrad, the cooling performance was very poor and therefore, two data points were omitted from the analysis.

In the following, we discuss the observed increase of the beam lifetime with increasing beam emittance in terms of tune shifts. The Coulomb force between charged particles in a beam causes repulsion, which leads to defocusing in both transverse planes and therefore to a reduction of the tune $Q$.

For a non-uniform charge distribution, the defocusing space-charge force is not linear with respect to the transverse coordinates. Therefore, each individual particle experiences a different tune shift. This betatron amplitude-dependent detuning, called tune spread, represents a certain area in the tune diagram. Assuming a Gaussian beam distribution, the incoherent tune shifts of the central particles in the beam, \textit{i.e.}, the maximal tune shifts, in the horizontal and vertical phase space are described by \cite{Schindl:2003ej}
\begin{linenomath}
\begin{equation}
\Delta Q^{\rm{inc}}_{x,y}=-\frac{r_{0}N}{\pi\beta_{\rm L}^2\gamma_{\rm L}^3}\cdot\frac{F_{x,y}G_{x,y}}{B_{\rm f}}\frac{1}{\epsilon_{x,y}+\sqrt{\epsilon_x\cdot \epsilon_y}}\hspace{0.15cm}.
\label{eq:tunespread}
\end{equation}
\end{linenomath}
Here $r_{0}$ is the classical proton radius, $N$ is the number of particles in the accelerator, $\beta_{\rm L}$ and $\gamma_{\rm L}$ are the Lorentz factors, $\epsilon_{x,y}$ denote the horizontal and vertical emittances, respectively, and $B_{\rm f}$ is the bunching factor. For an unbunched beam as used in the experiment $B_{\rm f}=1$. $G_{x,y}$ is a form factor depending on the particle distribution inside the beam. Here $G=2$ representing a Gaussian distribution was used.
The form factor $F_{x,y}$, which can be derived from Laslett's image coefficients for incoherent tune shifts~\cite{Laslett:1963zz}, was set to unity because the beam energy is small.

The tune measurement technique at COSY, based on the excitation of coherent transverse oscillations of the beam (see section~\ref{sec:tune}), however, is insensitive to incoherent tune shifts. 
The calculated incoherent tune shift $|\Delta Q^{\rm{inc}}_{x,y}|$ decreases with increasing beam emittance (see (\ref{eq:tunespread})), the associated area in the tune diagram shrinks, fewer betatron resonances are excited, and therefore the measured beam lifetime increases. This theoretical consideration is consistent with the results shown in figure~\ref{fig:tauvsemit}. For the smallest achieved beam emittances of about  $\epsilon=0.2$\,\textmu $m^{2}$, the maximum tune shift amounts to $|\Delta Q^{\rm{inc}}_{x,y}|\approx0.1$, thus with a nominal tune of $Q_{x,y}=3.6$ strong second-order betatron resonances at $Q_{x,y}=3.5$ are intersected.

\subsection{Contributions from  vacuum to  beam lifetime }
\label{sec:vacuum}
 In this section, the different contributions to the beam lifetime from the machine vacuum and the PAX target  are discussed. In order to minimize the beam losses due to the gas load from the ABS in the PAX target chamber and the adjacent up- and downstream sections, a dedicated vacuum system (see section~\ref{sec:Vacuum}) was implemented.

\begin{figure}[t]
\centering
\includegraphics[,width=\columnwidth]{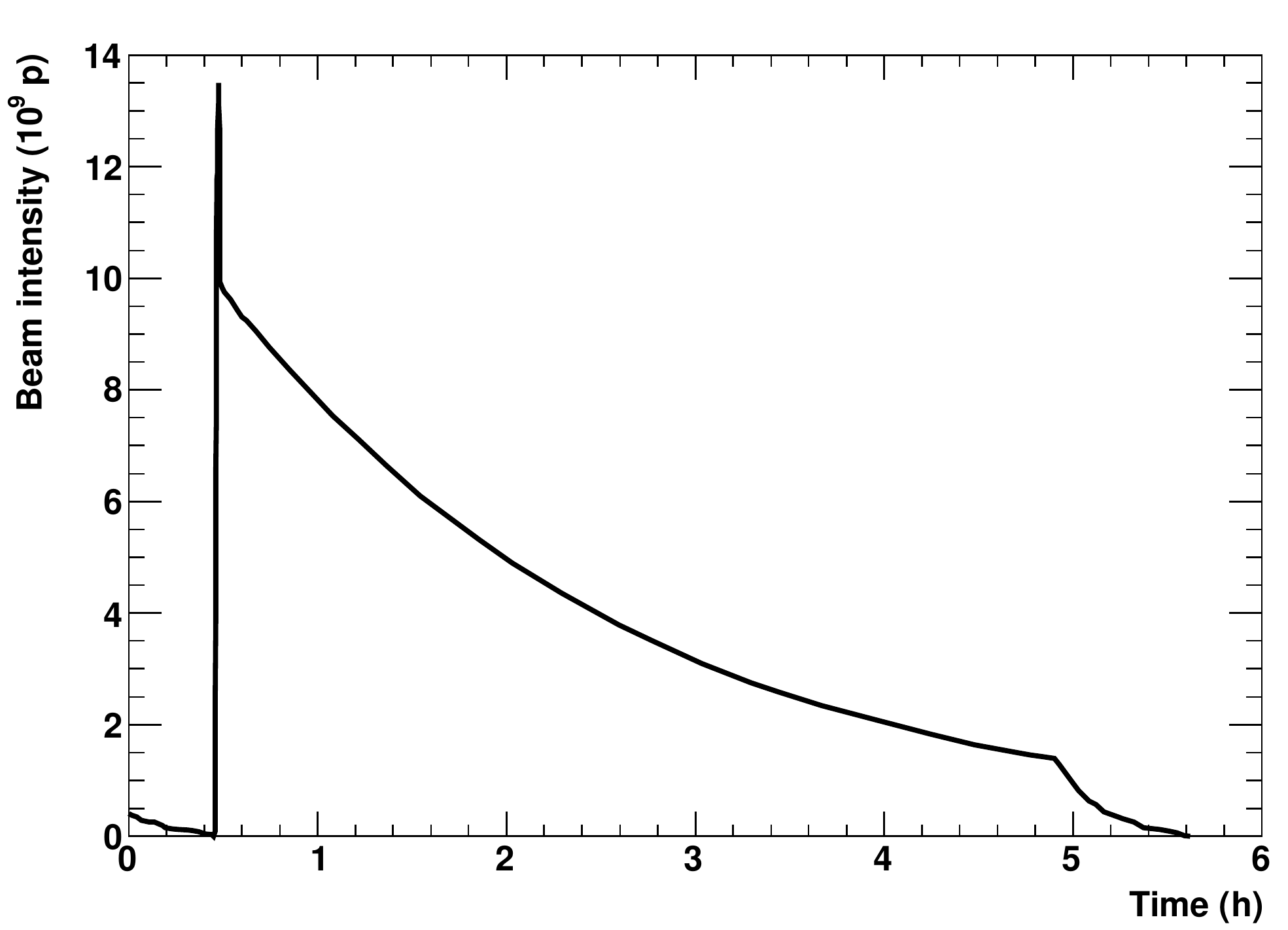}
\caption[]{Beam current as a function of time for a typical  spin filtering run, indicating  a beam lifetime of $\tau_\mathrm{b}=8000$\,s with the polarized hydrogen target switched on. At the end of the cycle, the much denser unpolarized deuterium cluster target is switched on to determine the resulting beam polarization. The total duration of the shown cycle amounts to $\approx 5.5$\,h.}
\label{fig:bct}
\end{figure}

The contributions  to the total beam lifetime can be written as
\begin{linenomath}
\begin{equation}
\frac{1}{\tau_{\rm b}}=
\frac{1}{\tau_{\rm cell}} +
\frac{1}{\tau_{\rm  low\beta}} +
\frac{1}{\tau_{\rm ring}}\, ,
\label{eq:tau_tot}
\end{equation}
\end{linenomath}
where $\tau_{\rm cell}$ denotes the single-scattering losses in the storage cell, $\tau_{{\rm low}\beta}$ those from the gas load elsewhere inside the low-$\beta$ section, and $\tau_{\rm ring}$ is the contribution from the ring,  independent of whether the PAX target was on or off.

After setting up the proton beam at $T_{p}=49.3$\,MeV (described in section~\ref{sec:beamprep}), a maximal beam lifetime of $\tau_{\rm ring}\approx12000$\,s was achieved without gas feed to the storage cell of the target setup. 
When the gas feed was switched on, typical total beam lifetimes of about $\tau_{\rm b}\approx8000$\,s were routinely provided during the spin-filtering experiments (see figure~\ref{fig:bct}).

The beam lifetime from single-scattering losses at the target, caused by those mechanisms that cannot be compensated by electron cooling, \textit{i.e.}, hadronic ($\sigma_{\rm{0}}$) and single Coulomb scattering ($\sigma_{\rm C}$), is given by
\begin{linenomath}
\begin{equation}
\tau_{\rm{cell}} = \left(\left[\sigma_{\rm C} + \sigma_{\rm{0}}\right]\cdot d_{\rm t}f\right)^{-1},
\label{eq:tau_cell}
\end{equation}
\end{linenomath}
 where $f\approx508$\,kHz denotes the revolution frequency. The total hadronic cross section $\sigma_{\rm{0}}=59.8$\,mb was extracted from the SAID database~\cite{said}, and the Coulomb loss cross section (see (\ref{eq:sc})) was determined from the machine acceptance angle at the target, $\Theta_\mathrm{acc}=6.45\pm 0.27$\,mrad (see table~\ref{tab:results}), yielding $\sigma_{\rm C}=677.6$\,mb.
The resulting beam lifetime from (\ref{eq:tau_cell}) yields $\tau_{\rm{cell}}=48500$\,s with $d_{\rm t}=5.5\cdot10^{13}\,$cm$^{-2}$ (see (\ref{eq:targetdensity})).

\begin{table}[t]
\renewcommand\arraystretch{1.5}
\centering
\begin{ruledtabular}
\caption{Contributions to the total beam lifetime of \mbox{$\tau_{\rm b} = 8000$\,s} during the spin-filtering experiments with polarized target with $d_{\rm t}=5.5\cdot10^{13}\,$cm$^{-2}$.}
   \begin{tabular}{p{6cm} c c}
 Losses due to & & \\ \hline
single-scattering in the storage cell & $\tau_{\rm cell}$           & 48500\,s\\
gas load elsewhere in the low-$\beta$ section & $\tau_{{\rm low}\beta}$     & 47500\,s\\
machine vacuum alone & $\tau_{\rm ring}$           & 12000\,s\\
 \end{tabular}
 \label{tab:beamlifetime}
\end{ruledtabular}
\end{table}
The contribution from single-scattering loss outside the cell in the low-$\beta$ section, determined from (\ref{eq:tau_tot}), yields $\tau_{{\rm low}\beta} = ({\tau_{\rm b}}^{-1} - \tau_{\rm ring}^{-1} - {\tau_{\rm cell}}^{-1})^{-1} =  47500$\,s.
 The three contributions to the total beam lifetime are summarized in table~\ref{tab:beamlifetime}. The results show that the total beam lifetime at COSY is dominated by the machine alone, whereas the target region contributes only one third.

\section{Spin filtering at 49.3\,MeV}
\label{sec:beamprep}
The goal of machine development was to provide a routine to set up COSY for the spin-filtering experiments. This routine (see section~\ref{sec:setup}) and the measurement cycles (see section~\ref{sec:meascycle}) are described below. Major requirements for the experiment were beam intensities of about $1\cdot10^{10}$ protons and long beam and polarization lifetimes. Dedicated cycles were set up to measure the beam polarization lifetime (see section~\ref{sec:pollifetime}), and the efficiency of the RF spin flipper (see section~\ref{sec:flipefficiency}) enabling the application of the cross-ratio method~\cite{Han65} within each cycle by reversing the beam polarization.

\subsection{Setting up the beam}
\label{sec:setup}
Based on the investigations described above, the following sequence of steps was applied  to provide long beam and polarization lifetimes together with high beam intensities.

\begin{enumerate}
 \item Setting up injection for protons at $T_{p}=45$\,MeV with standard COSY optics.
 \item Setting up electron cooling, including the transverse feedback system.
 \item First closed orbit correction.
 \item Switching on the low-$\beta$ section by increasing the current in the low-$\beta$ quadrupoles and simultaneously decreasing the current in the quadrupoles of the target telescope, while keeping the machine tunes constant (see section~\ref{sec:lowbeta}).
 \item Optimization of injection and electron cooling with the low-$\beta$ section switched on.
 \item Setting up stacking injection to increase the beam intensity.
 \item Second closed orbit correction.
 \item Setting up acceleration to $T_{p}=49.3$\,MeV with electron beam switched off.
 \item Setting up electron cooling without feedback with reduced electron current ($I_{e}=50$\,mA) for optimal beam lifetime.
 \item Change of machine optics to dispersion $D=0$ setting.
 \item Third closed orbit correction.
 \item Tune adjustment (using only QU1, QU5 and QU2, QU6, figure~\ref{fig:cosy}, to keep $D=0$) to provide optimal beam lifetime without transverse feedback.
 \item Adjustment of electron beam angle with respect to the proton beam to further optimize the beam lifetime (see section~\ref{sec:spacecharge}).
\end{enumerate}

Concerning step 9, it should be noted that it proved more effective to operate the transverse feedback system of the electron cooler (see section~\ref{sec:lattice}) at injection energy rather than at the experiment energy of $49.3$\,MeV. The excitation of coherent betatron oscillations in the beam, due to the lack of transverse feedback at experiment energy, was avoided by selecting a different working point ($Q_x= 3.62$ and $Q_y=3.64$) and by a reduction of the electron current.

\subsection{Measurement cycle}
\label{sec:meascycle}
\begin{figure*}[htb]
 \centering
 \includegraphics[angle=270, width=1\textwidth]{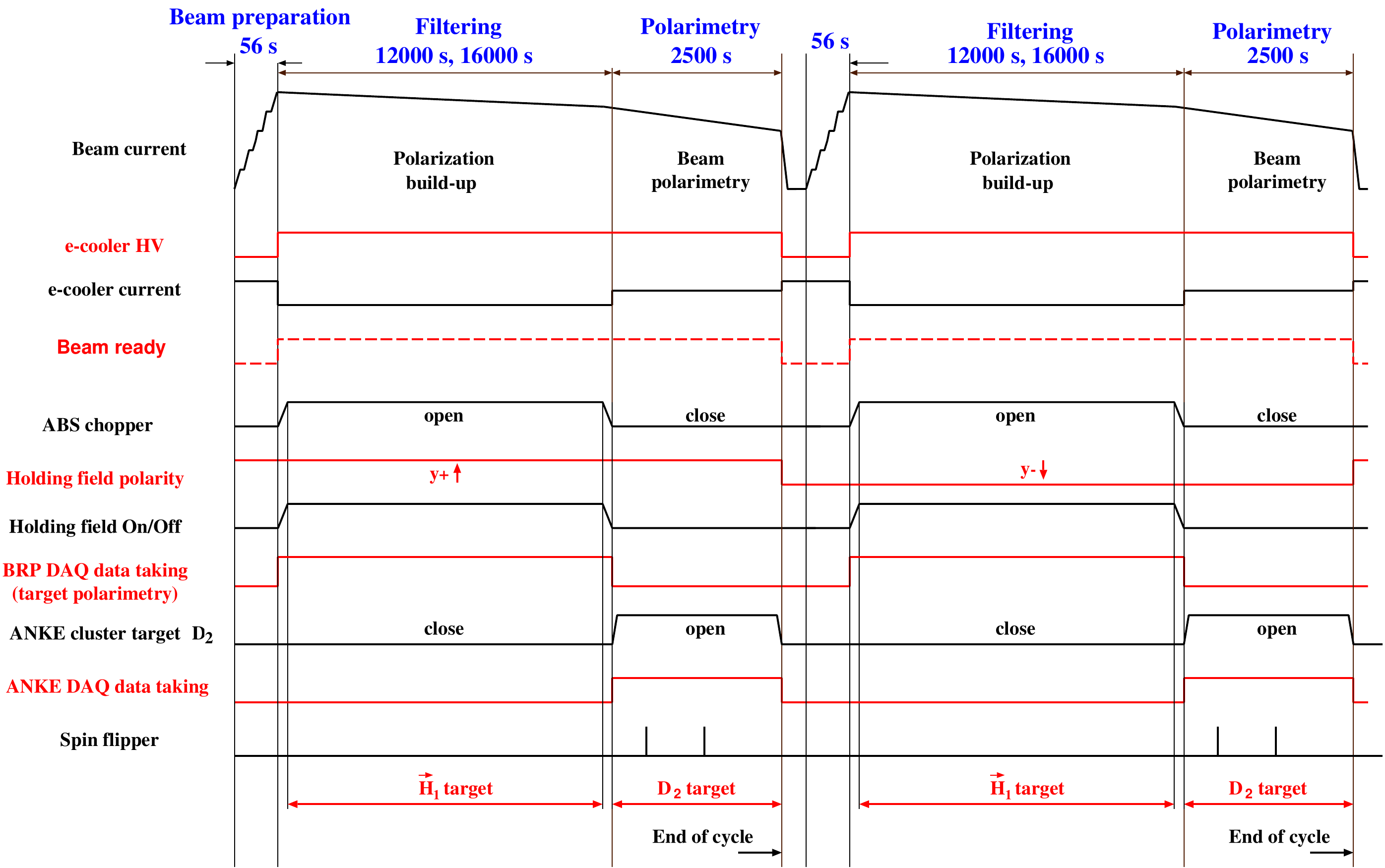}
\caption[]{Sequence of two spin-filtering cycles. During the polarization build-up the polarized internal target and the holding field were switched on, with the ABS chopper open, and the magnetic holding field along the $y$ axis. The BRP was used to measure the target polarization. After spin filtering the ABS chopper was closed and the ANKE cluster target together with the ANKE DAQ were switched on to measure the proton-beam polarization. The electron-cooler current was increased  to compensate for larger energy losses due to the thicker cluster target. During the measurement period the beam polarization was flipped several times to minimize systematic effects. The holding field polarity was reversed after each spin-filtering cycle.}
\label{fig:PAXSequence}
\end{figure*}
After the beam had been set up as described in section~\ref{sec:setup}, a dedicated spin-filtering cycle was implemented according to the sequence shown in figure~\ref{fig:PAXSequence}. 
The relevant figure of merit can be expressed as $\mathrm{FOM}=P^2(t)\cdot I(t)$~\cite{Ohlsen:1973wf} which led to the decision to use filtering periods of 1.5 and 2 beam lifetimes to yield the smallest overall uncertainties for the spin-dependent cross sections~\cite{Rathmann:2004pm}, while at the same time providing the time dependence of the polarization build-up. A cycle with no filtering but otherwise identical settings as in the spin filtering process was set up for systematic studies. 

The spin-filtering cycles were composed of three parts:
\begin{enumerate}
 \item \textbf{Beam preparation} (56\,s) \\
   An unpolarized proton beam was injected at a beam energy of $T_{p}=45$\,MeV, electron cooled ($I_{e}=150$\,mA) and subsequently accelerated to $T_{p}=49.3$\,MeV.
 \item \textbf{Spin-filtering} ($t_\mathrm{filter}= 0$\,s, $12000$\,s, and $16000$\,s)\\
   Polarized hydrogen atoms were injected into the storage cell at the PAX interaction point. The holding field coils, orienting the target polarization, were powered on in either $+y$ (up) or $-y$ (down) orientation for the duration of the spin-filtering period. Three different spin-filtering periods were used: $t_\mathrm{filter}= 12000$\,s and $t_\mathrm{filter}= 16000$\,s, corresponding to about 1.5 and 2 times the measured beam lifetime. For systematic reasons measurements with $t_\mathrm{filter}= 0$\,s were taken as well.
 \item \textbf{Beam polarimetry} (2500\,s)\\
   At the end of the spin-filtering period, the PAX polarized target was switched off, the ANKE deuterium cluster target was switched on, and the data acquisition for the determination of the  beam polarization was started (section~\ref{sec:beampolarimeter}). The current of the electron cooler was increased from $I_{e}=50$\,mA to 100\,mA. Reversing the beam polarization during this period, utilizing the spin flipper~\cite{PhysRevSTAB.7.024002},  allowed one to determine the induced beam polarization within each cycle, thereby reducing systematic errors.
\end{enumerate}

\subsection{Measurement of the polarization lifetime}
\label{sec:pollifetime}
In order to avoid depolarization of the beam during spin filtering, the betatron tunes were set far away from depolarizing resonances~\cite{Lehrach:2000vw}. These arise when the horizontal and vertical tunes, the orbit frequency, and the synchrotron frequency, or combinations thereof, are  synchronous with the spin tune. The  spin tune $\nu_{\rm{s}}$, the number of precessions of the spin vector around the vertical axis per beam revolution in the ring, is defined as
\begin{linenomath}
\begin{equation}
 \nu_{\rm{s}}=\gamma_{\rm L} G\,,
\end{equation}
\end{linenomath}
where $G=1.792847$~\cite{Chao:1999} is the proton anomalous magnetic moment, and $\gamma_{\rm L}$ the Lorentz factor. In a strong focusing synchrotron such as COSY, two different types of first-order spin resonances are excited. \textit{Imperfection resonances} are caused by magnetic field errors and misalignments of the magnets, for which the condition is given by $\gamma_{\rm L} G=k$, with $k\in\mathbb{N}$. \textit{Intrinsic resonances} are excited by horizontal fields due to vertical focusing. For these the condition is given by $\gamma_{\rm L} G=kP\pm Q_y$, where $P$ is the superperiodicity of the lattice, and $Q_y$ the vertical tune. \textit{Higher-order resonances} can depolarize a stored beam as well, when the condition $  \nu_s=k\pm lQ_{x}\pm m Q_{y}$, with $k,l,m\in \mathbb{Z}$ is fulfilled.

\begin{figure}[t]
 \centering
\includegraphics[width=\columnwidth]{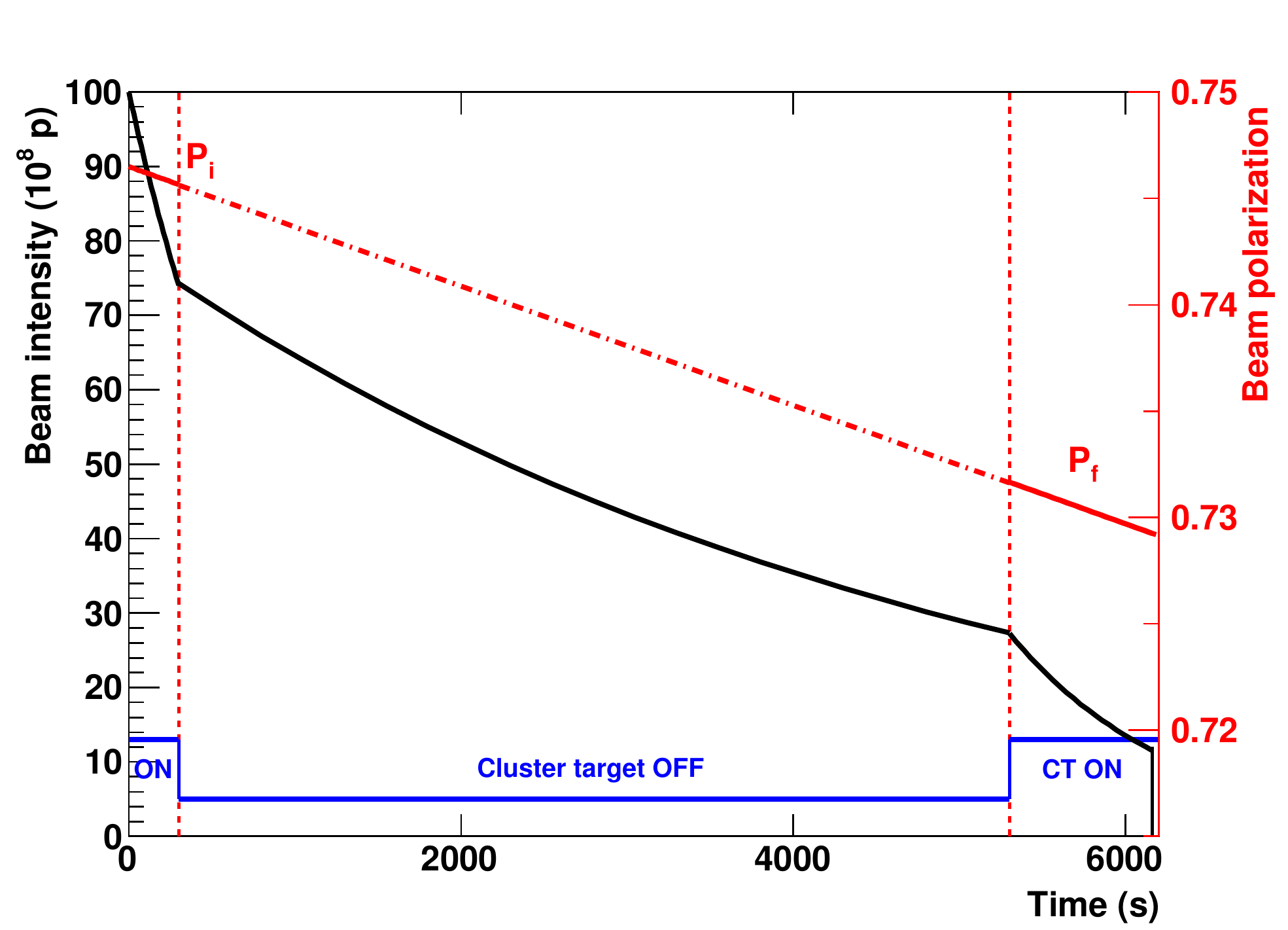}
\caption[]{Measurement cycle to determine the beam polarization lifetime. The beam polarization (red) is measured before ($P_{\rm i}$) and after ($P_{\rm f}$) a waiting period of 5000\,s. During the measurement, the deuterium cluster target (CT) of the ANKE polarimeter is switched on causing a faster decrease of the beam intensity (black). }
 \label{fig:beampol}
\end{figure}

The polarization lifetime $\tau_{\rm P}$ was measured in order to assess its effect on the final beam polarization after spin filtering. Figure~\ref{fig:beampol} shows the cycle setup schematically.
The beam was injected into COSY and accelerated to $T_{p}=49.3$\,MeV in exactly the same way as for the spin-filtering cycle, the only difference being that a polarized beam with $P\approx 0.75$ was injected, provided by the polarized ion source of COSY (see section~\ref{sec:COSY}).
The initial beam polarization $P_{\rm i}$ was determined during a time period of $t_{1}=300$\,s  using the beam polarimeter at the ANKE target place (see section~\ref{sec:beampolarimeter}). Subsequently, the cluster target was switched off for $t_{2}=5000$\,s in order to minimize beam losses. The measurement of the final polarization $P_{\rm f}$ lasted for $t_{3}=940$\,s. The durations of the measurement periods were optimized to yield the smallest relative errors in $\tau_{\rm P}$ and to achieve equal statistical errors of the beam polarization during both sequences. The beam polarization lifetime was determined by evaluating
\begin{linenomath}
\begin{equation}
 \tau_{\rm P}=\frac{-\Delta t}{\ln\left(\frac{\displaystyle P_{\rm f}}{\displaystyle P_{\rm i}}\right)}\,,
\end{equation}
\end{linenomath}
which exploits the exponential decay of the beam polarization as function of time~\cite{PhysRevE.56.3578}. The initial and the final beam polarizations were averaged over the measurement periods $t_{1}$ and $t_{3}$, respectively. Taking these measurement periods into account, the time difference between the two polarization measurements is given by $\Delta t = t_2+t'_1+t'_3=5496$\,s, where $t'_1$ and $t'_3$ account for the exponential decrease of the event rate within each measurement period. With $P_{\rm i}=0.746\pm 0.003$ and $P_{\rm f}=0.731\pm 0.003$~\cite{Bagdasarian:2012}, the determination of the polarization lifetime yielded
\begin{linenomath}
\begin{equation}
\tau_{\rm P}=(2.7\pm0.8)\cdot 10^5\,\mathrm{s}\,.
\end{equation}
\end{linenomath}
Therefore, the polarization losses during the spin-filtering experiments with filter times of $t_\mathrm{filter}=12000$\,s and $16000$\,s did not exceed $6\%$.

\subsection{Efficiency of  RF spin flipper}
\label{sec:flipefficiency}
During the polarization measurement period at the end of each filtering cycle, the beam polarization was flipped several times to enable the beam polarization to be determined within each cycle using the cross-ratio method~\cite{Han65}, whereby systematic errors are cancelled to first order.
\begin{figure}[t]
\centering
\includegraphics[width=\columnwidth]{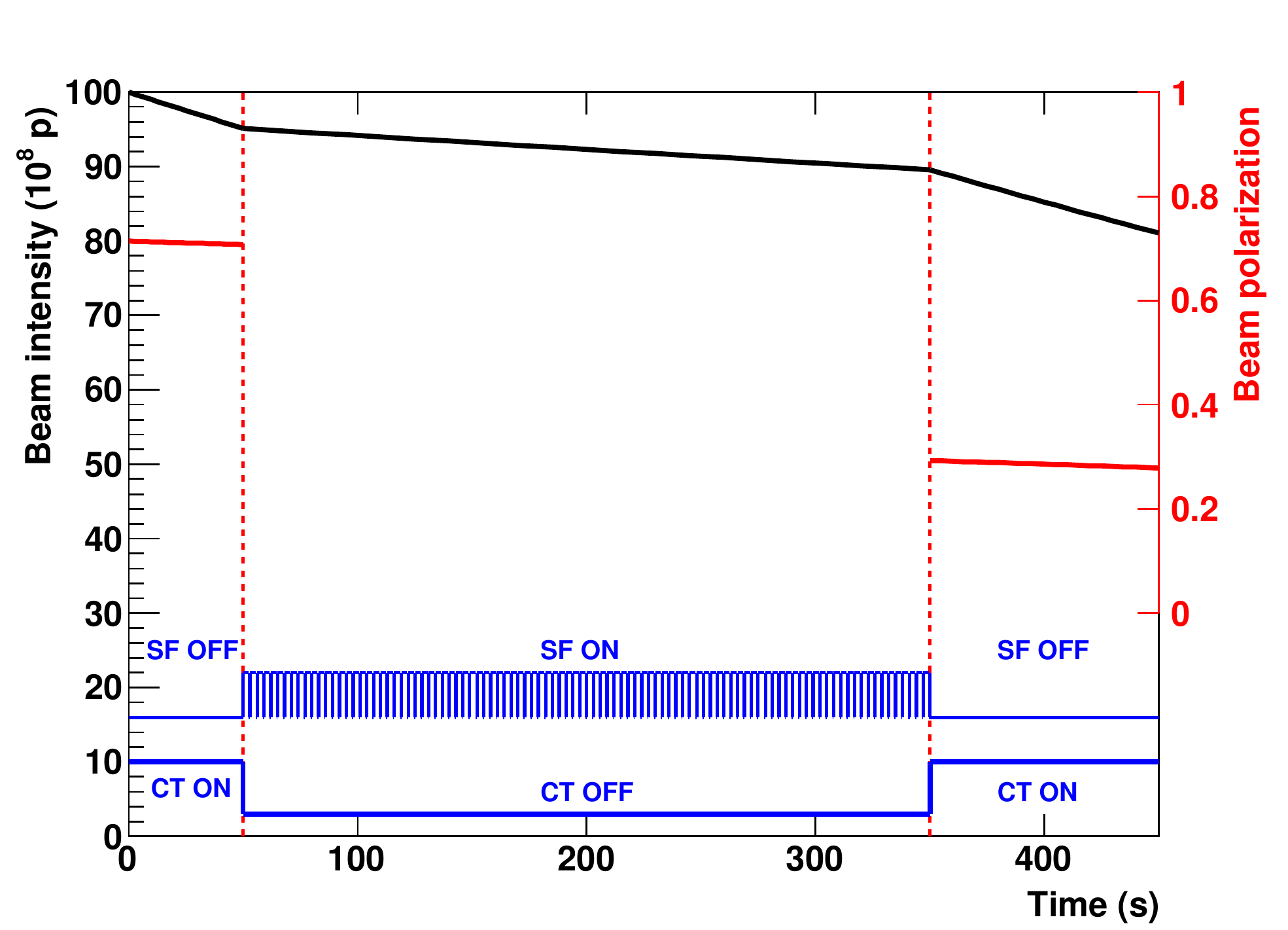}
\caption[]{Measurement cycle to determine the spin-flip (SF) efficiency. The beam polarization (red) is measured before ($P_{\rm i}$) and after ($P_{\rm f}$) $n_\mathrm{flip}=99$ spin flips. While these spin-flips are being executed, the deuterium cluster target (CT) of the ANKE polarimeter is switched off. }
 \label{fig:spinflip}
\end{figure}

The spin flips were generated using a Froissart-Stora frequency sweep induced with an RF solenoid~\cite{PhysRevSTAB.7.024002,Froissart:1960zz}. The RF frequency was swept over the precession frequency of the proton spin and flipped the spin resonantly at the frequency $f_{\rm {RF}}=f_0 \cdot (\gamma_{\rm L}G\pm k)$, which yielded $f_{\rm {RF}}=0.9620\,$MHz for $k=0$. The frequency ramp from $0.9605$\,MHz to $0.9625$\,MHz was carried out in $2.5\,\rm s$, therefore the effect on the duty cycle was negligible.

The spin flip efficiency,
\begin{linenomath}
\begin{equation}
\varepsilon_\mathrm{flip}=\exp\left(\frac{\ln\left(\frac{\displaystyle \mathopen|P_{\rm f}\mathclose|}{\displaystyle \mathopen|P_{\rm i}\mathclose|}\right)}{n_\mathrm{flip}}\right)\,,
\end{equation}
\end{linenomath}
where $P_{\rm i,f}$ are the initial and final polarizations, was determined in order to be able to correct for polarization losses and to adjust the number of flips $n_\mathrm{flip}$ within the measurement period. Since the change of sign of the polarization after each flip is not relevant for the presented analysis, the absolute values of $P_{\rm i,f}$ are used to determine the flip efficiency. 
 
A dedicated cycle was used to measure the efficiency of the spin flipper during commissioning and regularly during the experiment. This cycle, shown schematically in Fig.~\ref{fig:spinflip}, yielded the smallest relative error in $\varepsilon_\mathrm{flip}$. It begins with the injection and acceleration of a polarized proton beam ($P\approx 0.75$) to $T_{p}=49.3$\,MeV, followed by a polarization measurement lasting for about $50\,\rm s$. Subsequently, the cluster target was switched off and $n_\mathrm{flip}=99$ spin flips were performed within a time period of $300\,\rm s$. An odd number of spin flips was chosen to guarantee the presence of spin flips by observing a sign reversal of the final polarization. Finally, the beam polarization was measured again for about $100\,\rm s$. 

The three measurements of $\varepsilon_\mathrm{flip}$ during the experiment period, each lasting for about two hours, yielded in average
\begin{linenomath}
\begin{equation}
\varepsilon_\mathrm{flip}=0.9872\pm 0.0001\,.
\end{equation}
\end{linenomath}
During the initial spin-filtering measurements there were two spin flips. Thus, the polarization loss due to the spin-flipper never exceeded 3\%.

\section{Conclusion}
\label{sec:conclusion}
In this paper, we present the machine development for the spin-filtering experiments carried out at COSY~\cite{Augustyniak:2012vu}. The prime objective was to provide a long beam lifetime in the presence of a polarized hydrogen gas target. To this end, a dedicated low-$\beta$ section consisting of two quadrupole doublets was implemented at the PAX target place. The optimization of the beam lifetime included the search for optimal working points, closed orbit corrections, optimization of electron cooling, and the minimization of the $\beta$-functions at the PAX target.  

The low-$\beta$ insertion led to $\beta$-functions of $(\beta_x,\beta_y)=(0.31\pm 0.03\,\mathrm{m},0.46\pm 0.05\,\mathrm{m})$ at the center of the polarized storage cell target, resulting in a reduction of about a factor of ten compared to the situation before. Hence, single Coulomb scattering as the dominating loss mechanism for cooled beams was reduced by the same factor. In addition, this allowed us to use a narrow storage cell of diameter $d=9.6\,$mm and length $l=400$\,mm with an areal target-gas density of $d_{\rm t}=(5.5\pm 0.2)\cdot10^{13} \mathrm{atoms/cm^{2}}$. 

Special attention was given to the vacuum conditions in and around the target chamber through the installation of a sophisticated pumping system together with flow limiters at the entrance and exit of the chamber. The beam lifetime caused by the target region with an injected gas flow of $3.3\cdot10^{16}$ $\rm \vec{H}$/s contributed only one third to the total beam lifetime of $\tau_{\rm{b}}=8000$\,s, while the contribution of the machine itself was twice as large.

The machine acceptances, the beam widths, and the machine acceptance angle at the target were determined using a dedicated movable frame system, yielding \mbox{$A_x=(31.2 \pm 2.5)$\,\textmu m}, \mbox{$A_y=(15.7 \pm 1.8)$\,\textmu m}, \mbox{$(2\sigma_x,2\sigma_y)=(1.03\pm0.01\,\mathrm{mm},0.67\pm0.02\,\mathrm{mm})$}, and \mbox{$\Theta_\mathrm{acc}=(6.45\pm 0.27)$\,\rm mrad}. With the achieved $\beta$-functions, the  horizontal and vertical $2\sigma$ beam emittances were $\epsilon_{x}=(1.71\pm0.17)\,\text{\textmu m}$ and $\epsilon_{y}=(0.92\pm0.15)\,\text{\textmu m}$, respectively.

In order to improve the systematics of the spin-filtering experiment, an RF spin flipper was utilized to reverse the polarization of the stored beam after spin filtering. The  spin-flip efficiency determined in dedicated cycles  amounted to $\varepsilon_\mathrm{flip}=0.9872\pm 0.0001$, and the polarization loss due to the spin-flipper never exceeded 3\%. In addition, the polarization lifetime was determined in dedicated cycles, yielding $\tau_{\rm P}=(2.7\pm0.8)\cdot 10^5$\,s. Thus for the spin-filtering experiments at COSY with spin-filtering times of $t_\mathrm{filter}=12000$\,s and  $16000$\,s, the polarization loss due to a finite polarization lifetime did not exceed $6\%$.

The interplay of the  investigations presented in this paper fulfilled the demanding beam conditions for the first spin-filtering experiment at COSY. The presented results comprise a recipe for setting up a beam for spin-filtering experiments in a storage ring, directly applicable to the anticipated spin-filtering studies with antiprotons at the AD of CERN~\cite{AD_2009}.\\
\\

\begin{acknowledgments}
We are grateful to the COSY crew for their continuous support over several years and especially for providing good working conditions during the experiments.
We would like to acknowledge the invaluable help from the workshops and personnel of the Institut f\"ur Kernphysik (in particular J.~Sarkadi), the Zentralinstitut f\"ur Systeme der Elektronik (ZEA-2) and Zentralinstitut f\"ur Engineering und Technologie (ZEA-1) of Forschungszentrum J\"ülich (especially  
H.~Jagdfeld and H.~Straatmann), and by the workshop of the University of Ferrara (in particular V.~Carassiti)

Our thanks also go to the colleagues from CERN (P.~Chiggiato, J.~Hansen, P.~C.~Pinto, and I.~Wevers) for the test measurements and discussions concerning the vacuum design and the NEG coating.

The present work is supported by the ERC Advanced Grant POLPBAR (Grant Agreement 246980), the EU grants of Projects I3HP2 and I3HP3 (Grant Agreements 227431 and 283286), the COSY-FFE program (41853505-FAIR-009), and the Swedish Research Council (No. 624-2009-224 and No. 624-2010-5135).  

\end{acknowledgments}

\bibliographystyle{apsrev4-1}
\bibliography{newliteratur}

\end{document}